\documentclass[12pt, a4paper, reqno]{amsart}
\usepackage{amsmath, amsthm, amssymb, mathtools, microtype}
\usepackage[a4paper,margin=3cm]{geometry}
\usepackage[utf8]{inputenc}

\usepackage[hidelinks]{hyperref}
\usepackage[mathscr]{euscript}

\numberwithin{equation}{section}

\theoremstyle{plain}
\newtheorem{theorem}{Theorem}
\newtheorem{proposition}[theorem]{Proposition}
\newtheorem{lemma}[theorem]{Lemma}
\newtheorem{corollary}[theorem]{Corollary}
\theoremstyle{definition}

\newtheorem{definition}[theorem]{Definition}
\theoremstyle{remark}
\newtheorem{remark}[theorem]{Remark}

\newtheoremstyle{note}
{\bigskipamount}
{\bigskipamount}
{\footnotesize}
{}
{\footnotesize\itshape}
{.}
{.5em}
{}
\theoremstyle{note}
\newtheorem{digression}[theorem]{Digression} 

\DeclareMathOperator*{\res}{Res}

\DeclareMathOperator{\im}{Im}
\def\Z{\mathbb{Z}}	
\def\C{\mathbb{C}}	
\def\R{\mathbb{R}}	
\renewcommand{\leq}{\leqslant} 		
\renewcommand{\geq}{\geqslant}

\def\cA{\mathcal{A}}
\def\cD{\mathcal{D}}
\def\cL{\mathcal{L}}
\def\cN{\mathcal{N}}
\def\cS{\mathcal{S}}
\def\cT{\mathcal{T}}
\def\cV{\mathcal{V}}
\def\cW{\mathcal{W}}
\def\cZ{\mathcal{Z}}
\def\cf{\mathfrak{f}}
\def\KPt{\mathsf{t}}
\def\KPp{\mathsf{p}}
\def\cm{\mathfrak{m}}
\def\cs{\mathfrak{s}}
\def\cg{\mathfrak{g}}
\def\ch{\mathfrak{h}}
\def\ein{\mathrm{Ein}}
\def\cU{\mathcal{U}}
\def\cP{\mathcal{P}}
\def\I{\mathsf{i}}
\def\hS{\hat{S}}
\def\cR{\hat{R}}
\DeclareMathOperator{\End}{End}
\def\be{\mathbf{e}}
\def\re{\mathrm{Re}}

\begin{document}

\title[Hirota equations for extended nonlinear Schr\"odinger hierarchy]{Higher genera Catalan numbers and Hirota equations for extended nonlinear Schr\"odinger hierarchy}

\author{G. Carlet *}
\address{G.~C.: Institut de Mathématiques de Bourgogne, UMR 5584 CNRS, Université Bourgogne Franche-Comté, F-2100 Dijon, France}
\email{Guido.Carlet@u-bourgogne.fr}

\author{J. van de Leur}
\address{J.~v.d.L: Mathematical Institute, University of Utrecht, P.O. Box 80010, 3508 TA Utrecht, The Netherlands}
\email{J.W.vandeLeur@uu.nl}

\author{H. Posthuma}
\address{H.~P.: Korteweg-de Vries Institute for Mathematics, University of Amsterdam, Postbus 94248, 1090 GE Amsterdam, The Netherlands}
\email{H.B.Posthuma@uva.nl}	

\author{S. Shadrin}
\address{S.~S.: Korteweg-de Vries Institute for Mathematics, University of Amsterdam, Postbus 94248, 1090 GE Amsterdam, The Netherlands}
\email{S.Shadrin@uva.nl}	

\begin{abstract} We consider the Dubrovin--Frobenius manifold of rank $2$ whose genus expansion at a special point controls the enumeration of a higher genera generalization of the Catalan numbers, or, equivalently, the enumeration of maps on surfaces, ribbon graphs, Grothendieck's dessins d'enfants, strictly monotone Hurwitz numbers, or lattice points in the moduli spaces of curves. Liu, Zhang, and Zhou conjectured that the full partition function  of this Dubrovin--Frobenius manifold is a tau-function of the extended nonlinear Schr\"odinger hierarchy, 
an extension of a particular rational reduction of the  Kadomtsev--Petviashvili hierarchy. We prove a version of their conjecture specializing the Givental--Milanov method that allows to construct the Hirota quadratic equations for the partition function, and then deriving from them the Lax representation. 
\end{abstract}

\dedicatory{Dedicated to the memory of Boris Dubrovin}

\maketitle

\tableofcontents

\section*{Introduction}

This paper is devoted to the study of the integrable hierarchy associated with the Dubrovin--Frobenius manifold of rank $2$ given in the flat coordinates $t^1,t^2$ by 
\begin{align}
\label{eq:CatalanFrobeniusManifold1}
& \text{the metric }\eta_{\alpha\beta} = \delta_{\alpha+\beta, 3},  
\\ \label{eq:CatalanFrobeniusManifold2}
& \text{the prepotential } \textstyle F(t^1,t^2)=\frac 12 (t^1)^2 t^2 + \frac 12 (t^2)^2 \log t^2, \\
& \text{the unit vector field } e = \partial_{t^1}, 
\\ 
\label{eq:CatalanFrobeniusManifold3}
& \text{and the Euler vector field }E=t^1 \partial_{t^1} + 2 t^2 \partial_{t^2},
\end{align}
first introduced in~\cite[Example 1.1, Equation (1.24b)]{dub96}. Despite the fact that it is one of the first non-trivial examples of semi-simple Dubrovin--Frobenius manifolds, it was until recently not well-studied in the literature, since the enumerative meaning of its \emph{genus expansion} was unclear. 

\begin{digression}[on genus expansion] \label{dig:GenusExpansion}
In many examples, a Dubrovin--Frobenius manifold captures the primary genus $0$ part of the Gromov-Witten partition function of some target variety, or, more generally, the partition function of some naturally constructed cohomological field theory. 
In these cases, the enumerative meaning of the genus expansion is encoded in the all-genera descendent partition function. 
For instance, the rank~$2$ Dubrovin--Frobenius manifolds given in flat coordinates by the same metric as above and the prepotentials $\frac 12 (t^1)^2 t^2 + \frac 1{72} (t^2)^4$ and $\frac 12 (t^1)^2 t^2 + e^{t^2}$ (see again~\cite[Example 1.1]{dub96}) are related to the Witten $3$-spin class and the Gromov-Witten theory of $\C\mathbb{P}^1$, respectively, and there is an extensive literature studying these examples.

In general, there is a universal reconstruction procedure for the genus expansion of a semi-simple Dubrovin--Frobenius manifold. It can be given either by the universal Givental formula~\cite{Givental2001}, or, alternatively, as the tau-function that linearizes a special system of symmetries called the Virasoro constraints~\cite{DubrovinZhang2001}. Equivalence of these two approaches is proved in~\cite{DubrovinZhang2001}. This tau-function determines the tau-structure of a bi-Hamiltonian dispersive deformation of an integrable hierarchy of hydrodynamic type associated with the initial semi-simple Dubrovin--Frobenius manifold. 

Note that the construction of this hierarchy given in~\cite{DubrovinZhang2001} doesn't guarantee the regular (polynomial) dependence of the Poisson brackets and the densities of the Hamiltonians. The regularity of the first Poisson bracket and the densities of the Hamiltonians is proved in~\cite{BPS2012-1,BPS2012-2}, while the polynomiality of the second Poisson bracket is still an important open problem.
\end{digression}

Quite recently it was proved in~\cite{DNOPS2019,AnCheNoPe} that there exist a specialization of the logarithm of the partition function $\cD=\cD(\{t^i_d\}_{i=1,2;\, d\geq 0})$ associated with the Dubrovin--Frobenius manifold~\eqref{eq:CatalanFrobeniusManifold1}-\eqref{eq:CatalanFrobeniusManifold3} (say, consider the partition function given by the Givental formula) that  is the generating function of the \emph{generalized Catalan numbers} (see a table of their values in~\cite[Part III, Section 1.1]{AMM2005}) weighted by combinatorial factors. Note that since the underlying Dubrovin--Frobenius manifold has a singularity at $t^1=t^2=0$, we have to choose another reference point for the formal expansion of $\cD$ instead of the origin, and our choice throughout the paper is $t^1=0, t^2=1$. 

The generalized Catalan numbers (in some instances, up to some small combinatorial rescaling) are also studied under the names of strictly monotone Hurwitz numbers, enumerations of ribbon graphs, (rooted) maps on surfaces, Grothendieck's dessins d'enfants for strict Belyi functions, lattice points in the moduli spaces of curves, et cetera, see e.g.~\cite{GouldenJackson,Norbury,DumMul} for some references to the vast literature on this subject. 
This motivates us to have a closer look at this example of Dubrovin--Frobenius manifold.  Note that though it is known that there exists a specialization of $\cD$ to a generating function of generalized Catalan numbers, its explicit form is not available in the literature, so we give it below, see Theorem~\ref{thm:SpecializationOfD}.

\begin{digression}[on generalized Catalan numbers] \label{dig:Catalan} The generalized Catalan numbers enumerate graphs with $n\geq 1$ ordered vertices, connected by edges, with a fixed cyclic order of half-edges attached to each vertex, and with one distinguished half-edge at each vertex. For each such graph there is a unique, up to a homeomorphism, surface of genus $g\geq 0$, where this graph can be embedded such that its complement is a union of open disks. We call $g$ the genus of the graph. 
	
By $C_{g;k_1,\dots,k_n}$ we denote the number of such graphs of genus $g$ with $n$ vertices of indices $k_1,\dots,k_n$. 
In a dual language we can say that $C_{g;k_1,\dots,k_n}$ counts the number of ways (up to orientation preserving homeomorphisms) to glue a genus $g$ surface out of $n$ ordered polygons with $k_1,\dots,k_n$ sides, respectively, by identifying the pairs of sides, where each polygon has one distinguished side (these are the rooted maps). Obviously, for $g=0$ and $n=1$ $C_{0,k}$ is not equal to $0$ if and only if $k=2m$ is even, and in this case it is equal to the $m$-th Catalan number.  

A closely related concept is $D_{g;k_1,\dots,k_n}\coloneqq (k_1\cdots k_n)^{-1}\cdot C_{g;k_1,\dots,k_n}$. These numbers can be defined via enumeration of ribbon graphs, where each graph is counted with the weight equal to the inverse order of its automorphism group, or (not rooted) maps on surfaces, lattice points in the moduli spaces, and strictly monotone Hurwitz numbers / Grothendieck's dessins d'enfants for strict Belyi functions.  

Recently both $C_{g;k_1,\dots,k_n}$ and $D_{g;k_1,\dots,k_n}$ got a lot of attention since their generating functions serve as the basic examples for the Chekhov--Eynard--Orantin topological recursion and hypergeometric tau-function of the KP hierarchy. In particular, their relation to the Dubrovin--Frobenius manifold~\eqref{eq:CatalanFrobeniusManifold1}-\eqref{eq:CatalanFrobeniusManifold3} is a byproduct of their study in the context of topological recursion~\cite{DNOPS2019,AnCheNoPe}. 
\end{digression}

Consider the partition function $\cD$. The goal of this paper is to construct an integrable hierarchy for which this partition function would be a tau-function corresponding to the string solution. We prove that $\cD$ is a tau-function of the extended non-linear Schr\"odinger or AKNS~\cite{AKNS}  hierarchy defined in~\cite[Section 5]{cdz}, which can also be considered as an extension of a particular rationally reduced  KP or constrained KP hierarchy, see~\cite{BonoraXiong,Cheng,ChengLi,KonoSS,OS,Krichever,HelminckVdL,LiuZhangZhou} and references therein.

\begin{digression}[on special integrability] \label{dig:specialInt} It is now well known that the exponential of the generating function of the numbers $C _{g;k_1,\dots,k_n}$ is a tau function of the KP hierarchy~\cite{GouldenJackson}. It is an example of the so-called hypergeometric tau-function, let us denote it by $\cZ=\cZ(\{\KPt_d\}_{d\geq 1})=\cD|_{{t^1_d = (d+1)!\KPt_{d+1}, t^2_d=0,\, d\geq 0}}$. It is a natural general open question for the hypergeometric tau-functions what kind of further reduction of KP or lattice KP they would still satisfy. To this end, in the case of lattice KP a number of interesting examples is systematically studied in~\cite{Taka}. 
	
From that point of view, we do here one more step. Namely, we first start with $\cZ$, which has a very clear combinatorial enumerative meaning, and identify the reduction of KP for $\cZ$ as a special rational reduction of the KP hierarchy also known as the non-linear Schr\"odinger hierarchy~\cite{cdz} or AKNS hierarchy~\cite{AKNS}. 

It is known from~\cite{cdz} that in addition to the standard set of Hamiltonians generating the flows of $\partial/\partial t^1_d$, there is a an additional set of commuting Hamiltonians. The corresponding flows extend the tau-function $\cZ$, and this extension is identified with with $\cD$, where $\partial/\partial t^2_d$ are the flows of this additional set of Hamiltonians.
\end{digression}

Our construction consists of three big steps loaned from the existing literature and modified to fit our needs. First, we use the techniques of Givental, Milanov, Tseng, et al.~\cite{GiventalAn,GivenMil,m,mt08} to construct the Hirota quadratic equations for the partition function $Z$. 
Our general philosophy is to avoid using the superpotential of the Dubrovin--Frobenius manifold~\eqref{eq:CatalanFrobeniusManifold1}-\eqref{eq:CatalanFrobeniusManifold3}. 
The reason to proceed in this way is our intention to use this example as a departure point for the development of general structures producing Hirota equations and intrinsically existing for a Dubrovin--Frobenius manifold. Our construction of the periods relies on the well-known Proposition~\ref{proposition14}.
The asymptotic expansion at $\lambda \sim \infty$ in Proposition~\ref{fSf} is derived as a consequence. 
A more intrinsic general approach is given in~\cite{m-period}, where such asymptotic expansion, which is actually convergent, is used to define the periods. 

Second, we use the well-known method developed e.g. in~\cite{m, cvdl13} to pass from the Hirota equations to the Lax representation. Notice that the Hirota equations coincide with those of the extended Toda hierarchy, but with the primary times interchanged. The natural approach would be to follow the usual approach for KP reductions, but it is not clear how to apply the fundamental lemma in terms of pseudo-differential operators to obtain the Sato equations for the additional set of times. 
We therefore first recall (a particular case) of the computation in~\cite{cvdl13}  obtaining the Lax equations of the extended Toda hierarchy in the ``unnatural'' spatial variable.
In the third step we revisit in terms of the dressing operators the construction of~\cite{cdz} that allows to perform a change of the time corresponding to the spatial $x$-variable in order to re-organize the resulting hierarchy into the extended nonlinear Schr\"odinger hierarchy. 
In the last subsection we finally propose a direct but somehow not standard derivation of the pseudo-differential Sato equations from the Hirota quadratic equations.

This result, though quite non-trivial, is very much expected. Indeed, on the one hand it is already mentioned in~\cite{cdz} that on the level of the underlying Dubrovin--Frobenius manifolds the change of the time shifted by $x$ that turns the extended Toda hierarchy into the extended nonlinear Schr\" odinger hierarchy is reduced to a Legendre-type transformation that turns the Dubrovin--Frobenius manifold structure of the Gromov-Witten invariants of $\C\mathbb{P}^1$ into the one given by Equations~\eqref{eq:CatalanFrobeniusManifold1}-\eqref{eq:CatalanFrobeniusManifold3}. On the other hand, this result can be considered as the very first example that indirectly affirms a much more general conjecture of Liu, Zhang, and Zhou in~\cite{LiuZhangZhou}, which they posed for a different reason. This paper studies their extended 1-constrained KP case. The $n$-constrained case will be addressed in a forthcoming publication.

\begin{digression}[on Liu--Zhang--Zhou conjecture] The way Liu, Zhang, and Zhou arrive to their conjecture is quite different and very interesting. They study the so-called \emph{central invariants}~\cite{LiuZhang,dlz,cks} of the bi-Hamiltonian structures of a special class of the rationally constrained KP hierarchies and they show that all central invariants are constants equal to $1/24$. This is exactly the property that the bi-Hamiltonian structures of the Dubrovin--Zhang hierarchies associated to Dubrovin--Frobenius manifolds must have (as we mentioned above, the existence of the second bracket is an open conjecture, but for the definition of the central invariants one needs only its existence up to order $2$ in the dispersion parameter $\epsilon$, which is proved in~\cite{DubZha1Loop}). There is a direct relation between the dispersionless limits of this special class of hierarchies and the principal hierarchies of a particular family of Dubrovin--Frobenius manifolds (one for each rank $r\geq 2$). Based on this relation, Liu, Zhang, and Zhou conjecture that the Dubrovin--Zhang topological deformation of the corresponding principal hierarchies gives extensions of the aforementioned special class of the rationally contrained KP hierarchies.

Note that our result (that concerns the rank $2$ case in the setup of Liu--Zhang--Zhou) does confirm their conjecture but it does not fully resolve it in this case.  Indeed, we construct an integrable hierarchy for the genus expansion of the Dubrovin--Frobenius manifold~\eqref{eq:CatalanFrobeniusManifold1}-\eqref{eq:CatalanFrobeniusManifold3} going through the Hirota bilinear equations of Givental--Milanov. But there is no known identification of this approach and the approach of Dubrovin--Zhang via the topological deformation of the principle hierarchy (see~\cite{clps} for some first steps in that direction). Thus, though we obtain exactly the hierarchy that Liu, Zhang, and Zhou expect, it is not a proof of their conjecture in this case.
\end{digression}

In conclusion, let us mention that we believe that a detailed study of this example of Dubrovin--Frobenius manifold in the way we performed it is  quite helpful in the view of its possible generalizations (and a revision of similar results available in the literature). Indeed, we paid a special attention to specifying the convergence issues and emerging choices (for instance, calibration, choices of roots, etc.), which are quite often jammed in the literature though their effect on the resulting formulas is quite essential, as one can see from the detailed analysis in this paper.

\subsection*{Organization of the paper}

In Section~\ref{sec:FrobeniusManifolds} we introduce the Dubrovin--Frobenius manifold that we study in this paper and recall all essential structures related to it.  In Section~\ref{sec:PrincipleHierarchy} we study the structure of the principle hierarchy associated to this Dubrovin--Frobenius manifold, with a special attention to the possible choice of calibration. In Section~\ref{sec:GiventalQuantizationFormalism} we recall the Givental quantization formalism and define the all-genera partition functions (the ancestor potential and the descendent potential) associated to our Dubrovin--Frobenius manifold. In Section~\ref{sec:PeriodVectors} we study in detail the period vectors of our Dubrovin--Frobenius manifold, their values at a special point and their asymptotics. In Section~\ref{sec:VertexOperators} we introduce the associated vertex operators and study their structural properties. In Section~\ref{sec:HirotaAncestor} (in Section~\ref{sec:HirotaDescendant}, respectively) we prove the Hirota quadratic equations for the ancestor (descendent, respectively) potential of this Dubrovin--Frobenius manifold. We derive in Section~\ref{sec:LaxFormulation} the Lax formulation of the obtained integrable system, which we then identify with the extended nonlinear Schr\"odinger hierarchy.

\subsection*{Acknowledgments} 

G.~C., H.~P., and S.~S. were supported by the Netherlands Organization for Scientific Research. G.~C. is supported by the ANER grant ``FROBENIUS'' of the Region Bourgogne-Franche-Comté. The IMB receives support from the EIPHI Graduate School (contract ANR-17-EURE-0002).

\subsection*{Notation}

\begin{itemize}
\item $\I$ : the imaginary unit.
\item $\ch(n) := \sum_{k=1}^n k^{-1}$ : the $n$-th harmonic number, $\ch(0)=0$. 
\item $(a)_n := \Gamma(a+n) / \Gamma(a)$ : the Pochhammer symbol.
\item $\R_+$: the non-negative real axis as a subset of $\C$.
\item $\R_-$: the non-positive real axis as a subset of $\C$.
\item $e_1, \dots , e_n$: the canonical basis in $\C^n$.

\end{itemize}

We will often use, for $n\geq0$:
\begin{align}  
& \left( \frac12 \right)_n =\frac{\Gamma(1/2+n)}{\sqrt{\pi}} = \frac{(2n-1)!!}{2^n}= \frac{(2n)!}{4^n\, n!}, \\ 
& \left( \frac12 \right)_{-n} =\frac{\Gamma(1/2-n)}{\sqrt{\pi}} = \frac{(-2)^{n}}{(2n-1)!!} . 
\end{align}

\section{The Dubrovin--Frobenius manifold} \label{sec:FrobeniusManifolds}

\subsection{The Dubrovin--Frobenius manifold for Catalan numbers}

Let $M= \C \times \C^*$ with coordinates $(t^1, t^2)$. Let us define a charge $d=-1$ Dubrovin--Frobenius manifold structure on $M$ with potential 
\begin{equation}  
F(t) = \frac12 (t^1)^2 t^2 + \frac 12 (t^2)^2 \log t^2 .
\end{equation}
For a general introduction to the theory of Dubrovin--Frobenius manifolds refer to~\cite{dub96, dub98, dub99b}, or the more recent review in the first part of~\cite{cdg17}.

The unit and Euler vector fields, the metric, and the product on the tangent space are given by
\begin{equation}  
e= \frac{\partial }{\partial t^1}, \qquad  
E = t^1 \frac{\partial }{\partial t^1} + 2 t^2 \frac{\partial }{\partial t^2} , \qquad 
\eta = \begin{pmatrix}
0 &1\\ 1 & 0
\end{pmatrix}, \qquad 
\frac{\partial }{\partial t^2} \bullet \frac{\partial }{\partial t^2} = \frac1{t^2} \frac{\partial }{\partial t^1}.  
\end{equation}
The intersection form $g^{ij}=E^k c_{k}^{ij}$, where $c_{ij}^k$ are the structure constants of the product and the indexes are raised and lowered by the metric $\eta$, is equal to
\begin{equation}  
g = \begin{pmatrix}
2 & t^1 \\ t^1 & 2 t^2 
\end{pmatrix}.
\end{equation}
The discriminant $\Delta \subset M$ is the locus where the intersection form $g$ degenerates, that is
\begin{equation}  
\Delta = \{ t \in M \ | \ 4t^2=(t^1)^2 \} . 
\end{equation} 
We denote by $\Delta_\lambda \subset M \times \C$ the locus where the pencil $g  - \lambda \eta$ degenerates, which is 
\begin{equation}  
\Delta_\lambda = \{ (t, \lambda)   \in M\times \C \ | \ 4t^2=(t^1-\lambda)^2 \} . 
\end{equation} 
We have the following two standard endomorphisms on the tangent space, which in the flat trivialization read
\begin{equation}  
\mu = \begin{pmatrix}
1/2 & 0 \\ 0 & -1/2
\end{pmatrix}, \qquad 
\cU = \begin{pmatrix}
t^1 & 2 \\ 2t^2 & t^1
\end{pmatrix}
\end{equation}
respectively defined  by $\mu = (2-d)/2 -  \nabla E$ and $\cU = E \bullet $, with
\begin{equation}  
\eta \mu \eta = -  \mu, \qquad 
\eta  \cU \eta  = \cU^T.
\end{equation}

\subsection{The canonical coordinates}

Let us define two canonical coordinates charts on $M$. Let $V_1 = \{ (t^1, t^2 ) \in M | t^2 \not\in \R_- \} \subset M$ and $U_1 = \{ (u^1, u^2) \in \C^2 | \re (u^1 - u^2 ) > 0 \}$. The map $V_1 \to U_1$ given by 
\begin{equation}  
u^1 = t^1 + 2 \sqrt{t^2}, \qquad u^2 = t^1 - 2 \sqrt{t^2}
\end{equation}
is a diffeomorphism with inverse 
\begin{equation}  
t^1 = \frac{u^1 + u^2}2, \qquad t^2= \left( \frac{u^1 - u^2}4 \right)^2,
\end{equation}
where $\sqrt{t^2}$ is the principal branch of the square root on the cut plane $\C \setminus \R_-$. Let $\log(t^2)$ denote the principal branch of the logarithm. We always denote $(t^2)^\alpha = \exp ( \alpha \log t^2)$.
The same formulas give a diffeomorphism $V_2 \to U_2$, where $V_2 = \{ (t^1, t^2 ) \in M | t^2 \not\in \R_+ \}$ and $U_2 = \{ (u^1, u^2) \in \C^2 | \im (u^1 - u^2 ) > 0 \}$. To define the roots in this case we choose the branch of logarithm on $\C \setminus \R_+$ such that $\log(-1) = \pi \I$. 
Without further notice we will systematically work on $V_1$, the extension of our formulas to $V_2$ being obvious. 

The canonical coordinate vectors given by (over both coordinate charts)
\begin{equation}  
\frac{\partial }{\partial u^1} = \frac12 \frac{\partial }{\partial t^1} + \frac12 \sqrt{t^2} \frac{\partial }{\partial t^2}, \qquad 
\frac{\partial }{\partial u^2} = \frac12 \frac{\partial }{\partial t^1} - \frac12 \sqrt{t^2} \frac{\partial }{\partial t^2},
\end{equation}
are idempotents for the multiplication on the tangent spaces, and 
\begin{equation}  
e= \frac{\partial }{\partial u^1} + \frac{\partial }{\partial u^2} , \qquad 
E = u^1 \frac{\partial }{\partial u^1} + u^2 \frac{\partial }{\partial u^2} .
\end{equation}

\subsection{The normalised canonical frame}

Let use denote
\begin{equation}  
\Delta_i^{-1} = ( \frac{\partial }{\partial u^i}, \frac{\partial }{\partial u^i})
\end{equation}
so that $\Delta_1^{-1} = \frac12 \sqrt{t^2}$,  $\Delta_2^{-1} = - \frac12 \sqrt{t^2}$. The normalised canonical frame is defined as 
\begin{equation}  
\be_i = \Delta_i^{1/2} \frac{\partial }{\partial u^i} ,
\end{equation}
where we fix 
\begin{equation}  \label{eq:Deltas}
\Delta_1^{1/2} = \sqrt{2} (t^2)^{-1/4}, \qquad 
\Delta_2^{1/2} = \I \sqrt{2} (t^2)^{-1/4},
\end{equation}
and the transition matrix $\Psi$ from the normalised canonical frame to the flat frame, defined by
\begin{equation}  
\frac{\partial }{\partial t^\alpha} = \sum_j \mathbf{e}_j \Psi_{j\alpha } ,
\end{equation}
is given by 
\begin{equation}  \label{eq:PsiMatrix}
\Psi = (\Psi^{-1})^T \eta =
 \frac{1}{\sqrt{2}}
\begin{pmatrix}
 (t^2)^{1/4} & (t^2)^{-1/4}  \\
- \I(t^2)^{1/4}  & \I (t^2)^{-1/4}
\end{pmatrix}, \qquad 
\Psi^{-1} =\frac1{\sqrt{2}} \begin{pmatrix}
 (t^2)^{-1/4} & \I (t^2)^{-1/4} \\
  (t^2)^{1/4} &  -\I (t^2)^{1/4} 
\end{pmatrix}.
\end{equation}
In the following we will represent a tangent vector by the column matrix of its components in the relevant frame. The coefficients $V_{i,can} = \Psi_{i \alpha} V^\alpha_{flat}$ of a vector in the normalised canonical frame are obtained from the coefficients $V^i_{flat}$ in the basis of flat coordinated by left multiplication by the matrix $\Psi$. 
Note that the row index is $i$ in $\Psi_{i\alpha}$, and $\alpha$ in $(\Psi^{-1})^\alpha_i$.
Remark also that in the formulas above the branch of the roots depends on the chart used.

\section{The deformed flat connection and the principal hierarchy} \label{sec:PrincipleHierarchy}

\subsection{The deformed flat connection}
\label{sec:defflat}
Let $Y(t, z)$ be a two by two matrix valued function on $M \times \C$ which solves the deformed flatness equations 
\begin{equation}
\label{equ3}
- z \frac{\partial Y}{\partial z} = (\mu + \frac{\cU}{z}) Y, \qquad 
z\frac{\partial Y}{\partial t^\alpha} =  C_\alpha Y ,
\end{equation}
with $C_1=\mathbf{1}$ and 
\begin{equation}  
C_2 = \begin{pmatrix}
0 & (t^2)^{-1} \\ 1 & 0 
\end{pmatrix}.
\end{equation}
The columns of the fundamental matrix $Y$ are the gradients of a system of deformed flat coordinates $\tilde{t}_\alpha(t,z)$:
\begin{equation}  
Y^\beta_\alpha = \eta^{\beta\gamma}\frac{\partial \tilde{t}_\alpha}{\partial t^\gamma} .
\end{equation}
After fixing a branch of $\log z$, we look for solutions of the form
\begin{equation}  \label{eq:DefinitionOfS}
Y(t,z) = S(t,z) z^{-\mu} z^{-R}
\end{equation}
where $S = \sum_{k\geq 0} S_k z^{-k}$ is a matrix valued power series which converges in a small neighbourhood of $z=\infty$, with $S_0= \mathbf{1}$ and 
\begin{equation}  
R = \begin{pmatrix}
0 & 2 \\ 0 & 0
\end{pmatrix}.
\end{equation}
The matrix $S$ is uniquely fixed by setting $(S_1)_{1,2} = \psi + \log t^2$, see below, Section~\ref{sec:FormulasForS}. Here $\psi$ is the arbitrary constant that parametrizes the solutions of the resonant system~\eqref{equ3} near the Fuchsian singularity $z=\infty$.
The columns of the matrix $S$ are the gradients of analytic functions $\theta_\alpha$ on $M \times \C$ such that 
\begin{equation}  
(\tilde{t}_1 , \tilde{t}_2 ) = ( \theta_1 , \theta_2 ) z^{-\mu} z^{-R} ,
\end{equation}
and the $\tilde{t}_\alpha$ are said to form a Levelt system of deformed flat coordinates on $M$.

\subsection{The superpotential and the deformed flat coordinates}

The Dubrovin--Frobenius manifold structure on $M$ can be given in terms of the following superpotential $f: M \times \C^* \to \C$
\begin{equation}
f(t, \zeta)  = \zeta + t^1 +  t^2 \zeta^{-1} ,
\end{equation}
see~\cite{dub96}. Notice that  $\Delta_\lambda$ coincides with the set of points $(t , \lambda) \in M \times \C$ where the preimage of $\lambda$ via $f(t, \cdot)$ degenerates, namely is not given by two distinct points in $\C^*$. The superpotential induces the Frobenius manifold product via the identification of each tangent space with the local algebra $\C[\zeta, \zeta^{-1}]/(\partial_\zeta f)$.

We define the formal logarithm $\widetilde\log f$ as the formal Laurent series $\sum_{k\in\Z} a_k \zeta^k$, with coefficients given by 
\begin{equation}
\sum_{k\geq0} a_k \zeta^k = \frac12 \log(f \zeta) , \quad \zeta \sim 0, 
\end{equation}
\begin{equation}
\sum_{k<0} a_k \zeta^k = \frac12 \log(f /\zeta) , \quad \zeta \sim \infty . 
\end{equation}

\begin{proposition}
The Levelt system $\tilde{t}_\alpha(t,z)$ of deformed flat coordinates for the Catalan Dubrovin--Frobenius manifold $M$ is given by
\begin{equation} \label{levelt}
\tilde{t}_1 = \theta_1 z^{-\frac12}, \qquad 
\tilde{t}_2 = \theta_2 z^{\frac12} - \theta_1 2 z^{-\frac12} \log z,
\end{equation}
where 
\begin{equation}
\theta_\alpha =  \res g_\alpha \ d\zeta
\end{equation}
(here $\res$ denotes the formal residue applied to a formal Laurent series) and $g_\alpha(t,\zeta,z)$ are defined as 
\begin{equation}
\label{psi}
g_1 =  z \left( e^{f \slash  z} -1 \right), \qquad  
g_2 = 2 e^{f \slash z} \left( \widetilde\log f + \frac{\psi}2 - \ein(f \slash z) \right).
\end{equation}
\end{proposition}
The function $\ein(z)$ is the entire exponential integral defined by 
\begin{equation}
\ein(z) = \int_0^z ( 1-e^{-t}) \frac{dt}t.
\end{equation}
\begin{proof}
Substituting~\eqref{eq:DefinitionOfS} in~\eqref{equ3} we find the equations for the matrix $S$
\begin{equation}
-z S_z + [S, \mu] = \frac1z ( \cU S -  S R ), \qquad 
z S_{t^\alpha} = C_\alpha S
\end{equation}
and expressing $S$ in terms of the residue of the derivatives of $g_\alpha$ as
\begin{equation}
S^\alpha_\beta = \eta^{\alpha \gamma} \res \frac{\partial g_\beta}{\partial t^\gamma} d\zeta 
\end{equation}
we see that it is sufficient for the following two equations to be satisfied modulo terms with zero residue in $\zeta$
\begin{equation} \label{temp667}
z\frac{\partial^2 g_\gamma}{\partial t^\alpha \partial t^\beta} = c_{\alpha \beta}^\sigma \frac{\partial g_\gamma}{\partial t^\sigma}, \qquad 
-z \frac{\partial^2 g_\beta}{\partial z \partial t^\gamma} + \frac{\partial g_\beta}{\partial t^\gamma} (\mu_\beta + \mu_\gamma) - 
\frac1z \left( \cU^\rho_\gamma \frac{\partial g_\beta}{\partial t^\rho} -\frac{\partial g_\rho}{\partial t^\gamma} R^\rho_\beta  \right) =0.
\end{equation}
We define 
\begin{equation}
\frac{\partial \widetilde\log f}{\partial f} = \widetilde{f^{-1}} := \frac12 \left( (f^{-1})_0 + (f^{-1})_\infty \right)
\end{equation}
where $(f^{-1})_0$ and $(f^{-1})_\infty$ are the formal expansions of $f^{-1}$ at $\zeta =0$ and $\zeta=\infty$, respectively,
so that the chain rule holds 
\begin{equation}
\frac{\partial \widetilde\log f}{\partial t^\alpha} =  \frac{\partial \widetilde\log f}{\partial f} \frac{\partial f}{\partial t^\alpha}.
\end{equation}
It is easy to check that 
\begin{equation}
\frac{\partial g_\alpha}{\partial f} = \frac1z g_\alpha + 2 \widetilde{f^{-1}} \delta_\alpha^2, \qquad  
\frac{\partial^2 g_\alpha}{\partial f^2} = \frac1z  \frac{\partial g_\alpha}{\partial f} -2 \widetilde{f^{-2}} \delta_\alpha^2,  
\end{equation}
where $\widetilde{f^{-2}}$ is defined as the average of the formal expansions of $f^{-2}$ at $\zeta =0$ and $\zeta=\infty$ as in the case of $\widetilde{f^{-1}}$ above.

To prove the first equation in~\eqref{temp667} it is sufficient to observe that 
\begin{equation} \label{temp668}
z\frac{\partial^2 g_\gamma}{\partial t^\alpha \partial t^\beta} - c_{\alpha \beta}^\sigma \frac{\partial g_\gamma}{\partial t^\sigma} =
-2z  c_{\alpha \beta}^\sigma \frac{\partial f}{\partial t^\sigma}  \widetilde{f^{-2}} \delta_\gamma^2
+ z \frac{\partial^2 g_\gamma}{\partial f^2} \frac{\partial f}{\partial \zeta} K_{\alpha \beta} ,
\end{equation}
and to check that the right-hand side has vanishing residue. 
Here we have used the fact that the Frobenius multiplication is induced by the local algebra  $\C[\zeta, \zeta^{-1}]/(\partial_\zeta f)$, hence 
\begin{equation}
\frac{\partial f}{\partial t^\alpha} \frac{\partial f}{\partial t^\beta} = c_{\alpha \beta}^{\gamma} \frac{\partial f}{\partial t^\gamma} + K_{\alpha \beta} \partial_\zeta f   
\end{equation}
where $K_{\alpha \beta} = - (t^2)^{-2} \delta_\alpha^2 \delta_\beta^2 $.

The second equation in~\eqref{temp667} can be rewritten as 
\begin{equation} \label{temp666}
\frac{\partial }{\partial t^\alpha} \left( 
-z\frac{\partial g_\gamma}{\partial z} + \left(\frac32 + \mu_\gamma \right) g_\gamma - E(g_\gamma) + \frac2z g_1 \delta_\gamma^2  
\right) =0.
\end{equation}
One can directly verify the following quasi-homogeneity property of $g_\gamma$
\begin{equation}
\left( z\frac{\partial }{\partial z} + f \frac{\partial }{\partial f} - \mu_\gamma - \frac12 \right) g_\gamma =2 ( \frac{g_1}z +1)\delta_\gamma^2.
\end{equation}
Using the fact that $E(f) =f - \zeta \partial_\zeta f$ one has 
\begin{equation}
E(g_\gamma) = f \frac{\partial g_\gamma}{\partial f} - \zeta \frac{\partial g_\gamma}{\partial \zeta},  
\end{equation}
hence the formula~\eqref{temp666} is verified up to the $t^\alpha$-derivative of the term 
\begin{equation}
\frac{\partial (\zeta g_\gamma)}{\partial \zeta} -2 \delta_\gamma^2
\end{equation}
which has no residue. 
\end{proof}

\subsection{The calibration} \label{sec:FormulasForS}

The $S$ matrix is uniquely determined by the recursive formula
\begin{equation}  
k S_k  +S_k \mu - \mu S_k = \cU S_{k-1} -S_{k-1} R, \text{ for $k \geq 1$}
\end{equation}
with $S_0=\mathbf{1}$ and $(S_1)_{1,2} = \psi + \log t^2$, and
\begin{equation}  
R = \begin{pmatrix}
0 & 2 \\ 0 & 0
\end{pmatrix}.
\end{equation}
The first two terms are 
\begin{equation}  
S_1 = \begin{pmatrix}
t^1 & \psi + \log t^2 \\ t^2 & t^1 
\end{pmatrix}, \qquad  
S_2 = \begin{pmatrix}
\frac12 (t^1)^2 + t^2 & t^1 (\psi + \log t^2) \\
t^1 t^2 & \frac12 (t^1)^2 + t^2 (\psi + \log t^2 -1) 
\end{pmatrix}.
\end{equation}
We can write the $S$ matrix in terms of the superpotential as follows
\begin{equation}
S_k = \res \begin{pmatrix}
 \frac{f^k}{k !} \zeta^{-1} & \frac{2 f^{k-1}}{(k-1)!} \left( \widetilde\log	f + \frac{\psi}2 - \ch(k-1) \right) \zeta^{-1} \\
   \frac{f^k}{k !}  & \frac{2 f^{k-1}}{(k-1)!} \left( \widetilde\log f + \frac{\psi}2 - \ch(k-1) \right) 
\end{pmatrix} d\zeta .
\end{equation}
This expression can be derived from~\eqref{levelt} or checked by substitution in the recursive formula for $S_k$. 

For example, at the point given by $t^1=0$ and $t^2=1$ (we often consider specialization to this point below and call this the special point of $M$ denoted by $t_{sp}$) we can explicitly compute the $S$ matrix coefficients, which are given by
\begin{equation}  \label{eq:SMatrixSpecialPoint}
S_{2k} = \begin{pmatrix}
\frac{1}{(k!)^2} & 0 \\
0 & 
\frac{\psi + k^{-1} -2 \ch(k)}{k!(k-1)!}
\end{pmatrix}, \quad
S_{2k+1} = \begin{pmatrix}
0 & \frac{\psi-2\ch(k)}{(k!)^2} \\
\frac{1}{(k+1)! k!} & 0
\end{pmatrix}.
\end{equation}
The first few terms are 
\begin{equation}  \label{eq:S-matrixSpecial}
S_1 = \begin{pmatrix}
0 & \psi \\ 1 & 0  
\end{pmatrix}, \qquad 
S_2= \begin{pmatrix}
1 & 0 \\ 0 & \psi -1 
\end{pmatrix}, \qquad 
S_3 = \begin{pmatrix}
0 & \psi - 2 \\ \frac12 & 0 
\end{pmatrix},
\end{equation}
\begin{equation}  \notag
S_4= \begin{pmatrix}
\frac14 & 0 \\ 0 & \frac{\psi}2 -\frac54
\end{pmatrix}, \qquad 
S_5 = \begin{pmatrix}
0 & \frac{\psi}4 - \frac34 \\ \frac1{12} & 0 
\end{pmatrix}.
\end{equation}

\subsection{The principal hierarchy}

We can easily obtain the following explicit Hamiltonian form of the principal hierarchy, with Hamiltonian densities $h_{\alpha,p}$ given by the expansion of the  analytic part $\theta_\alpha$ of the deformed flat coordinates
\begin{equation} \notag 
\theta_\alpha = \sum_{p\geq0} h_{\alpha, p-1} z^{-p}.
\end{equation}
 For a general discussion of the principal hierarchy associated with a Frobenius manifold see~\cite{DubrovinZhang2001}.
\begin{proposition}
The principal hierarchy of the Catalan Dubrovin--Frobenius manifold is given in Hamiltonian form by
\begin{equation}
\frac{\partial t^i}{\partial t^{\alpha, p}} = \{ t^i (x), H_{\alpha,p} \}_1  
\end{equation}
where the Hamiltonians and their densities are given by
\begin{equation}
H_{\alpha,p} = \int h_{\alpha,p} \ dx, 
\qquad 
h_{\alpha, p} = \res	g_{\alpha,p} \ d\zeta ,
\end{equation}
the functions $g_{\alpha,p}(t,\zeta)$ are defined as
\begin{equation}
g_{1,p} = \frac{f^{p+2}}{(p+2)!} , \qquad 
g_{2,p} =  \frac{2 f^{p+1}}{(p+1)!} \left( \widetilde\log f - \ch(p+1) + \frac{\psi}2 \right),
\end{equation}
and the Poisson structure is 
\begin{equation}  
\{ t^i(x) , t^j(y) \}_1  = \eta^{ij} \delta'(x-y).
\end{equation}
\end{proposition}

Notice that here the following expansion of the entire exponential integral was used
\begin{equation} \notag
e^z\ein(z) = \sum_{p\geq1} \frac{\ch(p) z^p}{p!}.
\end{equation}

\subsection{The $R$ matrix} \label{sec:DefinitionOfR}

Let us represent the fundamental matrix $Y$ in the normalised canonical frame
\begin{equation}  
\tilde{Y} = \Psi Y.
\end{equation}
The $z$ part of the deformed flatness equations becomes
\begin{equation}  
-z\frac{\partial \tilde{Y}}{\partial z} = ( V + \frac{U}z ) \tilde{Y}, 
\end{equation}
where
\begin{equation}  
U := \Psi \cU \Psi^{-1} = \begin{pmatrix}
u_1 & 0 \\ 0 & u_2
\end{pmatrix}, \qquad 
V := \Psi \mu \Psi^{-1} = \begin{pmatrix}
0 & \I/2 \\ -\I/2 & 0
\end{pmatrix} .
\end{equation}
There exists a unique formal solution of this equation of the form 
\begin{equation}  
\tilde{Y}(t,z) = R(t,z) e^{U /z},
\end{equation}
where 
\begin{equation}  
R=\sum_{k\geq0} R_k z^{k}, \qquad  R_0 =\mathbf{1}.
\end{equation}
Indeed the coefficients $R_k$ are uniquely determined by the recursion relation
\begin{equation}  
[R_{k+1}, U] = (V+k) R_k . 
\end{equation}
Explicitly we have
\begin{equation} \label{RRmatrix}
R_k= 
\frac{\left(\frac12\right)_{k-1}\left(\frac12\right)_{k}}{(k)!}
 \begin{pmatrix}
\frac{(-1)^{k+1}}2& k \I \\[2mm]
(-1)^{k+1}k \I &-\frac{1}2
\end{pmatrix}
(u_2-u_1)^{-k} .
\end{equation}

\section{Givental quantization formalism and potentials} \label{sec:GiventalQuantizationFormalism}

In this section we introduce the quantization formalism of Givental~\cite{Givental2001,Givental2001-Q} (see also an exposition in~\cite{BakalovMilanov}) and explain his definitions of the so-called ancestor and descendent potentials associated to a Dubrovin--Frobenius manifold. We use the Givental formula for the descendent potential in order to link the Dubrovin--Frobenius manifold~\eqref{eq:CatalanFrobeniusManifold1}-\eqref{eq:CatalanFrobeniusManifold3} to the higher genera Catalan numbers.

\subsection{Symplectic loop space and quantization} 

Let $V$ be a $\C$-vector space with a non-degenerate symmetric bilinear form $(~,~)_V$, with basis $\{\phi_i\}$ and dual basis $\{\phi^i\}$. Let $\cV:=V((z))$ be the loop space of formal Laurent series with values in $V$, 
equipped with the symplectic form
\begin{equation}
\Omega(\cf,\cg) = \res_z(\cf(-z), \cg(z)) dz, 
\quad \cf, \cg \in \cV.
\end{equation}

\begin{remark}
In the following $V$ will be identified with the tangent space at a point of $M$ either via the flat trivialisation or via the normalised canonical frame. In the first case the basis is that of flat coordinate vector fields $\frac{\partial }{\partial t^i}$ and the bilinear form is the flat metric $\eta$ at the point, in the second case the $V$ is identified via the canonical basis with the Euclidean space together with the standard  inner product. 
\end{remark}

Darboux coordinates on $\cV$ can be defined as 
\begin{equation}
q^i_k = \Omega((-z)^{-k-1} \phi^i, \cdot)\,, \qquad 
p_{i,k} = \Omega(\cdot, \phi_i z^k) \,,\qquad k\geq 0\,,
\end{equation}
by which we can express any element $\cf \in \cV$ as
\begin{equation} %
\cf = \sum_{k\geq0} \left( q^i_k \phi_i z^k + p_{i,k} \phi^i (-z)^{-k-1} \right).
\end{equation}

Let $\cV_+:=V[[z]]$ be the space of formal Taylor series with values in $V$. Then we have a natural isomorphism $\cV\cong T^*\cV_+$. Let $\epsilon$ be a formal parameter. Consider the Fock space of functions on $\cV_+$ given by formal power series in the variables $q^i_k+\delta^i_1\delta^1_k$ with the coefficients in formal Laurent series in $\epsilon$. 

We consider linear and quadratic Hamiltonians on $\cV$. The standard (Weyl) quantization associates to them differential operators of order $\leq 2$ on the Fock space by the following rules:
\begin{gather}
(q^i_\ell){\hat{\ }} = \frac{1}{\epsilon} q^i_\ell,\quad 
(p_{i,\ell}){\hat{\ }} = \epsilon \frac\partial{\partial q^i_\ell}, \quad \\
(q^i_k q^j_\ell){\hat{\ }} = \frac{1}{\epsilon^2}q^i_k q^j_\ell, \quad 
(q^i_k p_{j,\ell}){\hat{\ }} = q^i_k \frac{\partial}{\partial q^j_\ell}, \quad
(p_{i,k} p_{j,\ell}){\hat{\ }} = \epsilon^2 \frac{\partial^2}{\partial q^i_\ell\partial q^j_\ell}\,.
\end{gather}

For instance, the linear Hamiltonian $h_\cf( \cdot ) = \Omega(\cf,\cdot)$ associated with a constant vector field $\cf= \sum_{l} I^{l} (-z)^l  \in \cV$ is given by
\begin{equation}
h_\cf =\sum_{l\geq 0} \left[ (-1)^{l+1} (I^{l},\phi^i) p_{i,l} + (I^{-(l+1)},\phi_i)  q^i_l \right]\,.
\end{equation}
It is convenient to denote $(I^k,\phi_i)$ (respectively, $(I^k,\phi^i)$) by $(I^k)_i$ (respectively,  $(I^k)^i$). The quantization of $h_\cf$ reads
\begin{equation}\label{eq:f-quantized}
\hat{\cf} \coloneqq (h_\cf){\hat{\ }} = \sum_{l\geq 0} \left[  \epsilon (-1)^{l+1} (I^l)^i \frac{\partial }{\partial q^i_l}  + \frac1{\epsilon}  (I^{-(l+1)})_i q^i_l \right]
\,.
\end{equation}
Note that in particular
\begin{equation}
[ \hat{\cf}_1 , \hat{\cf}_2 ] = \Omega(\cf_1,\cf_2)\,.
\end{equation}

We consider two Lie algebras, of purely positive and purely negative series in $z$, that is, either $\cm = \sum_{\ell\geq 1} m_\ell z^\ell$ or $\cm = \sum_{\ell\leq -1} m_\ell z^\ell$, $m_\ell \in \End(V)$, representing linear vector fields $\cm$ commuting with $z$.  These vector fields are  infinitesimally symplectic if $\cm^i_j (z) + \eta^{ik} \cm_k^l (-z)\eta_{lj} =0$. The Hamiltonian of $\cm$  is defined as $h_{\cm}(\cf)\coloneqq \frac 12 \Omega(\cm\cf,\cf)$, and $\hat{\cm}$ denotes its quantization,
\begin{equation}
\hat{\cm} \coloneqq (h_\cm){\hat{\ }}.
\end{equation}
For the operator $M=\exp(\cm)$ the symbol $\hat M$ denotes the operator $\exp(\hat{\cm})$.

For instance, the quadratic Hamiltonian associated with $\cs = \sum_{\ell\geq 1} s_\ell z^{-\ell}$ is given by
\begin{equation}
h_{\cs}(\cf) = \frac 12 \sum_{a,b\geq 0} (-1)^{b+1} q^i_a q^j_b (s_{a+b+1})^k_i \eta_{kj} - \sum_{a\geq0, \ell\geq 1} q^i_{a+\ell} p_{j,a} (s_\ell)^j_i\,,
\end{equation}
where $(s_{a+b+1})^k_i \eta_{kj} = (s_{a+b+1}\phi_i,\phi_j)$ and $(s_{a+b+1})^j_i = (s_{a+b+1}\phi_i,\phi^j)$. Its quantization reads
\begin{equation}\label{eq:log-S-hat}
(h_{\cs}){\hat{\ }} = \frac 1{2\epsilon^2} \sum_{a,b\geq 0} (-1)^{b+1} q^i_a q^j_b (s_{a+b+1})^k_i \eta_{kj} - \sum_{a\geq0, \ell\geq 1} q^i_{a+\ell} \frac{\partial}{\partial q^j_{a}} (s_\ell)^j_i\,.
\end{equation}

The quadratic Hamiltonian of the element $\mathfrak{r}:=\sum_{\ell\geq 1} r_\ell z^{\ell}$ is given by
\begin{equation}
h_\mathfrak{r}(\cf)=\frac{1}{2}\sum_{a,b\geq 0}(-1)^a p_{i,a}p_{j,b}(r_{a+b+1})^i_k\eta^{kj}-\sum_{a\geq 0,\ell\geq 1}q^i_ap_{j,a+\ell}(r_\ell)^j_i,
\end{equation}
leading to the quantization
\begin{equation} \label{eq:quantasationLogR}
(h_{\mathfrak{r}}){\hat{\ }}=\frac{\epsilon^2}{2}\sum_{a,b\geq 0}(-1)^a\frac{\partial}{\partial q^i_a}\frac{\partial}{\partial q^j_b}(r_{a+b+1})^i_k\eta^{kj}-\sum_{a\geq 0,\ell\geq 1}q^i_a\frac{\partial}{\partial q^j_{a+\ell}}(r_\ell)^j_i.
\end{equation}

\subsection{Symplectic transformations and potentials} Recall the series $S=S(t,z)$ defined by Equation~\eqref{eq:DefinitionOfS} and discussed in detail in Section~\ref{sec:FormulasForS}. It is a symplectic operator on $\cV$ for $\phi_i=\frac{\partial}{\partial t^i}$, $(\,,\,)_V= \eta$ that commutes with multiplication by $z$, so we can apply the quantization procedure described above and define $\hat S$. 

Recall the series $R=R(t,z)$ defined in Section~\ref{sec:DefinitionOfR}. Consider its action on the same $\cV$ in a different basis given by
\begin{equation}  
\mathbf{e}_j  = \sum_{\alpha}\frac{\partial }{\partial t^\alpha} (\Psi^{-1})^{\alpha }_j
\end{equation}
(note that $(\mathbf{e}_i,\mathbf{e}_i)_V=\delta_{ij}$). It is a symplectic operator on $\cV$ commuting with multiplication above as well, and the quantization procedure above defines the operator $\hat R$. 

Note that since the matrix $R$ is given in a different basis, then in order to apply Equation~\eqref{eq:quantasationLogR} one has to consider the operator $\Psi^{-1} R \Psi$. A better alternative is to use the basis $\mathbf{e}_i$ and the more natural variables $Q^i_a\coloneqq \sum_{\alpha}\Psi_{i\alpha} q^\alpha_a$.

We can now define the ancestor potential as
\begin{equation}
\mathcal{A}(\{q^1_a,q^2_a\}_{a\geq 0}):=\hat{\Psi}\hat{R}\prod_{i=1}^2\tau_{KdV}(\{Q^i_a\}_{a\geq 0}),
\end{equation}
where $\tau_{KdV}$ is the Witten--Kontsevich $\tau$-function for the KdV hierarchy in the variables with the dilaton shift (that is, with respect to the standard descendent variables $t_a$, $a\geq 0$, we have $Q^i_a = t_a - \delta_{1,a}$), and the operator $\hat{\Psi}$ recomputes the function $\hat{R}\prod_{i=1}^2\tau_{KdV}(\{Q^i_a\}_{a\geq 0})$ in the variables $q^i_a$.

Finally, the total descendent potential is defined as
\begin{equation}
\mathcal{D}:=C\hat{S}^{-1}\mathcal{A},
\end{equation}
where the extra factor $C$ is set to be (up to a multiplicative constant)
\begin{equation}
\log C(t^1,t^2) \coloneqq \int^{(t^1,t^2)} (R_1)^1_1 du^1 +(R_1)^2_2 du^2 
= -\frac 1{16} \log t^2.
\end{equation}

Note that the factor $C$ and operators $\hat S$, $\hat \Psi$, and $\hat R$ do depend on the point $(t^1,t^2)$ of the Dubrovin--Frobenius manifold. The coefficients of the ancestor potential $\cA$ also depend on the point  $(t^1,t^2)$ of the Dubrovin--Frobenius manifold, and we have to think of $\cD$ as a formal power series in $t^1_0-t^1, t^2_0-t^2, t^1_k, t^2_k$, $k\geq 1$, where the variables $q^i_a$, $i=1,2$, $a\geq 0$ of the descendent potential $\cD$ are related to the variables $t^i_a$ by the dilaton shift: $q^i_a = t^i_a - \delta^{i,1}\delta_{a,1}$. Givental proved in~\cite{Givental2001} that the total derivatives of $\cD$ with respect to $t^1$ and $t^2$ are equal to zero.

\subsection{Higher genera Catalan numbers} Recall the definition of the numbers $C_{g,k_1,\dots,k_n}$ given in Digression~\ref{dig:Catalan}.

\begin{theorem} \label{thm:SpecializationOfD} Assume $\psi=0$ and fix the point of expansion for $\cD$ to be $(t^1,t^2)=(0,1)$. We have: 
\begin{equation}
 \log \cD \big|_{\substack{t^2_0=1 \\
 		t^2_a = 0, a\geq 1}} = \sum_{g=0}^\infty \sum_{n=1}^\infty \frac{\epsilon^{2g-2}}{n!} \sum_{k_1,\dots,k_n\geq 0} C_{g,k_1+1,\dots,k_n+1} \prod_{i=1}^n \frac{t^1_{k_i}}{(k_i+1)!} 
\end{equation}	
\end{theorem}

\begin{proof} The first step of the proof is to rewrite the formula for the descendent potential in terms of the variables $\{t^1_a,t^2_a\}_{a\geq 0}$. To this end, we have to shift the variables:
	\begin{equation}
	\left(e^{-\frac{\partial}{\partial q^1_1}} C\hat S^{-1} e^{\frac{\partial}{\partial q^1_1}}\right)\Big|_{q^i_a \to t^i_a} \hat\Psi \left(e^{-\Psi^1_1\frac{\partial}{\partial Q^1_1}- \Psi^2_1\frac{\partial}{\partial Q^2_1}} \hat R e^{\frac{\partial}{\partial Q^1_1}+\frac{\partial}{\partial Q^2_1}}\right) \Big|_{Q^i_a \to T^i_a} \prod_{i=1}^2\tau_{KdV}(\{T^i_a\}_{a\geq 0},\epsilon^2),
	\end{equation}
	and in this formula the Witten--Kontsevich $\tau$-function $\tau_{KdV}$ are considered in the standard descendent variables. We can further rewrite this formula as 
	\begin{multline}
	\left(e^{-\frac{\partial}{\partial q^1_1}} C\hat S^{-1} e^{\frac{\partial}{\partial q^1_1}}\right)\Big|_{q^i_a \to t^i_a} \hat\Psi \left(e^{-\Psi^1_1\frac{\partial}{\partial Q^1_1}- \Psi^2_1\frac{\partial}{\partial Q^2_1}} \hat R e^{\Psi^1_1\frac{\partial}{\partial Q^1_1}+\Psi^2_1\frac{\partial}{\partial Q^2_1}}\right) \Big|_{Q^i_a \to T^i_a} \cdot
	\\ 
\cdot	e^{-\Psi^1_1\frac{\partial}{\partial T^1_1}- \Psi^2_1\frac{\partial}{\partial T^1_1}+\frac{\partial}{\partial T^1_1}+\frac{\partial}{\partial T^2_1}}
	\prod_{i=1}^2\tau_{KdV}(\{T^i_a\}_{a\geq 0},\epsilon^2),
	\end{multline}
	Recall that $\Psi^i_1 = \Delta_i^{-1/2}$, $i=1,2$. The dilaton equation implies that 
	\begin{equation}
	e^{-\Delta_i^{-1/2}\frac{\partial}{\partial T^i_1}+\frac{\partial}{\partial T^i_1}}\tau_{KdV}(\{T^i_a\}_{a\geq 0},\epsilon^2) = \tau_{KdV}(\{\Delta_i^{1/2}T^i_a\}_{a\geq 0},\Delta_i\epsilon^2), \qquad i=1,2.
	\end{equation}
	Thus the resulting formula for the descendent potential that is used in applications in the variables $\{t^i_a\}$ is given by
	\begin{align}\label{eq:GiventalFormula-TVar}
	& C{\,}^t\!\hat S^{-1}\hat\Psi {\,}^t\!\hat R\prod_{i=1}^2\tau_{KdV}(\{\Delta_i^{1/2}T^i_a\}_{a\geq 0},\Delta_i\epsilon^2),
	\end{align}
	where
	\begin{align}
	& {\,}^t\!\hat S \coloneqq 
	\left(e^{-\frac{\partial}{\partial q^1_1}} \hat S e^{\frac{\partial}{\partial q^1_1}}\right)\Big|_{q^i_a \to t^i_a} \,;\\ 
	& {\,}^t\!\hat R
	\coloneqq \left(e^{-\Psi^1_1\frac{\partial}{\partial Q^1_1}- \Psi^2_1\frac{\partial}{\partial Q^2_1}} \hat R e^{\Psi^1_1\frac{\partial}{\partial Q^1_1}+\Psi^2_1\frac{\partial}{\partial Q^2_1}}\right) \Big|_{Q^i_a \to T^i_a}\,.
	\end{align}
	Equation~\eqref{eq:GiventalFormula-TVar} is the standard expression for the Givental formula for the total descendent potential in the coordinates $\{t^i_a\}$ used in application, see e.g.~\cite{DSS,DOSS}. Its advantage is that at all steps of the computation of its coefficients one has to work with the formal power series. 
	
	At the second step we have to use some results from the theory of the Chekhov--Eynard--Orantin topological recursion~\cite{EO}. It is a recursive procedure that produces symmetric differentials from the small set of input data that consists of a Riemann surface $\Sigma$, two functions $x$ and $y$ defined on it, and a symmetric bi-differential $B$ on $\Sigma\times\Sigma$, (all these pieces of data are subjects to some extra conditions). This data is related to a choice of Dubrovin's superpotential for the Dubrovin--Frobenius manifolds~\cite{DNOPS2019}. We don't use the explicit formulation of the topological recursion itself, but we have to recall two results for the data given by the Riemann sphere $\mathbb{C}P^1$ with a global coordinate $z$ on it, functions $x=z+z^{-1}$, $y=z$, and the bi-differential $B(z_1,z_2) = dz_1dz_2/(z_1-z_2)^2$ from~\cite{Norbury,DumMul} and from~\cite{DNOPS2019,AnCheNoPe}. 
	
	It is proved in~\cite{Norbury,DumMul} that the topological recursion applied to this input data returns the symmetric $n$-differentials $\omega_{g,n}(z_1,\dots,z_n)$, $2g-2+n>0$, that expand near $z_1=\cdots=z_n=0$ in the variables $x_i=x(z_i)$, $i=1,\dots,n$, as 
	\begin{align}
	\omega_{g,n} & =  (-1)^n\sum_{k_1,\dots,k_n\geq 0 } C_{g,k_1+1,\dots,k_n+1} \prod_{i=1}^n \frac{dx_i}{x_i^{k_i+2}}
	\\ \notag &
	=(-1)^n\sum_{k_1,\dots,k_n\geq 0 } \frac{C_{g,k_1+1,\dots,k_n+1}}{\prod_{i=1}^n (k_i+1)!} \prod_{i=1}^n \frac{(k_i+1)!dx_i}{x_i^{k_i+2}}. 
	\end{align}

\begin{remark} This is one of the most studied cases of topological recursion, in particular, this equation follows also from the more general cases discussed in~\cite{ceo,EynBook,KazarianZograf,dops,DNOPS2019, AnCheNoPe,aceh,bdks}. Bear in mind, however, that some of this papers use a different convention for the topological recursion, which results in an extra overall $(-1)^n$ sign in the formulas for $\omega_{g,n}$ in some of these sources. 
\end{remark}

On the other hand, it is proved in~\cite{DNOPS2019,AnCheNoPe} that $\omega_n\coloneqq \sum_{g=0}^\infty \epsilon^{2g-2}\omega_{g,n}$ is given by 
\begin{equation} \label{eq:FirstFormulaWgn}
\omega_{n} = \sum_{\substack{i_1,\dots,i_n=1,2 \\ a_1,\dots,a_n\geq 0}} \left(\prod_{j=1}^n \frac{\partial}{\partial T^{i_j}_{a_j}}  \log{\,}^t\!\hat{R}\prod_{i=1}^2\tau_{KdV}(\{\Delta_i^{1/2}T^i_a\}_{a\geq 0},\Delta_i\epsilon^2)\right)\Bigg|_{T^i_a = 0} \prod_{j=1}^n d\Big(-\frac{d}{d x_j}\Big)^{a_j} \xi^{i_j} (z_j),
\end{equation}
where 
\begin{equation}
\xi^j(z) \coloneqq  \frac{dz}{d \sqrt{2(x(z)-x(p_j))}}\Bigg|_{z=p_j} \cdot \frac{1}{p_j-z}
\end{equation}
for $j=1,2$ and $p_1=1,p_2=-1$ being the critical points of $x$, and the whole expression~\eqref{eq:FirstFormulaWgn} is considered at the special point $(t^1,t^2) = (0,1)$ of the underlying Dubrovin--Frobenius manifold.  The choice of the square roots is required to be aligned with the choices made for $\Delta_1^{1/2},\Delta_2^{1/2}$. We have:
\begin{equation}
\frac{dz}{d\sqrt{2(x(z)-x(p_j))}}\Bigg|_{z=p_j} = \Delta_j^{-1/2}, \qquad j=1,2.
\end{equation}
An equivalent form of Equation~\eqref{eq:FirstFormulaWgn} in the flat frame is 
\begin{equation} \label{eq:SecondFormulaWgn}
\omega_{n} = \sum_{\substack{\alpha_1,\dots,\alpha_n=1,2 \\ a_1,\dots,a_n\geq 0}} \left( \prod_{j=1}^n \frac{\partial}{\partial t^{\alpha_j}_{a_j}}  \log\hat{\Psi}{\,}^t\!\hat{R}\prod_{i=1}^2\tau_{KdV}(\{\Delta_i^{1/2}T^i_a\}_{a\geq 0},\Delta_i\epsilon^2)\right)\Bigg|_{t^i_a = 0} \prod_{j=1}^n d \Big(-\frac{d}{d x_j}\Big)^{a_j} \tilde\xi^{\alpha_j} (z_j),
\end{equation}
where $\tilde{\xi}^\alpha = (\Psi^{-1})^\alpha_i \xi^i$, $\alpha=1,2$.  

\begin{remark} Note that the Dubrovin--Frobenius manifold that we consider here is in fact a Hurwitz Frobenius manifold in the terminology of Dubrovin. This means that Equation~\eqref{eq:SecondFormulaWgn} can also be derived from a general statement in~\cite{DNOPS2018}.
\end{remark}

In the third step of the proof, we recall that the action of ${\,}^t\!\hat S^{-1}$ for $2g-2+n>0$  amounts to the linear change of variables combined with the shift of the point of expansion given by
\begin{multline}
\left( \prod_{j=1}^n \frac{\partial}{\partial t^{\alpha_j}_{a_j}}  \log{\,}^t\!\hat{S}^{-1}\hat{\Psi}{\,}^t\!\hat{R}\prod_{i=1}^2\tau_{KdV}(\{\Delta_i^{1/2}T^i_a\}_{a\geq 0},\Delta_i\epsilon^2)\right)\Big|_{t^i_a=\delta^{i,2}\delta_{a,0}} =
\\ 
 = \sum_{\substack{0\leq \ell_j \leq a_j,\\j=1,\dots,n}} (S_\ell)^{\beta_j}_{\alpha_j} \left( \prod_{j=1}^n \frac{\partial}{\partial t^{\beta_j}_{a_j-\ell_j}}  \log\hat{\Psi}{\,}^t\!\hat{R}\prod_{i=1}^2\tau_{KdV}(\{\Delta_i^{1/2}T^i_a\}_{a\geq 0},\Delta_i\epsilon^2)\right)\Big|_{t^i_a = 0} .
\end{multline}
Note that at the special point $(t^1,t^2) = (0,1)$ we have $C(t^1,t^2) = 0$.
Therefore, the statement of the theorem is equivalent to the following identity:
\begin{align}
\res_{z=0} \frac{x^{k+1}}{(k+1)!} d \Big(-\frac{d}{d x}\Big)^{a} \tilde{\xi}^\alpha(z) =
\begin{cases}
0, & a> k \geq -1; \\
(S_{k-a})^\alpha_1, & k\geq a\geq 0. 
\end{cases} 
\end{align}
To this end, we just explicitly compute
\begin{align}
\xi^1(z) & = \frac{1}{\sqrt 2} \frac{1}{1-z}; & \xi^2(z) & = \frac{1}{\sqrt 2} \frac{\I}{1+z};
& \tilde\xi^1(z) & =  \frac{z}{1-z^2}; & \tilde\xi^2(z) & =  \frac{1}{1-z^2};
\end{align}
(recall~\eqref{eq:Deltas} and~\eqref{eq:PsiMatrix}), and then we see that indeed 
\begin{align}
& \res_{z=0} \frac{x^{k+1}}{(k+1)!} d \Big(-\frac{d}{d x}\Big)^{a} \frac{z}{1-z^2} =
\res_{z=0} \Big(\frac{d}{d x}\Big)^{a+1}\frac{x^{k+1}}{(k+1)!} \frac{zd(z+z^{-1})}{z^2-1}
\\ \notag
&
=\begin{cases}
0, & a> k \geq -1; \\
(m!)^{-2}, & k\geq a\geq 0,\ k-a=2m; \\
0, & k\geq a\geq 0,\ k-a=2m+1
\end{cases} 
=
\begin{cases}
0, & a> k \geq -1; \\
(S_{k-a})^1_1, & k\geq a\geq 0
\end{cases} 
\end{align}
(cf. Equation~\eqref{eq:SMatrixSpecialPoint}), and, analogously, 
\begin{align}
& \res_{z=0} \frac{x^{k+1}}{(k+1)!} d \Big(-\frac{d}{d x}\Big)^{a} \frac{1}{1-z^2} =
\res_{z=0} \Big(\frac{d}{d x}\Big)^{a+1}\frac{x^{k+1}}{(k+1)!} \frac{d(z+z^{-1})}{z^2-1}=
\\ \notag
&
=\begin{cases}
0, & a> k \geq -1; \\
0, & k\geq a\geq 0,\ k-a=2m; \\
(m!(m+1)!)^{-1}, & k\geq a\geq 0,\ k-a=2m+1
\end{cases} 
=
\begin{cases}
0, & a> k \geq -1; \\
(S_{k-a})^2_1, & k\geq a\geq 0.
\end{cases} 
\end{align}
This completes the proof of the theorem for $2g-2+n>0$. 

In the fourth step of the proof we have to discuss the unstable cases $(g,n)=(0,1)$ and $(g,n)=(0,2)$. Consider $(g,n)=(0,1)$. According to~\cite[Section 3.1]{Givental2001},  the coefficient of $\epsilon^{-2} t^1_a$ in $\log \cD |_{\substack{t^2_0=1; t^2_a = 0, a\geq 1}} $ is given by
\begin{equation}
[\epsilon^{-2} t^1_a] \log \cD |_{\substack{t^2_0=1\\t^2_a = 0, a\geq 1}} = \eta_{1\alpha} (S_{a+2})^\alpha_1.
\end{equation}
Then, using Equation~\eqref{eq:SMatrixSpecialPoint} we see indeed that 
\begin{align}
\eta_{1\alpha} (S_{a+2})^\alpha_1 &= 
\begin{cases}
0, & a = 2m; \\
\frac{1}{(m+1)!(m+2)!}, & a=2m+1
\end{cases}
\\ &=
\begin{cases}
0, & a = 2m; \\
\frac{1}{(2m+2)!}\frac{(2m+2)!}{(m+1)!(m+2)!}, & a=2m+1
\end{cases}
\\
&
=\frac{C_{0,a+1}}{(a+1)!}.
\end{align}
Consider $(g,n)=(0,2)$. According to~\cite[Section 3.1]{Givental2001},  the coefficient of $\epsilon^{-2} t^1_at^1_b$ in $\log \cD |_{\substack{t^2_0=1; t^2_a = 0, a\geq 1}}  $ is given by
\begin{align}
[\epsilon^{-2} t^1_at^1_b] \log \cD|_{\substack{t^2_0=1\\t^2_a = 0, a\geq 1}} = [z^a w^b] \frac{-\eta_{11} + \sum_{m,n=0}^\infty (S_m)^\mu_1 z^m  (S_n)^\nu_1 w^n \eta_{\mu\nu}}{z+w} .
\end{align}
Therefore, using explicit formulas~\eqref{eq:SMatrixSpecialPoint}, we have
\begin{equation}\label{eq:2ptGiventalExplicit}
(z+w)\sum_{a,b=0}^\infty [\epsilon^{-2} t^1_at^1_b] \log \cD |_{\substack{t^2_0=1\\t^2_a = 0, a\geq 1}}
=\sum_{p,q=0}^\infty \left(\frac{z^{2p}w^{2q+1}}{(p!)^2q!(q+1)!} +\frac{z^{2p+1}w^{2q}}{p!(p+1)!(q!)^2}\right).
\end{equation}
On the other hand, it is proven in~\cite{DumMul} that 
\begin{align}
\sum_{k_1,k_2\geq 0} \frac{C_{0,k_1+1,k_2+1}}{(k_1+1)!(k_2+1)!} \prod_{i=1}^2 \frac{(k_i+1)! dx_i}{x_i^{k_i+2}} 
& = d_1d_2 \log\left(\frac{z_1^{-1}-z_2^{-1}}{x_1-x_2}\right)
\\ \notag 
& =-d_1d_2\log(1-z_1z_2).
\end{align}
Therefore,
\begin{align}
& (z+w) \sum_{k_1,k_2\geq 0} \frac{C_{0,k_1+1,k_2+1}}{(k_1+1)!(k_2+1)!} z^{k_1}w^{k_2} 
\\ \notag 
& = \sum_{k_1,k_2\geq 0} (z^{k_1+1}w^{k_2}+z^{k_1}w^{k_2+1})\res_{z_1=0}\res_{z_2=0} 
\frac{x_1^{k_1+1}}{(k_1+1)!}\frac{x_2^{k_2+1}}{(k_2+1)!} \cdot-d_1d_2\log(1-z_1z_2)
\\ \notag 
& = \sum_{\substack{k_1,k_2\geq 0\\k_1+k_2\geq 1}} z^{k_1}w^{k_2}\res_{z_1=0}\res_{z_2=0} 
\frac{x_1^{k_1}}{(k_1)!}\frac{x_2^{k_2}}{(k_2)!} \cdot \big(dx_1d_2\log(1-z_1z_2)+dx_2d_1\log(1-z_1z_2)\big)
\\ \notag 
& = \sum_{\substack{k_1,k_2\geq 0\\k_1+k_2\geq 1}} z^{k_1}w^{k_2}\res_{z_1=0}\res_{z_2=0} 
\frac{x_1^{k_1}}{(k_1)!}\frac{x_2^{k_2}}{(k_2)!} \cdot \Big(\frac{1}{z_1}+\frac{1}{z_2}\Big) dz_1dz_2
\\ \notag 
& = \sum_{p.q\geq 0} \left(z^{2p}w^{2q+1} \frac{1}{(p!)^2 q!(q+1)!} + z^{2p+1}w^{2q} \frac{1}{p!(p+1)! (q!)^2} \right).
\end{align}
The latter expression coincides with the right hand side of Equation~\eqref{eq:2ptGiventalExplicit}, which proves the $(g,n)=(0,2)$ case of the theorem. 
\end{proof}

\begin{remark} Note that the computations done above for $2g-2+n>0$ are very close to the computations performed in~\cite{DOSS} (see also~\cite{FLZ}) for the Gromow-Witten theory of $\mathbb{C}P^1$ mentioned in Digression~\ref{dig:GenusExpansion}. This is explained by the following two facts. First, note that the $S$-matrix at the special point $(t^1,t^2)=(0,1)$ with $\psi=0$, see Equation~\eqref{eq:SMatrixSpecialPoint}, conjugated by $\eta$,  is equal to the $S$-matrix of the Dubrovin--Frobenius manifold given by the prepotential $\frac 12 (t^1)^2 t^2 + e^{t^2}$. Second, note that the $\tilde\xi^i$-functions in the above computation are obtained from the corresponding $\tilde\xi^i$-functions in the computations in~\cite{DOSS} by the interchanging of the superscripts. 
\end{remark}

\subsection{The descendent potential and the KP hierarchy} Consider the generating function $\cZ$ for the higher genera Catalan numbers 
\begin{equation}
\cZ \coloneqq \exp \left( \sum_{g\geq 0} \epsilon^{2g-2} \sum_{n\geq 1} \frac{1}{n!} \sum_{k_1,\dots,k_n\geq 1} C_{g,k_1,\dots,k_n} \prod_{i=1}^n \KPt_i \right)
\end{equation}
as a formal power series in the variables $\KPt_1,\KPt_2,\dots$. It is proved in~\cite[Theorem 5.2]{GouldenJackson} that $\cZ$ is a tau-function of the KP-hierarchy (more precisely, one should speak of $\hbar$-KP hierarchy in the sense of~\cite{TakaTake,NataZa}for $\hbar=\epsilon^2$, see~\cite{Andr}).
In particular, it is proved in~\cite[Theorem 5.2]{GouldenJackson} that $\cZ$ takes the following form:
\begin{equation}
	\cZ = \left( \sum_\lambda s_\lambda(\{\KPp_i\}_{i\geq 1}) s_\lambda(\{\tilde\KPp_i\}_{i\geq 1}) \prod_{(i,j)\in\lambda} (1+\epsilon(i-j))\right)\Bigg|_{\substack{\KPp_i = i\KPt_i/\epsilon,\ i\geq 1 \\
	\tilde\KPp_i = \delta_{i2}/\epsilon,\ i\geq 1}}\,.
\end{equation}
Here the sum is taken over all Young diagrams $\lambda$ including the empty one, and $s_\lambda$ are the Schur functions considered in the two copies of the power sums variables $\KPp_1,\KPp_2,\dots$ and $\tilde\KPp_1,\tilde\KPp_2,\dots$. This equation makes $\cZ$ a special case of the so-called hypergeometric family of KP tau-functions introduced and considered in~\cite{KMMM,OrlSch}. So, Theorem~\ref{thm:SpecializationOfD} has the following corollary.
\begin{corollary} $\cD|_{t^1_d = (d+1)! \KPt_{d+1}, t^2_0=1, t^2_d= 0, d\geq 1}$ is a KP tau-function.
\end{corollary}

Our main result can be considered as a refinement of this Corollary, cf.~Digression~\ref{dig:specialInt} in the Introduction. Namely, one of the ways to interpret the results that we prove in Sections~\ref{sec:HirotaDescendant} and~\ref{sec:LaxFormulation} is that $\cD|_{t^1_d = (d+1)! \KPt_{d+1}, t^2_0=1, t^2_d= 0, d\geq 1}$ appears to satisfy a particular rational reduction of the KP hierarchy~\cite{BonoraXiong,Cheng,ChengLi,KonoSS,Krichever,HelminckVdL,LiuZhangZhou} that possesses extra symmetries, which, once added explicitly form a second series of times in $\cD(\{t^i_d-\delta^{i,2}\delta_{d,0}\}_{i=1,2; d\geq 0})$.

\begin{remark} Note that though our approach to the construction of the rationally contrained KP hierarchy in the Hirota form does not use the results mentioned in this section, the change of variables $t^1_d = (d+1)!\KPt_{d+1}$, $d\geq 0$ that we have to apply to turn the natural variables of the total descendent potential into the standard KP variables emerges in a natural way. Indeed, in~Equation~\eqref{HQE185}, which is the formulation of the equations that we obtain in the Hirota form, the variables $q^1_d,\bar q^1_d$ are shifted by $\mp \epsilon d!/\lambda^{d+1}$, which matches precisely that in the standard formulation of the Hirota bilinear equations for KP hierarchy the corresponding shifts for $\KPt_{d+1},\bar \KPt_{d+1}$ would be $\mp \epsilon /((d+1)\lambda^{d+1})$.
\end{remark}

\begin{remark} The step from the standard KP hierarchy for $\cZ$ to its rational reduction fits very well into the context of a recent paper of Takasaki~\cite{Taka}, where he studies the possible reductions of the \emph{lattice} KP hierarchy for several families of hypergeometric tau-functions.
\end{remark}

\begin{remark} The function $\cZ$ in variables $\KPp_i$, $i\geq 1$ is expanded as 
\begin{equation}
	\cZ \coloneqq \exp \left( \sum_{g\geq 0} \epsilon^{2g-2} \sum_{n\geq 1} \frac{1}{n!} \sum_{k_1,\dots,k_n\geq 1} D_{g,k_1,\dots,k_n} \prod_{i=1}^n \KPp_i \right)
\end{equation}
(recall the definition of $D_{g,k_1,\dots,k_n}$ in Digression~\ref{dig:Catalan} in the Introduction). The obvious interpretations of the numbers $D_{g,k_1,\dots,k_n}$ that follow directly from the definition of the higher genera Catalan numbers is via the (weighted) enumeration of ribbon graphs, non-rooted maps, or Grothendieck dessins d'enfants for the strict Belyi functions, see e.g.~\cite{GouldenJackson,Norbury}. It is proved in~\cite{HarnadOrlov} that these numbers have also an interpretation in the framework of the theory of weighted Hurwitz numbers as the so-called $2$-orbifold strictly monotone Hurwitz numbers (see also~\cite{als} for a different proof).
\end{remark}

\section{The period vectors} \label{sec:PeriodVectors} 

\subsection{Definition and main properties}

We denote by $I^{(l)}_{e_i}(t, \lambda)$ for $l \in \Z$ and $i=1,2$ the period vector defined by its asymptotic behaviour near $\lambda \sim u^i$. Let us fix two cuts $L_i = u^i + e^{\I \pi/2} \R_+$ in the $\lambda$-plane, and the roots of $\lambda-u^i$ determined by the principal branch of the logarithm near each $u^i$ in the cut plane $\C \setminus \cup_i L_i$. We uniquely define the period vectors as follows.
\begin{proposition} \label{proposition14}
For each $l\in\Z$ and $i=1,2$ there exists a unique multivalued holomorphic solution 
$I_{e_i}^{(l)}(t,\lambda)$ defined on $(M \times \C) \setminus \Delta_\lambda$ with values in $\C^2$ 
of the equation
\begin{equation} \label{eqU}
(\cU- \lambda) \frac{\partial I^{(l)}}{\partial \lambda} = (\mu  + l +\frac12) I^{(l)} 
\end{equation}
such that 
\begin{equation} \label{normali}
I_{e_i}^{(l)} = \frac{(-1)^l}{\sqrt{2\pi}} \Gamma(l+1/2) (\lambda - u_i)^{-l-\frac12} ( \Psi^{-1} e_i + O(\lambda-u^i) )
\end{equation}
for $\lambda \in \C \setminus \cup_i L_i$, and such that the analytic continuation of $I_{e_i}^{(l)}$ along a small path $\gamma_i$ surrounding $u^i$ is equal to $- I_{e_i}^{(l)}$.
\end{proposition}
A general proof of this statement can be easily adapted from the case $l=0$ shown in~\cite[lemma 5.3]{dub99b}.

The period vectors thus defined satisfy also the following equation
\begin{equation} \label{dit}
\frac{\partial I^{(l)}}{\partial t^i} = - \frac{\partial }{\partial t^i} \bullet  \frac{\partial I^{(l)}}{\partial \lambda} ,
\end{equation}
and one can easily check that, in general
\begin{equation} \label{dat}
I^{(l+1)}_{e_i} = \frac{\partial I^{(l)}_{e_i}}{\partial \lambda \ \ \ } ,
\quad l \in \Z,
\end{equation}
\begin{equation}  
I^{(l-1)}_{e_i}(t, \lambda) = \int_{u^i}^\lambda I^{(l)}_{e_i}(t,\rho) \ d\rho, 
\quad l \leq 0.
\end{equation}
In the following, for $a = a_1 e_1 + a_2 e_2 \in \C^2$ we denote $I^{(l)}_a = a_1 I^{(l)}_{e_1} + a_2 I^{(l)}_{e_2}$ the corresponding solution of~\eqref{eqU}.

\begin{remark}[To be used in Section~\ref{sec:HirotaAncestor}] \label{rem:Imin1o} With this setup it is straightforward to check that the vector $I^{(-1)}_{e_1}-I^{(-1)}_{e_2}$ is constant on $M$, with the values of the components $(I^{(-1)}_{e_1})^1-(I^{(-1)}_{e_2})^1=-\pi\I$, $(I^{(-1)}_{e_1})^2-(I^{(-1)}_{e_2})^2=0$.
\end{remark}

\subsection{Monodromy} \label{sec:monodromy}

Let $\pi = \pi_1(\C \setminus \{ u^1, u^2\} )$ be the fundamental group of the pointed $\lambda$-plane with base point $\lambda_0$. Denote by $\gamma_i$, $i=1,2$, the  generators of $\pi$ corresponding to the two small loops around the points $u^i$ in counterclockwise direction connected to $\lambda_0$ by paths in the cut $\lambda$-plane. Moreover let us denote by $\gamma^* I$ the analytic continuation of a multivalued analytic function $I$ on the pointed plane $\C \setminus \{ u^1, u^2\}$ along the loop $\gamma \in \pi$. 

Denoting by $(,)_\lambda$ the flat pencil $g - \lambda \eta$ on the cotangent to $M$, let us define the symmetric bilinear form $<,>$ on $\C^2$ by 
$<a,b>:=(\eta_* I^{(0)}_{a} , \eta_* I^{0)}_b )_\lambda$, which is well-known to be independent of $t$ and $\lambda$. Here $\eta_*$ is the isomorphism from the tangent to the cotangent spaces to $M$ induced by the metric. In our case we have that the corresponding matrix $G_{ij} = <e_i, e_j>$ is
\begin{equation}  
G=-\frac12\begin{pmatrix}
1 & 1 \\ 1 & 1
\end{pmatrix}.
\end{equation}

We call reflection along $w \in \C^2$ the involution of $\C^2$ given by 
\begin{equation}  
v \mapsto v - 2 \frac{<v, w>}{<w,w>} w. 
\end{equation} 
Let us denote by $\gamma_i$ the reflection along $e_i$, and define the group homomorphism $\pi \to GL(\C^2)$ by sending the loop $\gamma_i$ to the reflection $\gamma_i$. We have that:
\begin{proposition} \label{prop:MonodromyOnI}
For each $\gamma \in \pi$ we have $\gamma^* I^{(l)}_a = I^{(l)}_{\gamma a}$.
\end{proposition}
This is just a reformulation of a general result, see~\cite[lemma 5.3]{dub99b}.

Explicitly the monodromy action is given by
\begin{proposition} \label{prop:monodromyC2}
The action of the generators of $\pi$ on the canonical basis of $\C^2$ is given by
\begin{align} \label{gamma}
&\gamma_1 e_1 = - e_1, \qquad & &\gamma_1 e_2 = e_2 - 2 e_1, \\
&\gamma_2 e_1 = e_1 - 2 e_2, \qquad & &\gamma_2 e_2 = - e_2 .
\end{align}
\end{proposition}

The action of $\gamma_1$, resp. $\gamma_2$, corresponds to the horizontal, resp. vertical, reflection w.r.t. the invariant subspace $V$ spanned by $e_1 - e_2$. Note that $\gamma_2 \gamma_1$ sends $a = a_1 e_1 + a_2 e_2$ to $a + 2 (a_1+a_2) (e_2-e_1)$, therefore acts on the affine line $a_1 +a_2 = b$ by translation by $-2b$ along $e_1-e_2$.
We also denote $\pi_i$ the projections to the invariants subspace, $\pi_1 a = a_2(e_2-e_1)$ and $\pi_2 a = a_1 (e_1 - e_2)$. 

We have therefore two types of orbits: the trivial orbits are given by the points of the invariant subspace, and the infinite orbits $\{ a, \gamma_1a \}  + \Z 2 (a_1 +a_2) (e_1-e_2)$ for $a = (a_1, a_2) \in \C^2$ not invariant, i.e., $a_1 +a_2 \not=0$. 

\begin{remark}
The representation of $\pi$ on the space $\C^2$ should not be confused with the action on the space of solutions of~\eqref{eqU}. Indeed we have that $I^{(0)}_{e_1} = I^{(0)}_{e_2}$ and consequently $I^{(l)}_{e_1} = I^{(l)}_{e_2}$ for $l\geq0$, so the period vectors span a one-dimensional subspace of the solution space.
For negative $l$ the two period vectors differ by polynomials in $\lambda$, so they are indeed a basis of the solution space of~\eqref{eqU}.
\end{remark}

\subsection{At a special point}

In the rest of this section we work at a special point $t_{sp}$ of $M$ given by $t^1=0$ and $t^2=1$. This corresponds to canonical coordinates $u^1= - u^2 = 2$. 
As can be easily checked, the two solutions of the system~\eqref{eqU} at the special point $t_{sp}$ are given by 
\begin{equation} \label{solution}
I^{(0)}_{e_1}(t_{sp}, \lambda) =  I^{(0)}_{e_2}(t_{sp}, \lambda) = 
\begin{pmatrix}
1 \\ \lambda/2
\end{pmatrix} 
(\lambda^2 -4)^{-\frac12}.
\end{equation}
This expression defines a holomorphic $\C^2$-valued function on the cut $\lambda$-plane given by $\C\setminus \left\{  (2 + i \R_+) \cup (-2 + i \R_+) \right\}$, where $\sqrt{\lambda^2 - 4} = \sqrt{\lambda-2} \sqrt{\lambda +2}$ and $\sqrt{\lambda \mp 2}$ denote the principal branches of the square root near $\pm 2$.

\subsection{Asymptotics of the period vectors for $\lambda \sim u^i$}
By asymptotic series near $u^i$ we mean a formal Laurent  series in $(\lambda -u^i)^{1/2}$. By asymptotic expansion of a multivalued function $g(\lambda)$ for $\lambda\sim u^i$ we mean an asymptotic series near $u_i$ which satisfies the asymptotic condition on a cut neighbourhood of $u_i$ for a choice of branch of $g(\lambda)$  and of $\sqrt{\lambda-u^i}$. Of course, according to the theory of normal formals of solutions near regular singularities, these asymptotic expansions are actually convergent. 

At the special point $t_{sp}$ we can easily compute the asymptotic expansions of the period vectors.
\begin{lemma} The asymptotic expansions of the period vector $I^{(l)}_{e_i}$ for $\lambda \sim u^i$ at the special point $t_{sp}$ are given by
\begin{equation} \label{asympt-u1} 
I^{(l)}_{e_1} (t_{sp},\lambda) \sim  \sum_{k \geq0} \frac{(-4)^{-k}}{k!} \begin{pmatrix}
\left( \frac12 \right)_k \left( \frac12 \right)_k\\ 
\frac{-2}{2k-1} \left( \frac12 \right)_{k+1} \left( \frac12 \right)_k 
\end{pmatrix} \partial_\lambda^{l-k+1}  \sqrt{\lambda-2} ,  \quad \lambda\sim u_1 =2  
\end{equation}
\begin{equation} \label{asympt-u2}
I^{(l)}_{e_2} (t_{sp},\lambda) \sim \I \sum_{k \geq0} \frac{4^{-k} }{k!} \begin{pmatrix}
\left( \frac12 \right)_k \left( \frac12 \right)_k \\
\frac2{2k-1} \left( \frac12 \right)_{k+1} \left( \frac12 \right)_k
\end{pmatrix}
 \partial_\lambda^{l-k+1} 
  \sqrt{\lambda+2} ,  \quad \lambda\sim u_2 =-2
\end{equation}
on the cut $\lambda$-plane $\C\setminus \left(  (2 + i \R_+) \cup (-2 + i \R_+) \right)$ with principal branches of the roots $\sqrt{\lambda \pm 2}$. 
\end{lemma}
\begin{proof}
The formulas for the case $l=0$ are obtained by a simple expansion of~\eqref{solution}. For other values of $l$ they follow by integration or derivation. 
\end{proof}

\subsection{Asymptotics of the period vectors for $\lambda \sim \infty$}
At the special point $t_{sp}$ of $M$ the period vectors $I_{e_1}^{(0)}=I_{e_2}^{(0)}$ have the following asymptotic expansion\footnotemark~ for $|\lambda| \sim \infty $ and $\arg \lambda \not= \pi/2$:
\begin{equation}  
I_{e_i}^{(0)}(t_{sp},\lambda) \sim I^{(0)}_{asy}(\lambda)
\end{equation}
where
\begin{equation}  
 I^{(0)}_{asy} :=
\frac12 \sum_{s\geq0} 
\frac{(2s)!}{s!s!} \lambda^{-2s-1}
\begin{pmatrix}
2 \\ \lambda
\end{pmatrix}
= \begin{pmatrix}
\frac1{\lambda} +\frac2{\lambda^3} + \cdots
\\ \frac12+\frac1{\lambda^2} +\frac3{\lambda^4} + \cdots
\end{pmatrix} .
\end{equation}
\footnotetext{To see this easily note that, once having fixed $\arg\lambda \not= \pi/2$, for $|\lambda|$ big enough one has  $$(\lambda-2)^{-1/2} (\lambda+2)^{-1/2} = \lambda^{-1} (1-4 \lambda^{-2})^{-1/2}.$$
However, for $\arg \lambda=\pi/2$ the right-hand side gets a minus signs, so also the asymptotic expansion above.
}
By taking derivatives we obtain the asymptotic expansions of the period vectors $I_{e_i}^{(l)}$ with $l>0$:
\begin{equation}  
 I_{e_i}^{(l)}(t_{sp},\lambda) \sim  I^{(l)}_{asy} := \partial_\lambda^l I^{(0)}_{asy} 
 \end{equation}
 where we have explicitly
 \begin{equation}  
  I^{(l)}_{asy}=
(-1)^l  \sum_{s\geq0}   \frac{(2s+l)!}{s! (s+1)!}\begin{pmatrix}
 s+1\\
  (2s+l+1)\lambda^{-1}
 \end{pmatrix}
 \lambda^{-2s-l-1} .
 \end{equation}

Let us define $I^{(-l)}_{formal}$ for $l>0$ as the iterated formal integration in $\lambda$ (without integration constants)  of the asymptotic expansion $I^{(0)}_{asy}$, i.e., 
\begin{equation}  
 I^{(-l)}_{formal} := \partial_\lambda^{-l} I^{(0)}_{asy}, \qquad l >0,
\end{equation}
where we have set $\partial_\lambda^{-1} \lambda^{-1} = \log \lambda$, $\partial_\lambda^{-1}( \lambda^p \log \lambda ) = \frac{\lambda^{p+1}}{p+1} \left(\log \lambda - \frac1{p+1}\right)$ for $p\geq0$, and 
  $\partial_\lambda^{-1} \lambda^{k} = \lambda^{k+1}/(k+1)$ for $k\not= -1$.
Explicitly we have that the two components of $ I^{(-l)}_{formal}$ are equal to 
\begin{equation} \label{formal1}
(I_{formal}^{(-l)})_1 = \sum_{0 \leq s \leq \frac{l-1}2} \frac{\lambda^{l-2s-1}(\log \lambda - \ch(l-2s-1))}{s!s!(l-2s-1)!} + \sum_{s \geq \frac{l}2} \frac{(2s-l)! \lambda^{l-2s-1}}{s!s! (-1)^l},
\end{equation}
\begin{equation} \label{formal2}
(I_{formal}^{(-l)})_2 = \frac12 \frac{\lambda^l}{l!}  
-\sum_{1\leq s \leq \frac{l}2} \frac{\lambda^{l-2s}(\log \lambda - \ch(l-2s))}{s!(s-1)!(l-2s)!} 
+\sum_{s \geq \frac{l+1}2} \frac{ (2s-l-1)! \lambda^{l-2s} }{s!(s-1)! (-1)^l}.
\end{equation}
In these expressions $\log \lambda$ denotes the principal branch of the logarithm on the complex plane $\C \setminus \I \R_+$ cut along the positive imaginary axis.

Finally we can give the asymptotic expansion of all the period vectors for $\lambda\sim \infty $.
\begin{proposition}\label{prop:asymptoticexpansionspecialpoint}
The period vectors $I^{(l)}_{e_i}(t_{sp},\lambda)$, $i=1,2$, at the special point $t_{sp}$ of $M$ have the following asymptotic expansions for $\lambda\sim\infty $ and $\arg \lambda \not= \pi/2$:
\begin{equation}
I^{(l)}_{e_i}(t_{sp},\lambda) \sim I^{(l)}_{asy}(\lambda) ,\quad \text{for } l\geq0,
\end{equation}
and 
\begin{equation} \label{asyneg}
I^{(l)}_{e_i}(t_{sp},\lambda) \sim  I^{(l)}_{formal}(\lambda) + P_i^{(l)}(\lambda), \quad \text{for }l <0 ,
\end{equation}
where
\begin{equation}
P_1^{(l)}(\lambda) := \begin{pmatrix}
\sum_{\substack{i,j\geq 0 \\ 2i+j=-l-1}} \frac{\ch(i)}{i!^2}\frac{\lambda^j}{j!} \\ 
-\sum_{\substack{i,j\geq 0 \\ 2i+1+j=-l-1}} \frac12\frac{\ch(i)+\ch(i+1)}{i!(i+1)!}\frac{\lambda^j}{j!}
\end{pmatrix},
\end{equation}
\begin{equation}
P_2^{(l)}(\lambda) := 
\begin{pmatrix}
\sum_{\substack{i,j\geq 0 \\ 2i+j=-l-1}} \frac{\ch(i) +\pi\mathsf{i}}{i!^2}
\frac{\lambda^j}{j!} \\ 
-\sum_{\substack{i,j\geq 0 \\ 2i+1+j=-l-1}} \frac12\frac{\ch(i)+\ch(i+1)+2\pi\mathsf{i}}{i!(i+1)!}\frac{\lambda^j}{j!}
\end{pmatrix} .
\end{equation}
\end{proposition}
\begin{proof}
Let us first compute the asymptotic expansion of $I^{(-1)}_{e_i}$ at the special point, by observing that
\begin{align}  
I^{(-1)}_{e_i} (t_{sp},\lambda) &= \int_{u^i}^{\lambda} I_{e_i}^{(0)}(t_{sp},\rho) d \rho \\
&= \int_{u^i}^\lambda 
\begin{pmatrix}
\frac1\rho \\ \frac12
\end{pmatrix} 
d\rho 
+ \int_{u^i}^\infty 
\left( I^{(0)}_{e_i} -\begin{pmatrix}
\frac1\rho \\ \frac12
\end{pmatrix}\right) d\rho 
+ \int_{\infty}^\lambda  
\left( I^{(0)}_{e_i} -\begin{pmatrix}
\frac1\rho \\ \frac12
\end{pmatrix}\right) d\rho.
\end{align}
In this expression the first integral contributes the linear and logarithmic terms appearing in the formal asymptotics plus some constants, while the last integral  exactly reproduces the negative powers of $\lambda$. Therefore $I^{(-1)}_{e_i} (t_{sp},\lambda)$ is asymptotic to $I^{(-1)}_{formal}$ plus a constant equal to
\begin{equation}  
\begin{pmatrix}
-\log u^i \\ - \frac{u^i}2 
\end{pmatrix}
+ \int_{u^i}^\infty 
\left( I^{(0)}_{e_i} -\begin{pmatrix}
\frac1\rho \\ \frac12
\end{pmatrix}\right) d\rho .
\end{equation}
Evaluating the integral one obtains exactly the constants given by 
\begin{equation}  
P^{(-1)}_1 = \begin{pmatrix}
0 \\ 0
\end{pmatrix}, \qquad 
P^{(-1)}_2 = \begin{pmatrix}
\pi \I \\ 0
\end{pmatrix},
\end{equation}
hence the formula~\eqref{asyneg} is valid for $l=-1$. 

Let us now show by induction that~\eqref{asyneg} is valid for $l\leq-2$. In this case we have that $\mu + l+ 1/2$ is invertible, so we can use~\eqref{eqU} to write at the special point $t_{sp}$ that
\begin{equation}  
I^{(l)}_{e_i}= (\mu + l +\frac12)^{-1}(\cU-\lambda)  I^{(l+1)}_{e_i}, 
\end{equation}
which is asymptotic to 
\begin{equation}  
(\mu + l +\frac12)^{-1}(\cU-\lambda) \left( I^{(l+1)}_{formal} + P_i^{(l+1)} \right)
\end{equation}
by inductive assumption. Proving that this asymptotic expansion is equal to $I^{(l)}_{formal} + P^{(l)}_i$ is equivalent to prove that 
\begin{equation} \label{toprove}
(\cU - \lambda) \partial_\lambda I^{(l)}_{formal} - (\mu+l+\frac12) I^{(l)}_{formal} + (\cU - \lambda) P_i^{(l+1)} - (\mu +l+\frac12) P^{(l)}_i
\end{equation}
is zero. Notice that $\partial_\lambda P^{(l)}_i = P^{(l+1)}_i$, therefore deriving this expression with respect to $\lambda$ amounts to send $l$ to $l+1$ in which case we know that it is zero by inductive assumption, by substituting~\eqref{asyneg} in~\eqref{eqU}. We conclude that~\eqref{toprove} is a constant. By induction it is also easy to see that $(\cU - \lambda) \partial_\lambda I^{(l)}_{formal} - (\mu+l+\frac12) I^{(l)}_{formal}$ is a polynomial, so to evaluate~\eqref{toprove} it is sufficient to set $\lambda=0$. 

Since we have defined $I^{(l)}_{formal}$ by formal integration without constant coefficient, we have that
\begin{equation}  
\left[ (\cU - \lambda) \partial_\lambda I^{(l)}_{formal} - (\mu+l+\frac12) I^{(l)}_{formal} \right]_{\lambda=0} 
\end{equation}
is given by minus the coefficient of $\lambda^{-1}$ in $I^{(l+1)}_{formal}$ which can be read off~\eqref{formal1} and~\eqref{formal2}, and which cancels with
\begin{equation}  
\left[ \cU P^{(l+1)}_i - (\mu+l+\frac12) P^{(l)}_i \right]_{\lambda=0} = 
\begin{pmatrix}
\frac1{(\frac{-l-1}{2})!^2} \quad  (\text{$l$ odd}) \\ 
\frac{-1}{(-\frac{l}2-1)! (-\frac{l}2)!} \quad (\text{$l$ even})
\end{pmatrix}  .
\end{equation} 
\end{proof}

\section{The vertex operators} \label{sec:VertexOperators}

In this section we use the period vectors of Section \ref{sec:PeriodVectors} to define certain vertex operators and we compute the action by conjugation of the $R$ and $S$ Givental group elements on them.

\subsection{Vertex loop space elements} \label{sec:VertexLoopSpaceElements}

For $a= (a_1, a_2) \in \C^2$ let us define
\begin{equation}
\cf_a(t,\lambda,z) = \sum_{l\in\Z}  I^{(l)}_a(t,\lambda) (-z)^l .
\end{equation}
The associated vertex operator is defined as the exponential of the quantisation of linear Hamiltonian $h_{\cf_a}$ of $\cf_a$, i.e.,
$\Gamma^a = e^{\widehat{\cf_a}}=e^{\widehat{(\cf_a)_-}}e^{\widehat{(\cf_a)_+}}$.

As a consequence of equations~\eqref{eqU}, ~\eqref{dit} and ~\eqref{dat} we have that 
\begin{corollary} 
The functions $\cf_a$ for $a\in \C^2$ satisfy
\begin{align}  
&-z \frac{\partial \cf_a}{\partial \lambda} = \cf_a , \qquad 
z \frac{\partial \cf_a}{\partial t^i} = \frac{\partial }{\partial t^i} \bullet \cf_a,  \qquad 
&(z\frac{\partial }{\partial z} + \lambda \frac{\partial }{\partial \lambda} + E ) \cf_a = (- \mu -\frac12 )\cf_a.   
\end{align}
\end{corollary}

\subsection{Asymptotics for $\lambda \sim u^i$}
Let us define the following formal $\C$-valued Laurent series in $z$ with coefficients which are multivalued functions on $\C^*$ corresponding to the vertex operator of the KdV hierarchy
\begin{equation}  
\cf_{KdV}(\lambda, z) = \sum_{l \in \Z} I^{(l)}_{KdV} (\lambda) (-z)^l, \qquad 
I^{(l)}_{KdV}(\lambda) = \partial_\lambda^l (2\lambda)^{-1/2}, 
\end{equation}
where $\partial_\lambda^\pm$ is the formal differentiation/integration  in $\lambda$ as above. As above we consider the principal branches of the roots on the cut $\lambda$-plane $\C \setminus \I \R_+$. 

\begin{proposition}
For $i=1,2$ we have the equality of asymptotic series at $\lambda \sim u^i$
\begin{equation} \label{fuiasy}
\cf_{e_i}(t,\lambda,z) = \Psi^{-1}(t) R(t,z) \cf_{KdV}(\lambda-u^i,z) e_i 
\end{equation}
\end{proposition}
\begin{proof}
By substitution it is easy to check that the formula holds at the special point $t_{sp}$. It is then sufficient to observe that the two sides of the equality satisfy the same equation in $u^i$. 
\end{proof}

\subsection{The functions $\cW_{a,b}(t,\lambda)$}
Given $a,b \in \C^2$ let us define the function
\begin{equation}
\cW_{a,b} (t,\lambda)  = (I_a^{(0)}(t,\lambda) , I_b^{(0)}(t,\lambda) ) ,
\end{equation}
which is clearly symmetric and linear in $a$ and $b$. In our case we have in particular
\begin{equation}  
\cW_{a,a} = (a_1+a_2)^2 \cW_{e_i,e_i}
\end{equation}
for $a= a_1 e_1 + a_2 e_2 \in \C^2$, and at the special point $\cW_{e_i,e_i}(t_{sp},\lambda) = \lambda (\lambda-2)^{-1}(\lambda+2)^{-1}$. 

Notice that $\cW_{\pi_i a ,b}$ is always vanishing, since $I^{(0)}_{\pi_i a} =0$.

The generators of the fundamental group $\pi$ of the pointed plane acts on the integral of $\cW_{a,a}$ as follows.
\begin{lemma}
For $a\in \C^2$ we have that 
\begin{equation}
\gamma_i^* \left( \int_{\lambda_0}^{\lambda} \cW_{a,a}d\rho \right) = 
\int_{\lambda_0}^\lambda \cW_{\gamma_i a, \gamma_i a}  d\rho 
+ \pi \I  (a_1+a_2)^2.
\end{equation}
\end{lemma}
\begin{proof}
By deforming the path of integration we can write 
\begin{equation} \label{gammaW}	
\gamma_i^* \int_{\lambda_0}^\lambda \cW_{a,a} d\rho - \int_{\lambda_0}^\lambda \cW_{\gamma_i a, \gamma_i a} d\rho= 
\int_{\lambda_0}^{u^i} \left( \cW_{a,a} - \cW_{\gamma_i a,\gamma_i a} \right) d\rho
+\lim_{r \to 0^+} \int_{C(u^i, r)} \cW_{a,a} d\rho
\end{equation}
where $C(u^i, r)$ is the circle with center at $u^i$ and radius $r$. In our case $\cW_{\gamma_i a,\gamma_i a} = \cW_{a,a}$ so we just need to evaluate the last integral, which is equal to $2\pi \I$ times 
the residue of $(a_1+a_2)^2 \cW_{e_i, e_i} $ at $\lambda=u^i$, which equals $\frac12 (a_1+a_2)^2$ by the normalisation~\eqref{normali}.
\end{proof}

\subsection{The functions $c_a(t,\lambda)$} \label{sec:DefinitionCa}
Let us consider an orbit of the action of the fundamental group $\pi$ of the pointed $\lambda$-plane on $\C^2$. The elements in such orbit can be parametrised as 
\begin{equation}  
\begin{pmatrix}
r \\ -r
\end{pmatrix} 
+ \frac{b}2 
\begin{pmatrix}
(-1)^k + 2 k \\ (-1)^k - 2 k
\end{pmatrix},
\quad k \in \Z.
\end{equation}
We define
\begin{equation}
c_a(t,\lambda) = d(a) \exp \left[ - \int_{\lambda_0}^{\lambda} \cW_{a,a}(t,\rho) d\rho  \right] .
\end{equation}
where $d(a)$ is a function defined on the chosen orbit such that $c_a$ is covariant under the action of $\pi$, i.e., $\gamma c_a = c_{\gamma a}$ for any $\gamma \in \pi$. This is equivalent to the following condition on the function $d(a)$
\begin{equation}  
d(\gamma_i a) = e^{\pi \I (a_1 + a_2)^2} d(a). 
\end{equation}
The function $d(a)$ is therefore fixed, up to multiplication by a constant, to be
\begin{equation}  
d(a_k) = e^{k \pi \I b^2},\quad b^2 \in \Z
\end{equation}
where $a_k$ is the element of the orbit parametrised as above.

\subsection{Splitting}
\begin{proposition}\label{prop:splittingGammaOperator}
Let $a=(a_1, -a_1)\in\C^2$, then for $\lambda \sim u^i$ we have the equality of asymptotic series in $\sqrt{\lambda-u^i}$ 
\begin{equation}
\Gamma^{a+ce_i} = \Gamma^a \Gamma^{ce_i} 
\end{equation}
\end{proposition}
\begin{proof}
Simply observe that $(\cf_a)_+ =0$, so 
\begin{equation}  
\Gamma^{a +c e_i} = e^{(\cf_a)_-} e^{(\cf_{ce_i})_-} e^{(\cf_{ce_i})_+} =\Gamma^a \Gamma^{c e_i} .
\end{equation}
\end{proof}

\subsection{Conjugation by $R$}
Let us now consider the conjugation of the vertex operator by the $R$ action of the Givental group.

\begin{proposition}\label{prop:conjwithR}
For $\lambda \sim u^i$ we have 
\begin{equation}
\cR^{-1} e^{\widehat{\Psi \cf_{c e_i}}} \cR = e^{\frac{c^2}2\int_{u^i}^{\lambda} \left(  \cW_{e_i, e_i} - \frac12 \frac1{\rho-u^i} \right)  d\rho } 
e^{\widehat{c \cf_{KdV}(\lambda-u^i,z) e_i }}
\end{equation}
\end{proposition}
\begin{proof}
We follow the proof given in~\cite{mt08}. Let us first recall the following consequence of BCH formula: for $\cf = \sum_l I^{(l)} (-z)^l \in \cV$ and $R(z) = \sum_{k\geq0} R_k z^k$ in the twisted loop group one has
\begin{equation}
\hat{R}^{-1} \widehat{e^{\cf}} \hat{R} = e^{\frac12 V \cf_-^2} \widehat{e^{R^{-1} \cf} },
\end{equation}
where the phase factor is given by
\begin{equation}  
\frac12 V\cf_-^2 = \frac 12\sum_{k,l\geq0} (V_{k,l} I^{(-l-1)}, I^{(-k-1)})
\end{equation}
where the matrices $V_{k,l} \in \End(V)$ are defined by 
\begin{equation}
\sum_{k,l \geq 0} V_{k,l} w^k z^l = \frac{1- R(w) R^*(z) }{w+z}.
\end{equation}
It follows that 
\begin{equation} \label{VVV1}
V_{k-1,l}+V_{k,l-1} = - R_k R^*_l + \delta_{k,0} \delta_{l,0},
\end{equation}
assuming that $V_{-1,l} = V_{k,-1}=0$. 

We need to compute the asymptotic expansion for $\lambda \sim u^i$ of the phase factor that in our case is given by 
\begin{equation} \label{VVV}
 V (\Psi \cf_{c e_i})_-^2=c^2\sum_{k,l \geq0} ( V_{k,l} \Psi I_{e_i}^{(-l-1)}, \Psi I_{e_i}^{(-k-1)} ) .
\end{equation}
Recall that each entry in $\Psi I^{(l)}_{e_i}$ is asymptotic to a formal Laurent series in $\sqrt{\lambda - u^i}$ for $\lambda \sim u^i$, with leading term of degree (at worst) $-2l-1$. This implies that the right-hand side of~\eqref{VVV} converges to a formal series in $\sqrt{\lambda - u^i}$, which in particular vanishes at $u^i$.  

Deriving~\eqref{VVV} by $\lambda$ and changing the indexes in the sums we obtain
\begin{equation}  
c^2  \sum_{k,l \geq 0} \left( (V_{k-1,l} + V_{k,l-1} )  \Psi I_{e_i}^{(-l)}, \Psi I_{e_i}^{(-k)} \right) ,
\end{equation}
which, by substituting~\eqref{VVV1}, equals $c^2$ times
\begin{equation} \label{VVV3}
(\Psi I^{(0)}_{e_i}, \Psi I^{(0)}_{e_i}) - \left( \sum_{l\geq0} R^*_l  \Psi I_{e_i}^{(-l)}, \sum_{k\geq0} R^*_k  \Psi I_{e_i}^{(-k)} \right).
\end{equation}
Note that~\eqref{fuiasy} gives
\begin{equation}  
\sum_{l\geq0} R^*_l  \Psi I_{e_i}^{(-l)} = I^{(0)}_{KdV}(\lambda-u^i) e_i,
\end{equation}
so the second term in~\eqref{VVV3} is equal to $-(2(\lambda-u^i))^{-1}$. By integration we obtain
\begin{equation}  
\frac12 V (\Psi \cf_{c e_i})_-^2=\frac{c^2}2 \int_{u^i}^\lambda \left[  \cW_{e_i,e_i} (t, \rho) - \frac12 \frac1{\rho - u^i} \right] d\rho . 
\end{equation}
\end{proof}

\subsection{Asymptotics at $\lambda \sim \infty$} \label{sec:asymptoticclambdainf}
Let us define the following formal Laurent series in $z$ with coefficients which are $\C^2$-valued functions over $\C \setminus \I \R_+$:
\begin{align} \label{eq:definition-f-infinity}
&\cf_{e_1,\infty }(\lambda,z) :=
\sum_{l\in\Z} \partial_\lambda^l 
\begin{pmatrix}
\partial_\lambda(\log\lambda + \frac\psi2) \\ \frac12
\end{pmatrix} (-z)^l , \\ 
&\cf_{e_2,\infty } (\lambda,z) :=
\cf_{e_1,\infty }(\lambda,z) + 
\sum_{l\in\Z} \partial_\lambda^{l+1} 
\begin{pmatrix}
 \pi \I\ \\ 0
\end{pmatrix} (-z)^l,  \label{fe2}
\end{align}
and for $a= (a_1, a_2) \in \C^2$ let $\cf_{a,\infty} = a_1 \cf_{e_1,\infty} + a_2 \cf_{e_2,\infty}$.
As above $\log \lambda$ denotes the principal branch of the logarithm over $\C \setminus \I \R_+$, $\partial_\lambda^{\pm1}$ is formal differentiation/integration in $\lambda$, and $\psi$ is the constant parametrising the calibration, defined in \S\ref{sec:defflat}.

\begin{proposition}
For any $a\in\C^2$, the asymptotic behaviour of $\cf_{a}$  for $|\lambda| \sim \infty$, $\arg \lambda \not= \pi/2$  is given by
\begin{equation} \label{fSf}
\cf_{a} (t, \lambda, z) \sim S(t, z)\ \cf_{a,\infty} (\lambda,z).
\end{equation}
\end{proposition}
\begin{proof}
We need to prove that 
\begin{equation} \label{expell}
I^{(l)}_{e_i}(t, \lambda) \sim \sum_{k=0}^\infty (-1)^k S_k I^{(l+k)}_{e_i,\infty }
\end{equation}
where $\cf_{e_i,\infty} = \sum_l I^{(l)}_{e_i,\infty } (-z)^l$. It is easy to check that this equality of asymptotic series holds at $t_{sp}$. The case $i=1$ follows by direct substitution, while in the case $i=2$ one can observe that the difference between the asymptotic expansions of the two period vectors is a polynomial in $\lambda$ which is exactly given by shifting $\log \lambda$ by $\pi \I$ as in the second term on the right-hand side of~\eqref{fe2}. 

Because of~\eqref{equ3}, both sides of~\eqref{expell} satisfy the same equation in $t^i$, therefore the must coincide on the whole $M$. 
\end{proof}

\begin{remark}
The asymptotic .... 
\end{remark}

\subsection{Conjugation by $S$} \label{sec:conjugacybyS}
Let us now consider the conjugation of the vertex operator by the $S$ action of the Givental group.  
We define $\cW_{a,b}^{\infty} = ( (I_{a,\infty}^{(0)}(t,\lambda) , I_{b,\infty}^{(0)}(t,\lambda) )$. Notice that $\cW_{e_i,e_i}^{\infty} = \lambda^{-1}$ for $i=1,2$. 

\begin{proposition} \label{prop:conjugationS}
For $a\in\C^2$ we have 
\begin{equation}
\hS \Gamma_\infty^{a} \hS^{-1} = e^{\frac{c(t)}{2} +\frac12 \int_\lambda^\infty  \left( \cW_{a,a} - \cW_{a,a}^\infty \right) d\rho} \Gamma^a ,
\end{equation}
with 
\begin{equation}
c(t) = \frac{(a_1+a_2)^2}{4}  (\log t^2 +\psi).
\end{equation}
\end{proposition}
\begin{proof}
It follows from the Baker-Campbell-Hausdorff formula that, for $\cf = \sum_l \cf_l z^l$ in the loop space $\cV$ and $S(z)$ in the twisted loop group, we have
\begin{equation}\label{eq:ConjugationByS}
\hat{S} \widehat{e^{\cf}} \hat{S}^{-1} = e^{\frac12 W(\cf_+,\cf_+)} \widehat{e^{S \cf}},
\end{equation}
where 
\begin{equation}
W(\cf_+,\cf_+)= \sum_{k,l\geq 0} (W_{k,l} \cf_l,\cf_k ),
\end{equation}
and the coefficients $W_{k,l}\in \End(V)$ are defined by the generating formula
\begin{equation} \label{defW}
\sum_{k,l\geq 0} W_{k,l} w^{-k} z^{-l} = \frac{S^*(w) S(z) -1}{w^{-1} + z^{-1} }.
\end{equation}

We need therefore to evaluate the phase factor $W((\cf_{a,\infty})_+, (\cf_{a,\infty})_+)$. We follow here the argument of~\cite[\S6.2]{mt08}. By definition, for $a,b \in \C^2$
\begin{equation} \label{hello101}
W((\cf_{a,\infty})_+, (\cf_{b,\infty})_+) 
= \sum_{k,l\geq 0} (-1)^{k+l}( W_{k,l} I^{(l)}_{a,\infty } , I^{(k)}_{b,\infty } ),
\end{equation}
therefore
\begin{equation} \label{derlam}
\frac{d}{d\lambda}W((\cf_{a,\infty})_+, (\cf_{b,\infty})_+)  = -\sum_{k,l\geq0} (-1)^{k+l} ((W_{k-1,l}+W_{k,l-1})  I^{(l)}_{a,\infty } , I^{(k)}_{b,\infty } ),
\end{equation}
where we assumed that $W_{-1,l}=W_{k,-1}=0$ and we used the fact that $\partial_\lambda  I^{(l)}_{a,\infty } =  I^{(l+1)}_{a,\infty }$. From~\eqref{defW} we get
\begin{equation}
W_{k-1,l}+W_{k,l-1} = S^*_k S_l -\delta_{k,0}\delta_{l,0},
\end{equation}
so~\eqref{derlam} equals
\begin{equation}
- ( \sum_{l\geq 0} S_l (-1)^l I^{(l)}_{a,\infty} , \sum_{k\geq 0} S_k (-1)^k I^{(k)}_{b,\infty} ) + ( I^{(0)}_{a,\infty}, I^{(0)}_{b,\infty} ). 
\end{equation}
By~\eqref{fSf}, we have
\begin{equation}
\frac{d}{d\lambda}W((\cf_{a,\infty})_+, (\cf_{b,\infty})_+)  = - \cW_{a,b} + \cW_{a,b}^\infty.
\end{equation}
From~\eqref{expell} it follows that the right hand-side has leading term of order $\lambda^{-2}$. Therefore we can formally integrate the asymptotic series and the integration constant  is given by the leading term $\frac{(a_1 +a_2)(b_1+b_2)}4(S_1)_{1,2}$ in the expansion of~\eqref{hello101}. We have shown that for $a, b \in \C^2$ we have the equality of asymptotic series for $\lambda\sim \infty $ 
\begin{equation}
W((\cf_{a,\infty})_+,(\cf_{b,\infty })_+)=\frac{(a_1 +a_2)(b_1+b_2)}4 (\log t^2 + \psi) + \int_\lambda^\infty \left[ \cW_{a,b}(t,\rho) -  \cW^\infty_{a,b}(t,\rho) \right]\ d\rho.
\end{equation}
The Proposition is proved.
\end{proof}

\section{The Hirota quadratic equations for the ancestor potential}\label{sec:HirotaAncestor}

In this section we define the ancestor Hirota quadratic equations and prove that the ancestor potential $\cA$ satisfies them.

\subsection{Definition of Hirota quadratic equations for the ancestor potential}  \label{sec:def-hirota-ancestor}
Recall that the total ancestor potential $\cA$ can be considered as a formal power series in the variables $q^i_\ell+\delta^i_1\delta^1_\ell$ for $i=1,2$ and $\ell \geq0$ whose coefficients are Laurent series in $\epsilon$, and it analytically depends on the point of $M$. 

Recall the discussion of the action of the monodromy group on $\mathbb{C}^2$ in Section~\ref{sec:monodromy}, in particular Proposition~\ref{prop:monodromyC2}. We choose a finite subset $A$ in an infinite orbit of the monodromy group defined as $A=\{a^+,a^-\}$, where $a^+=a_1e_1+a_2e_2$ and $a^-= (-a_1-2a_2)e_1 + a_2 e_2$. Note that 
\begin{align}
\gamma_1 a^{\pm} &= a^{\mp}; \\  
\gamma_2 a^{\pm} &= a^{\mp} \pm 2(a_1+a_2) (e_1-e_2).
\end{align}
In terms of the parametrization of the full orbit generated by $A$ given in Section~\ref{sec:DefinitionCa}, we have $b=a_1+a_2$ and $r=(a_1-a_2)/2$. Recall that we associate in Section~\ref{sec:DefinitionCa} to each element $a$ of this full orbit a function $c_a(t,\lambda)$ given on $a^{\pm}$, up to a multiplication by a non-zero constant, by
\begin{align} \label{eq:coeffDefAHQE}
c_{a^+} & = \exp \left[ - \int_{\lambda_0}^{\lambda} \cW_{a^+,a^+}(t,\rho) d\rho  \right]; \\  
c_{a^-} & = \exp \left[ -\pi \I (a_1 + a_2)^2  - \int_{\lambda_0}^{\lambda} \cW_{a^-,a^-}(t,\rho) d\rho  \right];
 \label{eq:coeffDefAHQE-b}
\end{align}
Recall also that to each point $a\in\mathbb{C}^2$ we associate in Section~\ref{sec:VertexLoopSpaceElements} a vertex operator $\Gamma^a$. 

Let $o=e_1-e_2$. 
Recall that $(I^{(-1)}_o)_1=0$, $(I^{(-1)}_o)_2=-\pi\I$ (see Remark~\ref{rem:Imin1o}, we use the scalar product to lower the indices). Let us define 
\begin{equation}
	\cN = \exp\left( - \sum_{(j,\ell)\not= (2,0)}  \frac{(I^{(-\ell-1)}_{o})_j q^j_\ell }{(I^{(-1)}_{o})_2}  \frac{\partial }{\partial q^2_0}  \right)=\exp\left( \sum_{\ell\geq 1, j=1,2}  \frac{(I^{(-\ell-1)}_{o})_j q^j_\ell }{\pi\I}  \frac{\partial }{\partial q^2_0}  \right).
\end{equation}

\begin{lemma} \label{lem:SingleValued}
	 If $b^2\in \mathbb{Z}$ and 
$\frac{b} \epsilon (q^2_0-\bar{q}^2_0) \in \mathbb{Z}$, the following expression is a single-valued function of $\lambda$:
\begin{equation} \label{A-HQE}
\cN \otimes \cN
\left( c_{a^+} \Gamma^{a^+} \otimes \Gamma^{-a^+} +c_{a^-} \Gamma^{a^-} \otimes \Gamma^{-a^-}\right) \left( \cA \otimes \cA \right) d\lambda .
\end{equation}
Here the two copies of $\cA$ depend on the variables $q^i_\ell$ and $\bar{q}^i_\ell$ respectively.
\end{lemma}

\begin{proof}
We need to prove that it is invariant under the action of the two generators $\gamma_1$, $\gamma_2$ of the fundamental group of the pointed complex plane.  Note that the coefficients of $\cN$ are single-valued functions in $\lambda$. Indeed, since $I^{(0)}_{o} = 0$, all $I^{(-\ell-1)}_{o}$, $\ell\geq 0$, are polynomials in $\lambda$. Recall also the action of the fundamental group of the $\Gamma^a$-operators (Proposition~\ref{prop:MonodromyOnI}) and the coefficients $c_a$ (Section~\ref{sec:DefinitionCa}).

For $\gamma_1$ we have:
\begin{align}
& \gamma_1^*\, \cN \otimes \cN
\left( c_{a^+} \Gamma^{a^+} \otimes \Gamma^{-a^+} +c_{a^-} \Gamma^{a^-} \otimes \Gamma^{-a^-}\right) \left( \cA \otimes \cA \right) d\lambda  \\  \notag 
& = \cN \otimes \cN
\left( c_{a^-} \Gamma^{a^-} \otimes \Gamma^{-a^-} +c_{a^+} \Gamma^{a^+} \otimes \Gamma^{-a^+}\right) \left( \cA \otimes \cA \right) d\lambda, 
\end{align}
so this expression is invariant under the action of $\gamma_1$. 

For $\gamma_2$ we first observe that the action of $\gamma_2$ on the vertex operator is given by
\begin{equation} 
\gamma_2^* \Gamma^{a^\pm} = \Gamma^{a^\mp \pm 2bo} = e^{\pm 2b \hat{\cf}_o}  \Gamma^{a^\mp} ,
\end{equation}
where we use the Baker-Campbell-Hausdorff formula and the fact that $(\cf_o)_+=0$ for the second equality (see~Proposition~\ref{prop:splittingGammaOperator}).  Note also (cf. Section~\ref{sec:DefinitionCa}) that 
\begin{equation}
\gamma_2^* c_{a^\pm} = c_{\gamma_2 a^{\pm} } = e^{\pm 2\pi\I b^2} c_{a^{\mp}}=c_{a^\mp}
\end{equation}
(here we use the condition $b^2\in\mathbb{Z}$ for the last equation). Therefore, the action of $\gamma_2$ on~\eqref{A-HQE} is 
\begin{align} \label{eq:firststepgamma2AncestorHQE}
& \gamma_2^*\, \cN \otimes \cN
\left( c_{a^+} \Gamma^{a^+} \otimes \Gamma^{-a^+} +c_{a^-} \Gamma^{a^-} \otimes \Gamma^{-a^-}\right) \left( \cA \otimes \cA \right) d\lambda  \\  \notag 
& = \cN \otimes \cN \left(  e^{2b\hat{f}_o} \otimes e^{-2b\hat{f}_o} c_{a^-} \Gamma^{a^-} \otimes \Gamma^{-a^-} +e^{-2b\hat{f}_o} \otimes e^{2b\hat{f}_o} c_{a^+} \Gamma^{a^+} \otimes \Gamma^{-a^+}\right) \left( \cA \otimes \cA \right) d\lambda.
\end{align}
Now note that $\hat{f}_o = \epsilon^{-1}\sum_{\ell\geq 0} (I^{(-\ell-1)}_o)_i q^i_\ell$ (cf.~Equation~\eqref{eq:f-quantized}), and, therefore, $\cN e^{\pm 2b \hat{f}_o} = e^{\pm 2b \epsilon^{-1}(I^{(-1)}_o)_2 q^2_0} \cN$. Therefore,~\eqref{eq:firststepgamma2AncestorHQE} is equal to
\begin{multline}
 \left( e^{\frac{2b} \epsilon (I^{(-1)}_o)_2 (q^2_0-\bar{q}^2_0)} \cN \otimes \cN  c_{a^-} \Gamma^{a^-} \otimes \Gamma^{-a^-} \right. + \\ \left. +e^{-\frac{2b} \epsilon (I^{(-1)}_o)_2 (q^2_0-\bar{q}^2_0)} \cN \otimes \cN  c_{a^+} \Gamma^{a^+} \otimes \Gamma^{-a^+}\right) \left( \cA \otimes \cA \right) d\lambda.
\end{multline}
Under the condition that $\frac{2b} \epsilon (I^{(-1)}_o)_2 (q^2_0-\bar{q}^2_0) \in 2\pi \I \mathbb{Z}$ (which can be simplified using $(I^{-1}_o)_2=-\pi\I$ to $\frac{b} \epsilon (q^2_0-\bar{q}^2_0) \in \mathbb{Z}$) this expression is equal to~\eqref{A-HQE}, which proves the invariance under the action of $\gamma_2$.
\end{proof}

Lemma~\ref{lem:SingleValued} implies that under the condition $b^2\in\mathbb{Z}$ expression~\eqref{A-HQE} can be considered as a formal power series in the variables $q^i_\ell+\delta^i_1\delta^1_\ell$ and $\bar q^i_\ell+\delta^i_1\delta^1_\ell$ for $i=1,2$ and $l \geq0$ and a Laurent series in $\epsilon$, the coefficients of whose restriction to  $\frac{b} \epsilon (q^2_0-\bar{q}^2_0) \in \mathbb{Z}$  are rational functions of $\lambda$ with possible poles at the points $\lambda=u^1,u^2,\infty$. 

\begin{definition} We say that the ancestor potential $\cA$ satisfies the \emph{ancestor Hirota quadratic equations} for the set $A=\{a^+,a^-\}$ if the aforementioned dependence on $\lambda$ is polynomial, that is, if there are no poles at $\lambda=u^1,u^2$.
\end{definition}

\subsection{Proof of the ancestor HQE} 
\begin{theorem}\label{thm:HirotaForAncestors} Let $b^2 =1$. 
Then the  ancestor potential $\cA$ satisfies the ancestor Hirota quadratic equation.
\end{theorem}

\begin{proof} Let us prove that~\eqref{A-HQE} is regular at $\lambda=u_1$ (the case of $\lambda=u_2$ is completely analogous). Note that $a^\pm = -a_2 o \pm b e_1$. According to Proposition~\ref{prop:splittingGammaOperator}, expression~\eqref{A-HQE} is equal to 
\begin{equation} \label{eq:ProofAncestorHQESplit}
\left(\cN \otimes \cN\right) \left(\Gamma^{-a_2 o} \otimes \Gamma^{a_2o}\right)
\left( c_{a^+} \Gamma^{be_1} \otimes \Gamma^{-be_1} +c_{a^-} \Gamma^{-be_1} \otimes \Gamma^{be_1}\right) \left( \cA \otimes \cA \right) d\lambda .
\end{equation}
Here the first two factors do not change the regularity in $\lambda$  at $\lambda=u^1$ (recall that $I^{(0)}_o=0$), so we have to show that 
\begin{equation}\label{eq:expressionwithoutNando}
\left( c_{a^+} \Gamma^{be_1} \otimes \Gamma^{-be_1} +c_{a^-} \Gamma^{-be_1} \otimes \Gamma^{be_1}\right) \left( \cA \otimes \cA \right) d\lambda .
\end{equation}
is regular in $\lambda$ at $\lambda=u_1$. 

Recall that $\cA = \hat{\Psi}^{-1} \hat {R}\, \tau_{KdV_1} \tau_{KdV_2} $, where the KdV tau-functions $\tau_{KdV_i}$ depend on the coordinates that correspond to the normalized canonical frame. In particular, the KdV hierarchy for these KdV tau-functions can be written as the regularity at $\lambda=u^i$, $i=1,2$, of the expression
\begin{multline} \label{eq:KdVlocal}
(\lambda-u^i)^{-\frac 12}\left( e^{\widehat{\cf_{KdV}(\lambda-u^i,z) e_i }}\otimes e^{-\widehat{\cf_{KdV}(\lambda-u^i,z) e_i }} -
\right. \\   \left. 
-e^{-\widehat{\cf_{KdV}(\lambda-u^i,z) e_i }}\otimes e^{\widehat{\cf_{KdV}(\lambda-u^i,z) e_i }}\right) \tau_{KdV_i} \otimes \tau_{KdV_i}  d\lambda.
\end{multline}
Proposition~\ref{prop:conjwithR} implies that expression~\eqref{eq:expressionwithoutNando} is equal to $\hat{\Psi}^{-1} \hat {R}\otimes \hat{\Psi}^{-1} \hat {R}$ applied to
\begin{multline}\label{eq:expressionRconj}
 \left(c_{a^+} e^{2b^2\int_{u^1}^{\lambda} \left(  \cW_{e_1, e_1} - \frac12 \frac1{\rho-u^1} \right)  d\rho } 
e^{\widehat{b\, \cf_{KdV}(\lambda-u^1,z) e_1 }}\otimes e^{-\widehat{b\, \cf_{KdV}(\lambda-u^1,z) e_1 }}
 \right. +\\   
 \left. +c_{a^-} e^{2b^2\int_{u^1}^{\lambda} \left(  \cW_{e_1, e_1} - \frac12 \frac1{\rho-u^1} \right)  d\rho } e^{\widehat{-b\, \cf_{KdV}(\lambda-u^1,z) e_1 }}\otimes e^{\widehat{b\, \cf_{KdV}(\lambda-u^1,z) e_1 }}\right) \cdot \\
\cdot  \left( \tau_{KdV_1} \tau_{KdV_2}  \otimes\tau_{KdV_1} \tau_{KdV_2}  \right) d\lambda .
\end{multline}
Let us compute the coefficients. We have:
\begin{align}
c_{a^+} e^{b^2\int_{u^1}^{\lambda} \left(  \cW_{e_1, e_1} - \frac12 \frac1{\rho-u^1} \right)  d\rho } & =  e^{-b^2\int_{\lambda_0}^{\lambda} \cW_{e_1,e_1}d\rho  + b^2\int_{u^1}^{\lambda} \left(  \cW_{e_1, e_1} - \frac12 \frac1{\rho-u^1} \right)  d\rho} \\  \notag 
& = e^{\frac{b^2}2 \log(\lambda_0-u^1)  + b^2\int_{u^1}^{\lambda_0} \left(  \cW_{e_1, e_1} - \frac12 \frac1{\rho-u^1} \right)  d\rho } (\lambda-u^1)^{-\frac{b^2}2};\\  
c_{a^-} e^{b^2\int_{u^1}^{\lambda} \left(  \cW_{e_1, e_1} - \frac12 \frac1{\rho-u^1} \right)  d\rho }  
& = e^{ -\pi \I b^2  - b^2 \int_{\lambda_0}^{\lambda} \cW_{e_1,e_1} d\rho +b^2\int_{u^1}^{\lambda} \left(  \cW_{e_1, e_1} - \frac12 \frac1{\rho-u^1} \right)  d\rho } \\ \notag 
& = e^{\frac{b^2}2 \log(\lambda_0-u^1)  + b^2\int_{u^1}^{\lambda_0} \left(  \cW_{e_1, e_1} - \frac12 \frac1{\rho-u^1} \right)  d\rho } e^{ -\pi \I b^2} (\lambda-u^1)^{-\frac{b^2}2}.
\end{align}
Under the assumption $b^2=1$ the coefficients, up to a common invertible factor that does not depend on $\lambda$, are equal to $\pm (\lambda-u^1)^{-\frac 12}$. 

Thus, under the assumption $b^2=1$ expression~\eqref{eq:expressionRconj} is equal up to an invertible factor 
\begin{equation}
e^{\frac{1}2 \log(\lambda_0-u^1)  + \int_{u^1}^{\lambda_0} \left(  \cW_{e_1, e_1} - \frac12 \frac1{\rho-u^1} \right)  d\rho } \tau_{KdV_2}\otimes \tau_{KdV_2}
\end{equation}
  that does not depend on $\lambda$ to the expression~\eqref{eq:KdVlocal} with $i=1$, whose regularity in $\lambda$ at $\lambda=u^1$ is part of the definition of $\cA$. This implies the regularity at $\lambda=u^1$ of~\eqref{eq:ProofAncestorHQESplit}.
\end{proof}

\begin{remark} The condition $b^2=1$ implies that $b=1$ or $b=-1$. Since $a^\pm = -a_2 o \pm be_1$, the set $A$ does not depend on this choice, and in the construction of the ancestor Hirota quadratic this choice only affects a common non-vanishing coefficient that we can ignore, cf.~Equation~\eqref{eq:coeffDefAHQE}.
\end{remark}

\begin{remark} With $b=1$ the restriction $\frac{b} \epsilon (q^2_0-\bar{q}^2_0) \in \mathbb{Z}$ reduces to $ (q^2_0-\bar{q}^2_0) \in \epsilon\mathbb{Z}$.
\end{remark}

\section{The Hirota quadratic equations for the descendent potential} \label{sec:HirotaDescendant}

In this section we define the descendent Hirota quadratic equations and  
prove that the de
scendant potential $\cD$ satisfies them. Note that we assumefrom now on  that $b=1$.

\subsection{The Hirota equation} 
Recall the definitions of $\cf_{e_i,\infty} = \sum_l I^{(l)}_{e_i,\infty } (-z)^l$, $i=1,2$, given in Section~\ref{sec:asymptoticclambdainf}, as well as more general $\cf_{a,\infty}$ and $I^{(l)}_{a,\infty }$ defined for any $a\in \C^2$. Recall also that for any $a,b \in \C^2$ we defined in Section~\ref{sec:conjugacybyS}  the function
$\cW^\infty_{a,b} (\lambda)  = (I_{a,\infty}^{(0)}(\lambda) , I_{b,\infty}^{(0)}(\lambda) ) $. 

Recall that $o=e_1-e_2$. Let us also define 
\begin{equation} \label{eq:def-N-infinity}
\cN_\infty = \exp\left( - \sum_{(j,\ell)\not= (2,0)} \frac{(I^{(-\ell-1)}_{o,\infty})_j q^j_\ell }{(I^{(-1)}_{o,\infty})_2}  \frac{\partial }{\partial q^2_0}  \right) = \exp\left( -
\sum_{\ell\geq 1} \frac{\lambda^{\ell}}{\ell!}q^2_\ell \frac{\partial }{\partial q_0^2} \right)\, .
\end{equation}
(The second equality here follows from Equations~\eqref{eq:definition-f-infinity}-\eqref{fe2}.) 

Let $A=\{a^+,a^-\}$ be the same set of points in $\C^2$ as in Section~\ref{sec:HirotaAncestor}, $a^{\pm} = -a_2o\pm e_1$  (that is, for $a^+=a_1e_1+a_2e_2$ we assume $a_1+a_2=1$). Let
\begin{align} \label{eq:coeffDefDescHQE}
c^\infty_{a^+} & = \exp \left[ - \int_{\lambda_0}^{\lambda} \cW^\infty_{a^+,a^+}(\rho) d\rho  \right]\,; \\  
c^\infty_{a^-} & = \exp \left[ -\pi \I - \int_{\lambda_0}^{\lambda} \cW^\infty_{a^-,a^-}(\rho) d\rho  \right]
\end{align}
(cf.~the analogous definitions in Sections~\ref{sec:DefinitionCa} and~\ref{sec:def-hirota-ancestor}). Since $I_{e_1,\infty}^{(0)} =I_{e_2,\infty}^{(0)}= (\frac 1\lambda, \frac 12)^t$, we have $\cW^\infty_{a^+,a^+}=\cW^\infty_{a^-,a^-}=\lambda^{-1}$ and, therefore, $c^\infty_{a^\pm} = \pm \frac{\lambda_0}{\lambda}$.

Consider the following expression:
\begin{equation} \label{eq:deschirota1}
\cN_\infty  \otimes \cN_\infty  \left( c_{a^+}^\infty  \Gamma^{a^+}_{\infty} \otimes \Gamma^{-a^+}_{\infty } +c_{a^-}^\infty  \Gamma^{a^-}_{\infty} \otimes \Gamma^{-a^-}_{\infty } \right) \left( \cD \otimes \cD	\right) d\lambda \,,
\end{equation}
where the two copies of $\cD$ depend on two different sets of variables, $q^i_\ell$ and $\bar{q}^i_\ell$, respectively.

\begin{lemma} For $ (q^2_0-\bar{q}^2_0) \in \epsilon\mathbb{Z}$ expression~\eqref{eq:deschirota1} is a single-valued function of $\lambda$.
\end{lemma}

\begin{proof} Since the whole expression is only defined as an asymptotic series  for $|\lambda|\sim \infty$, we have to check that the action of the monodromy $\gamma_\infty$ along the big circle is trivial. 
	
Note that $\cN_\infty$ and $c^\infty_{a^\pm}$ are single-valued in $\lambda$. For $\Gamma_\infty^{\pm a^\pm}$ the action of the monodromy $\gamma_\infty$ changes the branch of the logarithm in the definition of $\cf_{e_i,\infty}$. We have:
\begin{equation}
\gamma_\infty\cf_{e_i,\infty}= \cf_{e_i,\infty}+ \cf_\infty \,,
\end{equation}
where
\begin{equation}
\cf_\infty(\lambda,z) = 2\pi \I \sum_{l\in\Z} \partial_\lambda^{l+1} \begin{pmatrix}
1 \\ 0
\end{pmatrix}
(-z)^l
=2\pi \I \begin{pmatrix}
1 \\ 0
\end{pmatrix}
\sum_{l\geq0} \frac{\lambda^l}{l!} (-z)^{-l-1}
 \,.
\end{equation}
Thus $\gamma_\infty \Gamma_\infty^{e_i} =  e^{ \hat\cf_\infty} \Gamma_\infty^{e_i}$, and, therefore, for any constant $c$
\begin{equation}
\gamma_\infty \Gamma_\infty^{c a^{\pm}} =  e^{ \pm (c \hat\cf_\infty)} \Gamma_\infty^{c a^{\pm}} \,,
\end{equation}
where
\begin{equation}
\hat{\cf}_\infty = \frac{2\pi \I }{\epsilon} \sum_{k\geq 0}( \partial_\lambda^{-k} 1) q^2_k =  \frac{2\pi \I }{\epsilon} \sum_{k\geq 0} \frac{\lambda^k}{k!} q^2_k \,.
\end{equation}

Using this computation, we apply $\gamma_\infty$ to expression~\eqref{eq:deschirota1}, and we obtain:
\begin{equation} \label{eq:deschirota-monodromy}
\cN_\infty  \otimes \cN_\infty  \left( e^{ \hat\cf_\infty}\otimes e^{-\hat\cf_\infty} c_{a^+}^\infty  \Gamma^{a^+}_{\infty} \otimes \Gamma^{-a^+}_{\infty } + e^{ -\hat\cf_\infty}\otimes e^{\hat\cf_\infty}c_{a^-}^\infty  \Gamma^{a^-}_{\infty} \otimes \Gamma^{-a^-}_{\infty } \right) \left( \cD \otimes \cD	\right) d\lambda \,.
\end{equation}
Note that $\cN_\infty e^{ \pm \hat\cf_\infty} = e^{\pm \frac{2\pi\I}{\epsilon }q^2_0} \cN_\infty$. Therefore, \eqref{eq:deschirota-monodromy} is equal to 
\begin{multline} 
\left( e^{\frac{2\pi\I}{\epsilon }(q^2_0-\bar{q}^2_0)} \cN_\infty  \otimes \cN_\infty  \, c_{a^+}^\infty \Gamma^{a^+}_{\infty} \otimes \Gamma^{-a^+}_{\infty } + \right. \\ \left. + e^{-\frac{2\pi\I}{\epsilon }(q^2_0-\bar{q}^2_0)} \cN_\infty  \otimes \cN_\infty \,  c_{a^-}^\infty  \Gamma^{a^-}_{\infty} \otimes \Gamma^{-a^-}_{\infty } \right) \left( \cD \otimes \cD	\right) d\lambda \,,
\end{multline}
which coincides with~\eqref{eq:deschirota1} in the case $ (q^2_0-\bar{q}^2_0) \in \epsilon\mathbb{Z}$.
\end{proof}

\begin{definition}
We say that the descendent potential $\cD$ satisfies the descendent Hirota quadratic equation if the coefficients of expression~\eqref{eq:deschirota1} (expanded in $q^i_\ell+\delta^i_1\delta^1_\ell$, $\bar q^i_\ell+\delta^i_1\delta^1_\ell$, and $\epsilon$) are polynomial in $\lambda$.
\end{definition}

\subsection{Hirota equations for the descendent potential}
\begin{theorem}
	The descendent potential $\cD$ satisfies the descendent Hirota quadratic equation. 
\end{theorem}

\begin{proof} Recall that $\cD = C \hat{S}^{-1} \cA$, where all three factors on the right hand side depend on the point of $M$, but their product is independent. Note that the factor $C$ does not depend on $\lambda$ and is constant in $q^i_\ell,\bar{q}^i_\ell$ (in particular, it commutes with all operators involved in expression~\eqref{eq:deschirota1}). For the operator $ \hat{S}^{-1} $ we use the following two lemmata:

\begin{lemma} \label{lem:S-conjugationGamma}
We have the equality of the asymptotic series for $\lambda\sim\infty $
\begin{equation}
c^\infty_{a^\pm} (\hat{S} \otimes \hat{S} )( \Gamma^{a^\pm}_\infty \otimes \Gamma^{-a^\pm}_\infty ) (\hat{S}^{-1} \otimes \hat{S}^{-1} )  = F\cdot c_{a^\pm}  ( \Gamma^{a^\pm} \otimes \Gamma^{-a^\pm} ),
\end{equation}
where $F=\exp\left(\frac{1}{4}  (\log t^2 +\psi)+\int_{\lambda_0}^\infty \left( \cW_{a^\pm,a^\pm} - \cW_{a^\pm,a^\pm}^\infty \right) d\rho\right)$.
\end{lemma}
\begin{proof} Recall Proposition~\ref{prop:conjugationS}. Since the sum of coordinates of $a^\pm$ is equal to $\pm1 $, it implies that 
\begin{equation}
c^\infty_{a^\pm} (\hat{S} \otimes \hat{S} )( \Gamma^{a^\pm}_\infty \otimes \Gamma^{-a^\pm}_\infty ) (\hat{S}^{-1} \otimes \hat{S}^{-1} )  = c^\infty_{a^\pm} e^{\frac{1}{4}  (\log t^2 +\psi)+\int_\lambda^\infty  \left( \cW_{a^\pm,a^\pm} - \cW_{a^\pm,a^\pm}^\infty \right) d\rho} (\Gamma^{a^\pm} \otimes \Gamma^{-a^\pm} )\,.
\end{equation}
Now observe that 
\begin{multline}
c^\infty_{a^\pm} e^{\frac{1}{4}  (\log t^2 +\psi)+\int_\lambda^\infty  \left( \cW_{a^\pm,a^\pm} - \cW_{a^\pm,a^\pm}^\infty \right) d\rho} 
 = \\
 = \pm e^{\frac{1}{4}  (\log t^2 +\psi)+\int_\lambda^\infty  \left( \cW_{a^\pm,a^\pm} - \cW_{a^\pm,a^\pm}^\infty \right) d\rho- \int_{\lambda_0}^{\lambda} \cW^\infty_{a^\pm,a^\pm}(\rho) d\rho}
\\     
= \pm e^{\frac{1}{4}  (\log t^2 +\psi)+\int_{\lambda_0}^\infty \left( \cW_{a^\pm,a^\pm} - \cW_{a^\pm,a^\pm}^\infty \right) d\rho- \int_{\lambda_0}^{\lambda} \cW_{a^\pm,a^\pm}(\rho) d\rho}
 = F\cdot c_{a^\pm}
\end{multline}
(for the last equality here recall the definition of $c_{a^\pm}$ given in Equations~\eqref{eq:coeffDefAHQE}-\eqref{eq:coeffDefAHQE-b}).
\end{proof}
Note that the factor $F$ does not depend on $\lambda$ and is constant in $q^i_\ell,\bar{q}^i_\ell$ (in particular, it commutes with $\cN_\infty$). Note also that the factor $F$ doesn't depend on the choice of the sign $\pm$ in $a^\pm $ since $\cW^\infty_{a^+,a^+}=\cW^\infty_{a^-,a^-}$ and $\cW_{a^+,a^+}=\cW_{a^-,a^-}$.

\begin{lemma} \label{lem:S-conjugationN}
	We have the equality of the asymptotic series for $\lambda\sim\infty $
\begin{equation}
\cN_\infty \hat{S}^{-1} = QO \cN,
\end{equation}
where $Q$ is an exponential of a linear combination of terms $\epsilon^{-2} q^i_\ell q^j_m$, whose coefficients are polynomial in $\lambda$ and depend on the point of $M$, and $O$ is the exponential of a linear vector field in $q^i_\ell$ that does not contain differentiation $\partial / \partial q^2_0$ and whose coefficients depend on the point of $M$.
\end{lemma}
\begin{proof} Typically, we commute the operators using the quantisation rules. However, $\cN$ and $\cN_\infty$ are not obtained by quantization, so we have to go into a detailed analysis of the commutation of these operators with $\hat{S}$. 
	
Recall the structure of $\log \hat{S}$ (see Equation~\eqref{eq:log-S-hat}). It can be split into two summands,  a linear combination of all terms of the type $\epsilon^{-2}q^i_k q^j_\ell$ and a linear combination of the terms  $q^i_k{\partial}/{\partial q^j_\ell}$, with the coefficients depending on the point of $M$. Consider $\cN_\infty \hat{S}^{-1}$. Using the Baker-Campbell-Hausdorff formula we extract all quadratic terms $\epsilon^{-2}q^i_k q^j_\ell$ in $\hat{S}^{-1}$ to a separate exponential and commute it through $\cN_\infty$ to the left. This gives the coefficient $Q$. 

Now we have to compute 
\begin{align}\label{eq:commN-1}
&
\cN_\infty\exp\left( \sum_{\substack{\ell\geq 1\\a\geq 0}} q^i_{a+\ell} \frac{\partial}{\partial q^j_{a}} (s_\ell)^j_i\right) = 
\\  
\notag
&
\exp\left( - \sum_{\substack{(j,\ell)\\ \not= (2,0)}} \frac{(I^{(-\ell-1)}_{o,\infty})_j q^j_\ell }{(I^{(-1)}_{o,\infty})_2}  \frac{\partial }{\partial q^2_0}  \right)
\exp\left( \sum_{\substack{\ell\geq 1\\a\geq 0}} q^i_{a+\ell} \frac{\partial}{\partial q^j_{a}} (s_\ell)^j_i\right) \,.
\end{align}
To this end note that (cf. Equation~\eqref{eq:ConjugationByS})

\begin{align}\label{eq:commN-2}
&
\exp\left(- \sum_{\substack{\ell\geq 1\\a\geq 0}} q^i_{a+\ell} \frac{\partial}{\partial q^j_{a}} (s_\ell)^j_i\right) \exp\left( - \sum_{\substack{(j,\ell)\\ \not= (2,0)}} \frac{(I^{(-\ell-1)}_{o,\infty})_j q^j_\ell }{(I^{(-1)}_{o,\infty})_2}  \frac{\partial }{\partial q^2_0}  \right)
\exp\left( \sum_{\substack{\ell\geq 1\\a\geq 0}} q^i_{a+\ell} \frac{\partial}{\partial q^j_{a}} (s_\ell)^j_i\right) 
\\  \notag 
& 
= \exp\left( - \sum_{(k,\ell)\not= (2,0)} \sum_{p \geq 0} \frac{(I^{(-\ell-1)}_{o,\infty})_k (S^{-1}_p)^k_j q^j_{\ell+p} }{(I^{(-1)}_{o,\infty})_2}  \frac{\partial }{\partial q^2_0}  \right)
\\ \notag  
& 
= \exp\left( - \sum_{(k,\ell)} \sum_{p \geq 0} \frac{(I^{(-\ell-1)}_{o,\infty})_k (S^{-1}_p)^k_j q^j_{\ell+p} }{(I^{(-1)}_{o,\infty})_2}  \frac{\partial }{\partial q^2_0}  
+\sum_{p \geq 0} \frac{(I^{(-1)}_{o,\infty})_2 (S^{-1}_p)^2_j q^j_{p} }{(I^{(-1)}_{o,\infty})_2}  \frac{\partial }{\partial q^2_0}  
\right)
\\  \notag 
& = \exp\left( - \sum_{(j,\ell)} \frac{\sum_{p\geq 0} (-1)^p (S_p I^{(-(\ell+p)-1+p)}_{o,\infty})_j q^j_{\ell+p} }{(I^{(-1)}_{o,\infty})_2}  \frac{\partial }{\partial q^2_0}
+\sum_{p \geq 0} \frac{(I^{(-1)}_{o,\infty})_2 (S^{-1}_p)^2_j q^j_{p} }{(I^{(-1)}_{o,\infty})_2}  \frac{\partial }{\partial q^2_0}  
\right)\,.
\end{align}

Using Equation~\eqref{expell} and the observations that $I^{(\ell)}_{o,\infty}=0$ and $I^{(\ell)}_{o}=0$ for $\ell \geq 0$, we can rewrite this expression as 
\begin{equation}\label{eq:commN-3}
 \exp\left(\sum_{p \geq 1} (S^{-1}_p)^2_j q^j_{p} \frac{\partial }{\partial q^2_0}  
\right)\cN\,.
\end{equation}
Now note that the BCH formula implies
\begin{equation}\label{eq:commN-4}
\exp\left( \sum_{\substack{\ell\geq 1\\a\geq 0}} q^i_{a+\ell} \frac{\partial}{\partial q^j_{a}} (s_\ell)^j_i\right) =
\exp\left( \sum_{\substack{\ell\geq 1,\, a\geq 0 \\ (j,a)\not=(2,0)}} q^i_{a+\ell} \frac{\partial}{\partial q^j_{a}} (s_\ell)^j_i\right)
 \exp\left(-\sum_{p \geq 1} (S^{-1}_p)^2_j q^j_{p} \frac{\partial }{\partial q^2_0}  
\right)\,.
\end{equation}

Let $O$ denote the first factor on the right hand side of Equation~\eqref{eq:commN-4}. Then Equations~\eqref{eq:commN-1}, \eqref{eq:commN-2}, \eqref{eq:commN-3}, and \eqref{eq:commN-4}, collected together, imply that
\begin{equation}
\cN_\infty\exp\left( \sum_{\substack{\ell\geq 1\\a\geq 0}} q^i_{a+\ell} \frac{\partial}{\partial q^j_{a}} (s_\ell)^j_i\right) = O\cN\,.
\end{equation} 
\end{proof}
Using these two lemmata, we can rewrite Equation~\eqref{eq:deschirota1} as the asymptotic series expansion for $\lambda\sim \infty$ of the following expression:
\begin{equation} \label{eq:deschirota2}
C^2 F\cdot(Q\otimes Q)(O\otimes O) (\cN \otimes \cN)
\left( c_{a^+} \Gamma^{a^+} \otimes \Gamma^{-a^+} +c_{a^-} \Gamma^{a^-} \otimes \Gamma^{-a^-}\right) \left( \cA \otimes \cA \right) d\lambda \,.
\end{equation}

Note that the operator $C^2F\cdot  (Q\otimes Q)(O\otimes O) $ does not contain derivatives with respect to $q^2_0$ and $\bar{q}^2_0$. Therefore, the restriction $ (q^2_0-\bar{q}^2_0) \in \epsilon\mathbb{Z}$ can be applied to this operator and to rest of the formula simultaneously. 
Note also that it is an invertible operator that preserves the polynomiality property in $\lambda$, that is, the coefficients of this expression restricted to $ (q^2_0-\bar{q}^2_0) \in \epsilon\mathbb{Z}$ (and expanded in $q^i_\ell+\delta^i_1\delta^1_\ell$, $\bar q^i_\ell+\delta^i_1\delta^1_\ell$, and $\epsilon$) are polynomial in $\lambda$ if and only if the same property holds for the asymptotic expansion of 
\begin{equation} \label{eq:deschirota3}
(\cN \otimes \cN)
\left( c_{a^+} \Gamma^{a^+} \otimes \Gamma^{-a^+} +c_{a^-} \Gamma^{a^-} \otimes \Gamma^{-a^-}\right) \left( \cA \otimes \cA \right) d\lambda \,,
\end{equation}
which is indeed the case by Theorem~\ref{thm:HirotaForAncestors}.
\end{proof}

\subsection{Explicit form of the Hirota equations}
The goal of this Section is to work out an explicit form of the Hirota quadratic equations for the descendent potential (which is the equations of the polynomiality of the $1$-form given by Equation~\eqref{eq:deschirota1} at $q^2_0-\bar q^2_0 = \epsilon k$, $k\in\Z$)

\begin{proposition} For any value of the calibration parameter $\psi$ the descendent potential $\cD$ satisfies the following equations: 
\begin{align}
  0 = \res\limits_{\lambda=\infty} \textstyle \lambda^{n-1}d\lambda \Big[
\lambda^k& \exp(\frac{k\psi}2)  \textstyle \exp \big( \frac 1\epsilon \sum_{\ell\geq 0} \frac 12 \frac{\lambda^{\ell+1}}{(\ell+1)!} (q^1_\ell -\bar q^1_\ell)
- \frac 1\epsilon \sum_{\ell\geq 1}  \frac{\lambda^\ell}{\ell!} \ch(\ell)  (q^2_\ell-\bar q^2_\ell) \big)\times 
 \notag \\ \label{HQE185}
& \textstyle 
\cD\big(\big\{q^1_\ell - \epsilon \frac{\ell!}{\lambda^{\ell+1}}\big\}_{\ell \geq 0},\
q^2_0-\frac{\epsilon}{2} -\sum_{\ell\geq 1}\frac{\lambda^{\ell}}{\ell!}q^2_\ell,\ \big\{
q^2_\ell\big\}_{\ell\geq 1}\big) \times 
\\ \notag
& \textstyle 
\cD\big(\big\{\bar q^1_\ell + \epsilon \frac{\ell!}{\lambda^{\ell+1}}\big\}_{\ell \geq 0},\
\bar q^2_0+\frac{\epsilon}{2} -\sum_{\ell\geq 1}\frac{\lambda^{\ell}}{\ell!}\bar q^2_\ell,\ \big\{
\bar q^2_\ell\big\}_{\ell\geq 1}\big)
\\ \notag
\textstyle 
-\lambda^{-k}& \exp(-\frac{k\psi}2) 
 \textstyle 
\exp \big( -\frac 1\epsilon \sum_{\ell\geq 0} \frac 12 \frac{\lambda^{\ell+1}}{(\ell+1)!} (q^1_\ell -\bar q^1_\ell)
+\frac 1\epsilon \sum_{\ell\geq 1}  \frac{\lambda^\ell}{\ell!} \ch(\ell)  (q^2_\ell-\bar q^2_\ell) \big)\times
\\ \notag
& \textstyle 
\cD\big(\big\{q^1_\ell + \epsilon \frac{\ell!}{\lambda^{\ell+1}}\big\}_{\ell \geq 0},\
q^2_0+\frac{\epsilon}{2} -\sum_{\ell\geq 1}\frac{\lambda^{\ell}}{\ell!}q^2_\ell,\ \big\{
q^2_\ell\big\}_{\ell\geq 1}\big) \times 
\\ \notag
& \textstyle 
\cD\big(\big\{\bar q^1_\ell - \epsilon \frac{\ell!}{\lambda^{\ell+1}}\big\}_{\ell \geq 0},\
\bar q^2_0-\frac{\epsilon}{2} -\sum_{\ell\geq 1}\frac{\lambda^{\ell}}{\ell!}\bar q^2_\ell,\ \big\{
\bar q^2_\ell\big\}_{\ell\geq 1}\big)
\Big]\Big|_{q^2_0-\bar q^2_0 = k\epsilon} \,.
\end{align}
for any $k\in\Z$ and for any $n{\geq 0}$.
\end{proposition}

\begin{proof}
Recall Equations~\eqref{eq:definition-f-infinity} and~\eqref{fe2}. For $a^{\pm} = \pm e_1 - a_2o$ they imply that
\begin{align}
\cf_{a^\pm,\infty} & = \pm \sum_{\ell \in \Z} \partial_\lambda^\ell 
\begin{pmatrix}
\partial_\lambda (\log \lambda + \frac{\psi}{2}) \\
\frac 12
\end{pmatrix}
(-z)^\ell
+ a_2 \sum_{\ell\in\Z}  \partial_\lambda^{\ell +1} 
\begin{pmatrix}
\pi\I \\ 0
\end{pmatrix}
(-z)^\ell 
\\ \notag
& = \sum_{\ell \geq 0} (-z)^{-1-\ell} 
\begin{pmatrix}
\frac{\lambda^\ell}{\ell!}
\Big(a_2 \pi \I \pm (\log \lambda - \ch(\ell) + \frac{\psi}{2} )  \Big)\\
\pm \frac 12 \frac{\lambda^{\ell+1}}{(\ell+1)!}
\end{pmatrix} 
\\ \notag
& \phantom{ =\ }
\pm \sum_{\ell\geq 0} (-z)^\ell \frac{(-1)^\ell \ell!}{\lambda^{\ell+1}} 
\begin{pmatrix}
1 \\ 0
\end{pmatrix} 
\pm (-z)^0 \frac 12
\begin{pmatrix}
0 \\ 1
\end{pmatrix} \,.
\end{align}
Therefore,
\begin{align}
\hat{\cf}_{a^\pm,\infty} = & \pm \frac 1\epsilon \sum_{\ell\geq 0} \frac 12 \frac{\lambda^{\ell+1}}{(\ell+1)!} q^1_\ell 
\pm \frac 1\epsilon \sum_{\ell\geq 0}  \frac{\lambda^\ell}{\ell!} \Big(\log \lambda - \ch(\ell) + \frac{\psi}{2} \Big) q^2_\ell 
+ \frac 1\epsilon a_2 \pi\I \sum_{\ell\geq 0}\frac{\lambda^\ell}{\ell!} q^2_\ell 
\\ \notag
& \mp \epsilon \sum_{\ell\geq0}  \frac{\ell!}{\lambda^{\ell+1}} \frac{\partial}{\partial q^1_\ell}
\mp  \frac \epsilon 2 \frac{\partial}{\partial q^2_0} \,.
\end{align}

Recall also Equation~\eqref{eq:def-N-infinity}:
\begin{equation} \label{eq:def-N-infinity-2nd}
\cN_\infty = \exp\left( -
\sum_{\ell\geq 1} \frac{\lambda^{\ell}}{\ell!}q^2_\ell \frac{\partial }{\partial q_0^2} \right)\, .
\end{equation}
We have:
\begin{align}
\cN_\infty \hat{\cf}_{a^\pm,\infty} = & \Bigg[
\pm \frac 1\epsilon \sum_{\ell\geq 0} \frac 12 \frac{\lambda^{\ell+1}}{(\ell+1)!} q^1_\ell 
\mp \frac 1\epsilon \sum_{\ell\geq 1}  \frac{\lambda^\ell}{\ell!} \ch(\ell)  q^2_\ell 
\pm \frac 1\epsilon \Big(\log \lambda + \frac{\psi}{2} \Big) q^2_0
+ \frac 1\epsilon a_2 \pi\I q^2_0 \\ 
\notag & 
 \mp \epsilon \sum_{\ell\geq0}  \frac{\ell!}{\lambda^{\ell+1}} \frac{\partial}{\partial q^1_\ell}
 \mp  \frac \epsilon 2 \frac{\partial}{\partial q^2_0}
\Bigg] \cN_\infty 
\end{align}

Note that for $q^2_0-\bar q^2_0 = \epsilon k$, $k\in\mathbb{Z}$, we have:
\begin{equation}
\exp\Big(\pm \frac 1\epsilon \Big(\log \lambda + \frac{\psi}{2} \Big) (q^2_0-\bar q^2_0)
+ \frac 1\epsilon a_2 \pi\I (q^2_0-\bar q^2_0)\Big)=
\exp(k\pi\I a_2)\exp\Big(\pm \frac{k\psi}{2}\Big) \lambda^{\pm k} 
\end{equation}
Since that the factor $\exp(k\pi\I a_2)$ doesn't depend on the choice of the point $a^\pm$ and doesn't depend on $\lambda$, it doesn't affect the polynomiality in $\lambda$ and can be omitted. For the same reason we can replace in~\eqref{eq:deschirota1} the coefficients $c^\infty_{a^\pm}$ by $\pm \lambda^{-1}$.
Modulo these not important factors, the full expression~\eqref{eq:deschirota1} can be rewritten as
\begin{align} \label{HQE}
\textstyle  \frac{d\lambda}{\lambda} \Big[
\lambda^k \exp(\frac{k\psi}2) & \textstyle \exp \big( \frac 1\epsilon \sum_{\ell\geq 0} \frac 12 \frac{\lambda^{\ell+1}}{(\ell+1)!} (q^1_\ell -\bar q^1_\ell)
- \frac 1\epsilon \sum_{\ell\geq 1}  \frac{\lambda^\ell}{\ell!} \ch(\ell)  (q^2_\ell-\bar q^2_\ell) \big)\times 
\\ \notag
& \textstyle 
\cD\big(\big\{q^1_\ell - \epsilon \frac{\ell!}{\lambda^{\ell+1}}\big\}_{\ell \geq 0},\
 q^2_0-\frac{\epsilon}{2} -\sum_{\ell\geq 1}\frac{\lambda^{\ell}}{\ell!}q^2_\ell,\ \big\{
 q^2_\ell\big\}_{\ell\geq 1}\big) \times 
\\ \notag
& \textstyle 
\cD\big(\big\{\bar q^1_\ell + \epsilon \frac{\ell!}{\lambda^{\ell+1}}\big\}_{\ell \geq 0},\
\bar q^2_0+\frac{\epsilon}{2} -\sum_{\ell\geq 1}\frac{\lambda^{\ell}}{\ell!}\bar q^2_\ell,\ \big\{
\bar q^2_\ell\big\}_{\ell\geq 1}\big)
\\ \notag
\textstyle 
-\lambda^{-k} \exp(-\frac{k\psi}2) 
& \textstyle 
\exp \big( -\frac 1\epsilon \sum_{\ell\geq 0} \frac 12 \frac{\lambda^{\ell+1}}{(\ell+1)!} (q^1_\ell -\bar q^1_\ell)
+\frac 1\epsilon \sum_{\ell\geq 1}  \frac{\lambda^\ell}{\ell!} \ch(\ell)  (q^2_\ell-\bar q^2_\ell) \big)\times
\\ \notag
& \textstyle 
\cD\big(\big\{q^1_\ell + \epsilon \frac{\ell!}{\lambda^{\ell+1}}\big\}_{\ell \geq 0},\
q^2_0+\frac{\epsilon}{2} -\sum_{\ell\geq 1}\frac{\lambda^{\ell}}{\ell!}q^2_\ell,\ \big\{
q^2_\ell\big\}_{\ell\geq 1}\big) \times 
\\ \notag
& \textstyle 
\cD\big(\big\{\bar q^1_\ell - \epsilon \frac{\ell!}{\lambda^{\ell+1}}\big\}_{\ell \geq 0},\
\bar q^2_0-\frac{\epsilon}{2} -\sum_{\ell\geq 1}\frac{\lambda^{\ell}}{\ell!}\bar q^2_\ell,\ \big\{
\bar q^2_\ell\big\}_{\ell\geq 1}\big)
\Big]\Big|_{q^2_0-\bar q^2_0 = k\epsilon} \,.
\end{align}
This expression is polynomial in $\lambda$ if and only if the residues of its products with $\lambda^n$, $n\geq 0$, at $\lambda=\infty$ are equal to zero, which completes the proof of the proposition.
\end{proof}

\section{The Lax formulation of the Catalan hierarchy} \label{sec:LaxFormulation}
In this section we would like to find a Lax representation for the integrable hierarchy associated with the descendent HQE. Such hierarchy has as natural spatial variable the time $X=q^1_0$. Proceeding directly as in the rational reduction of KP is quite straightforward as long as we don't consider the equations for the ``logarithmic'' times $q_\ell^2$. Noticing that the descendent HQEs are identical to those of the Extended Toda hierarchy~\cite{cdz, cvdl13} under exchange of the two sets of times, we first proceed in deriving the Lax form of the equations as in~\cite{cvdl13} using as space variable the time $q_0^2 = x$, and then obtain indirectly a Lax representation in terms of pseudo-differential operators in the proper space variable $X=q^1_0$  using the approach in section $5$ of~\cite{cdz}. Finally we reconsider the Hirota equations and directly derive the Lax equations in pseudo-differential operator form.

\subsection{Lax representation with difference operators} \label{8.1}
Here we quickly repeat with slight modifications the derivation of the Lax representation of the extended Toda hierarchy following~\cite{cvdl13} in the case $N=M=1$. 

Let us consider the multivalued one form on the $\lambda$ plane
\begin{equation}
\omega_\infty := \left( c_{a^+}^\infty  \Gamma^{a^+}_{\infty} \otimes \Gamma^{-a^+}_{\infty } +c_{a^-}^\infty  \Gamma^{a^-}_{\infty} \otimes \Gamma^{-a^-}_{\infty } \right) \left( \cD \otimes \cD	\right) d\lambda \,.
\end{equation}
The Hirota equation~\eqref{eq:deschirota1} can be equivalently formulated by saying that 
\begin{equation} \notag
(\cN_\infty  \otimes \cN_\infty ) \omega_\infty
\end{equation}
is regular at $\lambda \sim \infty$. To obtain the Lax representation we have to switch to an equivalent HQE which is $\partial_x$-operator valued, as in~\cite{m}. We therefore introduce the operators
\begin{equation}  
\Gamma_\infty^{\delta \#} = e^{ x \frac{\partial }{\partial q_0^2}} 
e^{\sum_{l>0} \frac{\lambda^l}{l!} q^2_l \partial_x}, \qquad 
\Gamma_\infty^{\delta} = e^{-\sum_{l>0} \frac{\lambda^l}{l!} q^2_l \partial_x}   e^{ x \frac{\partial }{\partial q_0^2}} . 
\end{equation}
Since we have that
\begin{equation}  
\Gamma_\infty^{\delta \#} = e^{\sum_{l>0} \frac{\lambda^l}{l!} q^2_l \partial_x} e^{ x \frac{\partial }{\partial q_0^2}}  \cN_\infty,  \qquad 
\Gamma_\infty^{\delta} = e^{ x \frac{\partial }{\partial q_0^2}}  e^{-\sum_{l>0} \frac{\lambda^l}{l!} q^2_l \partial_x} \cN_\infty, 
\end{equation}
it follows that 
\begin{equation}  
(\cN_\infty \otimes \cN_\infty) \omega_\infty  \text{ regular }  \iff 
(\Gamma_\infty^{\delta \#}  \otimes \Gamma_\infty^{\delta} ) \omega_\infty \text{ regular } .
\end{equation}

Along the lines of~\cite{cvdl13} by carefully commuting the operators we can prove that
\begin{equation}  
\Gamma_\infty^{\delta \#}   \Gamma_\infty^{a^\pm} \cD = 
\cD' \cW^\pm \lambda^{\pm \frac{q^2_0 +x}\epsilon } 
e^{( \pm\frac\psi{2\epsilon} -(a_1-1) \frac{\pi \I}\epsilon ) (q^2_0 +x)}
\end{equation}
and
\begin{equation}  
\Gamma_\infty^{\delta}   \Gamma_\infty^{-a^\pm} \cD = 
\lambda^{\mp \frac{q^2_0 +x}\epsilon } 
e^{( \mp\frac\psi{2\epsilon} +(a_1-1) \frac{\pi \I}\epsilon ) (q^2_0 +x)}
\cW^{*\pm}  \cD',
\end{equation}
where we define
\begin{equation}  
\cW^\pm = \cP^\pm \exp \left({\pm \frac1{2\epsilon} \sum_{l \geq 0} \frac{\lambda^{l+1}}{(l+1)!} q^1_l  
+ \sum_{l >0} \frac{\lambda^l}{l!} (\partial_x \mp \frac{\ch_l}{\epsilon} ) q^2_l } \right)
\end{equation}
\begin{equation}  
\cW^{* \pm} =  \exp \left({\mp \frac1{2\epsilon} \sum_{l \geq 0} \frac{\lambda^{l+1}}{(l+1)!} q^1_l  
-\sum_{l >0} \frac{\lambda^l}{l!} (\partial_x \mp \frac{\ch_l}{\epsilon} ) q^2_l } \right) \cP^{* \pm}
\end{equation}
and
\begin{equation}  
\label{8.9}
\cP^\pm = \frac{e^{\widehat{(\cf_\infty^{a^\pm})_+} }\cD'}{\cD'(x-\frac{\epsilon}2)}, \qquad 
\cP^{* \pm} = \frac{e^{\widehat{(\cf_\infty^{-a^\pm})_+} }\cD'}{\cD'(x+\frac{\epsilon}2)}.
\end{equation}
Here $\cD'(q,x) = \cD(q)_{| q^2_0 \to q^2_0 + x}$.

Substituting, we find that the HQE are equivalent to the regularity of 
\begin{equation}  
\bigl[ \cW^+(q)  \cW^{* +} (\bar{q}) \lambda^{k} e^{\psi k/2 } - \cW^-(q) \cW^{*-}( \bar{q}) \lambda^{-k} e^{-\psi k /2} \bigr] \frac{d\lambda}{\lambda} 
\end{equation}
where $(q-\bar{q})^2_0 =k\epsilon$. In residue form we have 
\begin{equation}  
\res_\lambda \bigl[ \cW^+(q)  \cW^{* +} (\bar{q}) \lambda^{k} e^{\psi k/2 } - \cW^-(q) \cW^{*-}( \bar{q}) \lambda^{-k} e^{-\psi k /2} \bigr] \lambda^{n-1} d\lambda = 0 , \quad  n \geq0 .
\end{equation}

We now convert this expression in a bilinear equation for difference operators. A difference operator is defined as a Laurent series in the formal variable $\Lambda$. Multiplication of such operators, when defined, is given by $\Lambda^s f(x) =  f(x+s\epsilon) \Lambda^s$. 
Given a difference operator $A = \sum_s a_s \Lambda^s = \sum_{s}  \Lambda^s \tilde{a}_s$ the left and right symbols are respectively defined as $\sigma_l(A) = \sum_s a_s \lambda^s$ and $\sigma_r(A) = \sum_s \tilde{a}_s \lambda^s$. Recall that 
\begin{equation} \notag
\res_\lambda \sigma_l(A) \sigma_r(B) \frac{d \lambda}\lambda = \res_\Lambda AB,
\end{equation}
where $\res_\Lambda A := a_0$, for a proof see \S 3.2 in~\cite{cvdl13}.

Let us define operators $W^{\pm}$ and $W^{* \pm}$ by 
\begin{equation}  
\sigma_l(W^+)= \cW^+ , \qquad 
\sigma_r(W^{* +} ) = \cW^{* +},
\end{equation}
\begin{equation}  
\sigma_l(W^-)= \cW^-|_{\lambda \to \lambda^{-1} e^{-\psi}} , \qquad 
\sigma_r(W^{* -} ) = \cW^{* -}|_{\lambda \to \lambda^{-1} e^{-\psi}} 
\end{equation}
which implies
\begin{equation}  
W^+ = P^+ \exp \left({ \frac1{2\epsilon} \sum_{l \geq 0} \frac{\Lambda^{l+1}}{(l+1)!} q^1_l  
+ \sum_{l >0} \frac{\Lambda^l}{l!} (\partial_x - \frac{\ch_l}{\epsilon} ) q^2_l } \right),
\end{equation}
\begin{equation}  
W^{* +} =  \exp \left({ -\frac1{2\epsilon} \sum_{l \geq 0} \frac{\Lambda^{l+1}}{(l+1)!} q^1_l  
- \sum_{l >0} \frac{\Lambda^l}{l!} (\partial_x - \frac{\ch_l}{\epsilon} ) q^2_l } \right) P^{* +},
\end{equation}
\begin{equation}  
W^- =  P^- \exp \left({ -\frac1{2\epsilon} \sum_{l \geq 0} \frac{\Lambda^{-l-1} e^{-(l+1)\psi}}{(l+1)!} q^1_l  
+ \sum_{l >0} \frac{\Lambda^{-l} e^{-l\psi}}{l!} (\partial_x + \frac{\ch_l}{\epsilon} ) q^2_l } \right),
\end{equation}
\begin{equation}  
W^{* -} =  \exp \left({ +\frac1{2\epsilon} \sum_{l \geq 0} \frac{\Lambda^{-l-1} e^{-(l+1)\psi}}{(l+1)!} q^1_l  
- \sum_{l >0} \frac{\Lambda^{-l} e^{-l\psi}}{l!} (\partial_x + \frac{\ch_l}{\epsilon} ) q^2_l } \right) P^{* -}.
\end{equation}
Here the operators $P^\pm$ and $P^{* \pm}$ have been defined by 
\begin{equation}  
\sigma_l(P^+)= \cP^+, \qquad 
\sigma_r(P^{* +}) = \cP^{* +}
\end{equation}
\begin{equation}  
\sigma_l(P^-)= \cP^-|_{\lambda \to \lambda^{-1} e^{-\psi}} , \qquad 
\sigma_r(P^{* -} ) = \cP^{* -}|_{\lambda \to \lambda^{-1} e^{-\psi}} .
\end{equation}
Note that $P^+$ and $P^{* +}$ are power series in negative powers of $\Lambda$ with leading term equal to $1$, while $P^-$ and $P^{* -}$ are power series in positive powers of $\Lambda$. 
We have that
\begin{equation}  
\res_\Lambda [ W^+(q) \Lambda^{n} W^{*+} (\bar{q}) \Lambda^{-k} ] = \res_\Lambda [ W^-(q) \Lambda^{-n} e^{-n\psi} W^{*-} (\bar{q}) \Lambda^{-k} ] .
\end{equation}
Notice that here $q^2_0=\bar{q}^2_0$ .
Since this holds for $k\in\Z$ and there is no $k$ dependence in the square bracket we finally find
\begin{equation}   \label{WWeq}
W^+(q) \Lambda^{n} W^{*+} (\bar{q})= W^-(q) \Lambda^{-n} e^{-n \psi} W^{*-} (\bar{q}) .
\end{equation}

For $q=\bar{q}$ we get
\begin{equation}  
P^+ \Lambda^n P^{* +} = P^- \Lambda^{-n} e^{-n \psi} P^{* -} , 
\end{equation}
which implies for $n=0$ that $P^{*\pm} = (P^\pm)^{-1}$, consequently for $n=1$ we obtain the constraint 
\begin{equation}  \label{dress-eth}
P^+ \Lambda (P^+)^{-1} = P^- \Lambda^{-1} e^{-\psi} (P^-)^{-1} =: L 
\end{equation}
where $L$ is a difference operator of the form $L= \Lambda+ v + e^u \Lambda^{-1}$.

We can easily express the coefficients in the Lax operator in terms of the total descendent potential as
\begin{equation}  
\label{v}
v = (\Lambda^{1 /2} - \Lambda^{-1 /2} ) \epsilon \frac{\partial \log \cD'}{\partial q^1_0}  ,
\end{equation}
\begin{equation}  
u= \Lambda^{-1 /2}  (\Lambda + \Lambda^{-1} -2 ) \log \cD' - \psi .
\end{equation}

Let us now obtain the Sato equations by differentiating with respect to $q^i_l$ the bilinear equation~\eqref{WWeq} and setting $\bar{q}=q$. We obtain
\begin{equation}  
\label{Sato-shift}
\epsilon\frac{\partial P^\pm}{\partial q^i_l}  = \mp ( A^i_l)_\mp P^\pm
\end{equation}
where 
\begin{equation}  
A^1_l = \frac{L^{l+1}}{(l+1)!}, \qquad 
A^2_l = 2 \frac{L^l}{l!} \left( \log L - \ch(l) \right) .
\end{equation}
The logarithm of $L$ is defined, following~\cite{cdz}, as 
\begin{equation}  \label{deflog}
\log  L = \sum_{k\in\mathbb{Z}}w_k \Lambda^k= \frac\epsilon2  \left( P^-_x (P^-)^{-1} - P^+_x (P^+)^{-1} \right) .
\end{equation}

\begin{remark}
The dressing~\eqref{dress-eth} of $L$ by $P^+$ defines an injective map
\begin{equation}
\C[v,e^u][v_k, u_k; k\geq 0][[\epsilon]]_0  \to \C[p_i; i\geq1][p_{i,k}; i,k\geq1][[\epsilon]]_0
\end{equation}
by the substitutions 
$v \mapsto p_1 - p_1(x+\epsilon)$ and 
$e^u \mapsto p_2 - p_2(x+\epsilon) - p_1 (p_1 - p_1(x+\epsilon))$, where $p_i$ are the coefficients in $P^+$. 
The subscript denotes the degree zero homogeneous part of the formal power series ring, where the degree is $k$ for $u_k$, $v_k$ and $p_{i,k}$, is $-1$ for $\epsilon$ and zero for the remaining generators. 
It is clear that the coefficients in $L^p = P^+ \Lambda^p (P^+)^{-1}$ are elements of $\C[v,e^u][v_k, u_k; k\geq 0][[\epsilon]]_0$, or equivalently in its image via the above injection. 
It was proved in~\cite{cdz} that the coefficients $w_k$ defined by
\begin{equation}
P^+ \epsilon\partial_x (P^+)^{-1} = \epsilon \partial_x - \epsilon P_x^+ (P^+)^{-1} =  \epsilon \partial_x  + 2 \sum_{k\leq-1}  w_k \Lambda^k 
\end{equation}
are also in the image of such injection, so they define unique elements $w_k$ in $\C[v,e^u][v_k, u_k; k\geq 0][[\epsilon]]_0$. 
One can prove that the coefficients $w_k$ for $k\geq0$ defined by
\begin{equation}
-P^- \epsilon\partial_x (P^-)^{-1} =-\epsilon \partial_x +\epsilon P^-_x (P^-)^{-1} =-\epsilon \partial_x + 2 \sum_{k\geq 0} w_k \Lambda^k 
\end{equation}
as elements of $\C[q_0, q_0^{-1}, q_i; i\geq1][q_{i,k}; i\geq0, k\geq1][[\epsilon]]_0$ are given by 
\begin{equation} \label{w0wk}
w_0=\frac{\psi}2+\frac\epsilon 2\Lambda(\Lambda-1)^{-1} \frac{\partial u}{\partial x} =\frac{\psi}2+ \frac{u}2 +\dots , \qquad 
w_k = e^{-u(x+\epsilon)} \cdots e^{-u(x+\epsilon k)} w_{-k}(x+\epsilon k)
\end{equation}
for $k\geq1$.
We have therefore that the coefficients $w_k$ of $\log L$ are uniquely differential polynomials in 
\begin{equation}
\cA := \C[u, v,e^{\pm u}][v_k, u_k; k\geq 0][[\epsilon]]_0.
\end{equation}
\end{remark}

\begin{remark}
To prove~\eqref{w0wk} we define, as in ~\cite{m}, a linear anti-involution $\sigma$ on the space of difference operators by
\begin{equation}
\sigma(a(x)\Lambda^k)=Q(e^\psi \Lambda)^{-k}a(x) Q^{-1}, \quad \mbox{where } Q= \frac{ \cD'(x+\frac{\epsilon}2)}{\cD'(x-\frac{\epsilon}2)},
\end{equation}
is the coefficient of $\Lambda^0$ in $P^-$. We extend this anti-involution on the space of formal operators in both $\Lambda$ and $\epsilon \partial_x$, by
\begin{equation}
\sigma(\epsilon\partial_x)=Q(-\epsilon\partial_x-\psi)Q^{-1}=-\epsilon\partial_x-\psi +\epsilon Q^{-1}\frac{\partial Q}{\partial x} .
\end{equation}
Note that from \eqref{8.9} it follows that
\begin{equation}
\cP^- = Q \cP^{* +}, \qquad 
\cP^+ = Q \cP^{* -}, 
\end{equation}
from which it is easy to derive that 
\begin{equation}
\sigma(P^\pm)=Q(P^{\mp})^{-1},\quad 
\sigma(L)=L
\end{equation}
and
\begin{equation}
\sigma(P^\pm \epsilon \partial_x (P^\pm)^{-1})=- P^\mp \epsilon\partial_x (P^\mp)^{-1}- \psi .
\end{equation}
Thus $\sigma(\log L)=\log L$ and therefore
\begin{equation}\label{wk-k}
w_{-k}   = e^{-k \psi} \frac{Q(x)w_{k}(x-\epsilon k)}{Q(x-\epsilon k)} .
\end{equation}
From the relations 
\begin{equation}
\label{w0}
w_0 =  \frac{\epsilon}2 Q^{-1} \frac{\partial Q}{\partial x}, \qquad 
e^{u+\psi} = \frac{Q}{Q(x-\epsilon)},
\end{equation}
the equations~\eqref{w0wk} easily follow. For example we have
\begin{equation}
w_{-1}=\frac\epsilon2 \left((\Lambda-1)^{-1} \frac{\partial v}{\partial x}\right), \qquad 
w_1=\frac\epsilon2 \left(\Lambda e^{-u}(\Lambda-1)^{-1} \frac{\partial v}{\partial x}\right).
\end{equation}
\end{remark}

Finally, it follows from Sato equations and from the commutativity of $L$ and $A^i_l$ that $L$ satisfies the Lax equations: 
\begin{equation}  \label{laxxx} 
\epsilon \frac{\partial L}{\partial q^i_l} = [ (A^i_l)_+ , L ] .
\end{equation}
They define commuting derivations of $\cA$.

\subsection{Lax representation via change of variable}
Let us introduce a variable $X$ by the shift $q_0^1 \mapsto q_0^1 + X$. The Lax equations for the time $q_0^1$ read:
\begin{equation}
\epsilon v_X = (\Lambda-1) e^u, \qquad 
\epsilon u_X = (1- \Lambda^{-1}) v.
\end{equation}
Denoting $\phi = v(x-\epsilon)$ and $\rho = e^u$, notice that we can express the $x$ derivatives of $u$ and $v$ as $X$ differential polynomials in $u$, $v$ or equivalently in $\rho$, $\phi$. By substitution in the evolutionary equations implied by~\eqref{laxxx} we obtain the desired hierarchy of equations having $X$ as space variable. A Lax representation for this hierarchy, called extended NLS, was outlined in~\cite{cdz}. Here we give an equivalent presentation using dressing operators. 

Denote by $\tilde{P}$ the pseudo-differential operator in the variable $X$ obtained by substituting $\Lambda$ by $\epsilon\partial_X$ in the dressing operator $P^+$ in the previous section, namely
\begin{equation}
P^+ = \sum_{k=-\infty}^0 p_k \Lambda^k, \qquad 
\tilde{P} = \sum_{k=-\infty}^0 p_k (\epsilon\partial_X)^k.
\end{equation}

Let 
\begin{equation}
\cL = \tilde{P} \epsilon \partial_X \tilde{P}^{-1}, \qquad 
\cS = \tilde{P}\epsilon \partial_X \Lambda^{-1} \tilde{P}^{-1}, \qquad  
\cT  = \tilde{P}(\epsilon \partial_X)^2 \Lambda^{-1} \tilde{P}^{-1}
\end{equation}
where it should be noted that the last two operators above are both pseudo-differential in the variable $X$ and difference in the variable $x$. 

We define $\log	\cL$ as in ~\cite{cdz} by
\begin{equation} \label{logdef}
\log \cL = \sum_{k \in \Z} \tilde{w}_k \Lambda^k  ((\epsilon \partial_X - \phi) \Lambda^{-1})^k 
\end{equation}
where $\tilde{w}_k$ are differential polynomials in the variables $\rho$, $\phi$ and their $X$  derivatives obtained from $w_k$ via the substitutions mentioned above.  Of course also the $x$ derivatives of $\phi$ appearing as a result of the shifts have to be expressed in terms of $X$ derivatives of the variables $\rho$ and $\phi$.

\begin{remark}
In the formula for $\log \cL$ notice we have 
\begin{align} \label{subS}
&\Lambda^k  ((\epsilon \partial_X - \phi) \Lambda^{-1})^k = 
(\epsilon \partial_X - \phi(x+\epsilon k)) \cdots 
(\epsilon \partial_X - \phi(x+\epsilon)), \quad k\geq 0 , \\
&\Lambda^{-k}  ((\epsilon \partial_X - \phi) \Lambda^{-1})^{-k} = 
(\epsilon \partial_X - \phi(x+\epsilon (-k+1)))^{-1} \cdots 
(\epsilon \partial_X - \phi(x))^{-1}, \quad k\geq 1 , 
\end{align} 
where all $x$ derivatives have to be expressed as differential polynomials in the $X$-derivatives by the substitutions above. The operator $(\epsilon \partial_X - \phi)^{-1}$ is the inverse of $\epsilon \partial_X -\phi$ in the algebra of usual pseudo-differential operators, explicitly
\begin{equation}
(\epsilon \partial_X - \phi)^{-1} = \sum_{l \geq0} (\epsilon \partial_X)^{-1} (\phi (\epsilon \partial_X)^{-1})^{l}.
\end{equation}
Notice however that the infinite sum of positive powers of $k$ in~\eqref{logdef} leads to non-convergent infinite sums in front of every positive power of $\epsilon \partial_X$.

To give a well-defined meaning to the formal operator~\eqref{logdef} we consider it as a pseudo-differential operator in $\epsilon \partial_X - \phi$. Notice that the product rule for pseudo-differential operators gives  the formally equivalent rule 
\begin{equation}
(\epsilon \partial_X - \phi)^k f = \sum_{l\geq 0} \binom{k}{l}  f^{(l)} (\epsilon \partial_X - \phi)^{k-l}.
\end{equation}
Because when written in terms of $\epsilon \partial_X - \phi$ the substitutions~\eqref{subS} now involve in each factor corrections of strictly positive degree in $\epsilon$, it follows that $\log \cL$ is a well-defined operator of the form 
\begin{equation}
\log \cL = \sum_{k\in \Z} a_k  (\epsilon \partial_X - \phi)^k
\end{equation}
where $a_k$ are differential polynomials in $\rho$, $\phi$ and their $X$-derivatives. Moreover the multiplication of such operator, which involves both infinite positive and negative powers of $\epsilon \partial_X - \phi$,  by an operator with only finite number of arbitrary powers of $\epsilon \partial_X - \phi$ (see the form of $\cL$ below) is also well-defined.
We conclude that the Lax equations below provide derivations in the ring of differential polynomials in $\rho$, $\phi$ and their $X$-derivatives.
\end{remark}

\begin{proposition}
The operator $\tilde{P}$ satisfies the following Sato equations 
\begin{equation} \label{Sato-par}
\epsilon\frac{\partial \tilde{P}}{\partial q_l^i} = -(\tilde{A}_l^i )_- \tilde{P} 
\end{equation}
where 
\begin{equation} \label{deftal}
\tilde{A}_\ell^1 = \frac{\cL^{\ell+1}}{(\ell +1 ) !} ,\qquad 
\tilde{A}_\ell^2 = \frac2{\ell !}  ( \log \cL - \ch(\ell))\cL^{\ell}.
\end{equation}
Moreover the operators $\cS$ and $\cT$ are given by
\begin{equation} \label{STeq}
\cS = (\epsilon \partial_X - \phi) \Lambda^{-1} ,\qquad 
\cT = ((\epsilon \partial_X)^2 - \epsilon \partial_X \phi + \rho) \Lambda^{-1} 
\end{equation}
so, in particular they do not contain negative powers of $\epsilon \partial_X$, and
 the Lax operator is equal to their ratio
\begin{equation}
\cL = \cT \cS^{-1} = \epsilon \partial_X + \rho (\epsilon \partial_X - \phi)^{-1}.
\end{equation}
\end{proposition}

\begin{proof}
The Sato equation~\eqref{Sato-shift} for $q_0^1$ and the dressing~\eqref{dress-eth} of $\Lambda$ can be written respectively as 
\begin{equation} \label{twoeqs}
\epsilon \frac{\partial P^+}{\partial X\ }  =  - L_- P^+  = (-e^u \Lambda^{-1}) P^+ , \qquad 
(\Lambda + v + e^u \Lambda^{-1} ) P^+ = P^+ \Lambda.
\end{equation}
Moving all shifts $\Lambda$ on the right the first equation becomes 
\begin{equation}
\epsilon\frac{\partial P^+}{\partial X} =
-e^u P^+(x-\epsilon) \Lambda^{-1} .
\end{equation}
Notice that in this expression all power of $\Lambda$ appear on the right of the remaining coefficients, therefore we can replace them with powers of $\epsilon \partial_X$, obtaining
\begin{equation} \label{PX1}
\epsilon\frac{\partial  \tilde{P}}{\partial X} =  -e^u \tilde{P}(x-\epsilon)  (\epsilon\partial_X)^{-1} .
\end{equation}
Similarly we can rewrite the second equation above in terms of pseudo-differential operators as 
\begin{equation} \label{PX2}
\tilde{P}(x+\epsilon) \epsilon\partial_X + v \tilde{P} + e^u \tilde{P}(x-\epsilon) (\epsilon\partial)^{-1} = \tilde{P} \epsilon\partial_X .
\end{equation}
Substituting~\eqref{PX1} in~\eqref{PX2} we get
\begin{equation}
\tilde{P} \epsilon \partial_X \Lambda^{-1} \tilde{P}^{-1} = \Lambda^{-1} (\epsilon\partial_X -v) = (\epsilon\partial_X -\phi)\Lambda^{-1} 
\end{equation}
and similarly from~\eqref{PX2} we find
\begin{equation}
\tilde{P} (\epsilon\partial_X)^2 \Lambda^{-1} \tilde{P}^{-1} = (e^u + \epsilon\partial_X (\epsilon\partial_X -\phi) ) \Lambda^{-1} ,
\end{equation}
thus proving~\eqref{STeq}.
A similar computation shows that Sato equations~\eqref{Sato-shift} can be rewritten in terms of pseudo-differential operators as~\eqref{Sato-par} where the operators $\tilde{A}_\ell^i$ are obtained by substituting $\Lambda^k$ with $\Lambda^k \cS^k$ as follows
\begin{equation} \label{deftas}
A_\ell^i = \sum_k a_{\ell, k}^i \Lambda^k   \mapsto 
\tilde{A}_\ell^i = \sum_k a_{\ell, k}^i \Lambda^k \cS^k .
\end{equation}
This follows from the observation that Sato equations are written as 
\begin{equation}
\epsilon \frac{\partial P^+}{\partial q_\ell^i} =- \sum_{k<0} a_{\ell, k}^i \Lambda^k P^+ = -\sum_{k<0} a_{\ell, k}^i P^+(x+\epsilon k) \Lambda^k
\end{equation}
so imply
\begin{equation}
\epsilon \frac{\partial \tilde{P}}{\partial q_\ell^i} =- \sum_{k<0} a_{\ell, k}^i \Lambda^k \tilde{P}(x-\epsilon k) (\epsilon\partial_X)^k  = -\sum_{k<0} a_{\ell, k}^i \Lambda^k \tilde{P} \Lambda^{-k}(\epsilon\partial_X)^k.
\end{equation}
Multiplication on the right by $\tilde{P}^{-1}$ gives 
\begin{equation}
\epsilon \frac{\partial \tilde{P}}{\partial q_\ell^i}\tilde{P}^{-1} = -
\sum_{k<0} a_{\ell, k}^i \Lambda^k \tilde{P} \Lambda^{-k}(\epsilon\partial_X)^k \tilde{P}^{-1} =- \sum_{k<0} a_{\ell, k}^i \Lambda^k \cS^k,
\end{equation}
which proves our assertion. Notice  moreover that the projection $(\cdot)_-$ commutes with the substitution  $\Lambda^k$ $\mapsto$ $\Lambda^k \cS^k$. 

To complete the proof we need to show that the operators $\tilde{A}_\ell^i$ defined by the substitution~\eqref{deftas} coincide with those give by formulas~\eqref{deftal}. Let us first consider the case $i=1$. From the dressing we know 
\begin{equation}
L^\ell P^+ = P^+ \Lambda^\ell.
\end{equation}
Denoting
\begin{equation}
L^\ell = \sum_k b^\ell_k \Lambda^k
\end{equation}
we have
\begin{equation}
\sum_k b_k^\ell \Lambda^k P^+ = P^+ \Lambda^\ell ,
\end{equation}
which, by the same argument above, gives
\begin{equation}
\sum_k b_k^\ell \Lambda^k \tilde{P} \Lambda^{-k} (\epsilon\partial_X)^k = \tilde{P} (\epsilon\partial_X)^\ell
\end{equation}
and finally 
\begin{equation}
\sum_k b_k^\ell \Lambda^k\cS^k = \cL^\ell .
\end{equation}

For the case $i=2$ we have to use a slightly different argument because we don't have a definition of $\log \cL$ in terms of dressing operators in the pseudo-differential operator case. 
The definition~\eqref{logdef} of $\log \cL$ amounts to substituting $\Lambda^k$ with $\Lambda^k \cS^k$ in $\log L$ for ETH as given in~\eqref{deflog}. We have that the product $\log L \cdot L^\ell$ after such substitution is equal to
\begin{equation}
\sum_j w_j \Lambda^j \sum_k b_k^\ell \Lambda^k \cS^{k+j}
= \sum_j w_j \Lambda^j \left(\sum_k b_k^\ell \Lambda^k \cS^{k} \right) \cS^j
=\sum_j w_j \Lambda^j   \cL^\ell  \cS^j.
\end{equation}
Because $\cS$ and $\cL$ commute this expression equals $\log \cL \cdot \cL^\ell$ which completes the proof. 
\end{proof}

\begin{corollary}
The Lax operator $\cL$ satisfies the following equations
\begin{equation}
\epsilon\frac{\partial \cL}{\partial q_\ell^i} = [-(\tilde{A}_\ell^i)_- , \cL ]  
\end{equation}
for $i=1,2$, $\ell \geq0$.
\end{corollary}

\begin{remark}
Clearly the operators $\cS$ and $\cT$ satisfy the same Lax equations as $\cL$. 
\end{remark}

\subsection{Lax representation from the Hirota equations}
In this section we want to approach the problem of deriving the Lax equations from the HQE directly, namely using a fundamental lemma to convert it into equations for pseudo-differential operators, like in the usual derivation of rational reductions of KP. Obtaining the Sato equations for the new times $q_\ell^2$ is not straightforward.

\subsubsection{Preliminaries}
Let us define
\begin{equation}
P(X,x,q,\lambda) := {\cP^+}_{{\Big|}\substack{x\to x+\frac\epsilon2 \\ q_0^1 \to q_0^1 + X}} ,
\qquad 
\tilde{P}(X,x,q,\lambda) := {\cP^{*+}}_{{\Big|}\substack{x\to x-\frac\epsilon2 \\ q_0^1 \to q_0^1 + X}},
\end{equation}
or, more explicitly 
\begin{equation} \label{defPP}
P(X,x,q,\lambda) = \frac{ e^{- \epsilon\sum_{\ell\geq0} \frac{\ell !}{\lambda^{\ell+1}}\frac{\partial }{\partial q_\ell^1}  }\cD''}{\cD''} ,
\qquad 
\tilde{P}(X,x,q,\lambda) = \frac{ e^{ \epsilon\sum_{\ell\geq0} \frac{\ell !}{\lambda^{\ell+1}}\frac{\partial }{\partial q_\ell^1}  }\cD''}{\cD''}
\end{equation}
where $\cD''$ equals the total descendent potential $\cD$ with dependences on $X$ and $x$ introduced via the shifts $q_0^1 \mapsto q_0^1 + X$, $q^2_0 \mapsto q^2_0 + x$. 
Notice that for consistency with the usual KP reductions here the symbols $P(\lambda)$ resp. $\tilde{P}(\lambda)$ are shifted by $\pm \epsilon\slash 2$ w.r.t $\cP^+$ resp. $\cP^{*+}$. 
Let us also define
\begin{gather} \label{defQ}
Q(X,x,q) = \frac{{\cD''}_{|x\to x+\epsilon}}{\cD''}, \qquad 
R(X,x,q) = Q(X, x-\epsilon, q)^{-1} , \\
\phi(X,x,q) = -\epsilon R(X,x,q)^{-1} \frac{\partial R(X,x,q)}{\partial X} , \\ 
\rho(X,x,q) = e^{-\psi} Q(X,x,q) R(X,x,q).
\end{gather}

We now rewrite the HQE ~\eqref{HQE} in an equivalent form which is more convenient for the use of the fundamental lemma for pseudo-differential operators. 

\begin{lemma}
The HQE~\eqref{HQE} is equivalent to 
\begin{multline}
\Big[ 
\exp \Big( \frac 1\epsilon \sum_{\ell\geq 0} \frac{\lambda^{\ell+1}}{(\ell+1)!} (q^1_\ell -\bar q^1_\ell)
- \frac 2\epsilon \sum_{\ell\geq 1}  \frac{\lambda^\ell}{\ell!} \ch(\ell)  (q^2_\ell-\bar q^2_\ell) \Big)
\times \\
 \qquad \times P(X,q,\lambda) \exp\Big(  \sum_{\ell\geq 1}  \frac{\lambda^\ell}{\ell!} (q^2_\ell-\bar q^2_\ell) \partial_x   \Big)
e^{-(k-1)\epsilon \partial_x}
\lambda^k  \tilde{P}(\bar{X},\bar{q},\lambda  ) e^{-\epsilon \partial_x}
e^{\frac{\lambda}{\epsilon}(X-\bar{X})} \Big]_{\leq0} =  \\
=e^{-k \psi} \Big[ Q(X,q) e^{\epsilon \partial_x} \tilde{P}(X,q,\lambda) 
\exp\Big(  \sum_{\ell\geq 1}  \frac{\lambda^\ell}{\ell!} (q^2_\ell-\bar q^2_\ell) \partial_x   \Big) \times \\
\times e^{-(k+1)\epsilon \partial_x}
\lambda^{-k}  \tilde{P}(\bar{X},\bar{q},\lambda  ) Q^{-1}(\bar{X},\bar{q})
 \Big]_{\leq0} .
\end{multline}
\end{lemma}

\begin{remark}
Note that in this formula: (a) we have $q_0^2 = \bar{q}_0^2$ and $P$, $\tilde{P}$ and $Q$ depend on $x$ via the shift $x \to x+ \epsilon$ in~\eqref{defPP} and~\eqref{defQ}; 
(b) the projections refer to the powers of $\lambda$, so we have the equality of the coefficients of $\lambda^n$ in the right-hand side and left-hand side for $n\leq0$; 
(c) the expression has values in power series in $\partial_x$, however the dependence on $\partial_x$ can be removed by right multiplication by 
\begin{equation}
\exp\Big( - \sum_{\ell\geq 1}  \frac{\lambda^\ell}{\ell!} (q^2_\ell-\bar q^2_\ell) \partial_x   \Big)
e^{k\epsilon \partial_x}
\end{equation}
which depends only on nonnegative powers of $\lambda$ so preserves the equality.
\end{remark}

\begin{proof}
Starting from~\eqref{HQE}, notice that we can multiply it by  
\begin{equation}
\exp(\frac{-k\psi}2)  
\exp \big( \frac 1\epsilon \sum_{\ell\geq 0} \frac 12 \frac{\lambda^{\ell+1}}{(\ell+1)!} (q^1_\ell -\bar q^1_\ell)
- \frac 1\epsilon \sum_{\ell\geq 1}  \frac{\lambda^\ell}{\ell!} \ch(\ell)  (q^2_\ell-\bar q^2_\ell) \big)
\end{equation} 
since it contains only nonnegative powers of $\lambda$. We can then introduce the dependence on $X$, $x$ by the shifts $q_0^1 \to q_0^1 + X$, $\bar{q}_0^1 \to \bar{q}_0^1 + \bar{X}$, $q_0^2 \to q_0^2 + x$. Rewriting the shifts of the variables appearing in $\cD$ as shift operators gives the desired formula above.
\end{proof}

\subsubsection{The Fundamental Lemma}
In the following let $P(X,\epsilon\partial_X)$, $Q(X,\epsilon\partial_X)$ be pseudo-differential operators and $P(X,\lambda)$, $Q(X, \lambda)$ the corresponding symbols, e.g. 
\begin{equation}
P(X,\epsilon\partial_X) = \sum_{k} p_k(X)(\epsilon \partial_X)^k, \qquad 
P(X, \lambda) = \sum_k p_k(X) \lambda^k.
\end{equation}
Recall that the residue $\res_{\partial_X}$ of a pseudo-differential operator coincides with the residue $\res_{\lambda=\infty} d\lambda$ of its symbol and the adjoint is defined by
\begin{equation}
P(X,\epsilon\partial_X)^* = \sum_{k} (-\epsilon \partial_X)^{k} p_k(X).
\end{equation}
\begin{lemma}
The equality holds:
\begin{equation}
\label{fl}
 \res_{\lambda=\infty}  P(X,\lambda)  Q(\bar X,-\lambda) e^{(X-\bar X)\frac\lambda\epsilon}d\lambda=\epsilon
\res_{\partial_X} \, P(X,\epsilon\partial_X) Q^*(X,\epsilon\partial_X)e^{u \partial_X}\big|_{u=X-\bar X}\,.
\end{equation}
\end{lemma}
\begin{proof}
It suffices to show that \eqref{fl}  holds for $P(X,\epsilon\partial_X)=(\epsilon\partial_X)^k$ and $Q(X,\epsilon\partial_X)= B(X)(-\epsilon\partial_X)^\ell$ for $k+\ell<0$.
In that case  we use Taylor's formula for the left-hand side, which is equal to 
\[
\begin{aligned}
\res\limits_{\lambda=\infty}d\lambda\&\lambda^{k+\ell}\sum_{m=0}^\infty &\frac{(\bar X-X)^m}{m!}\frac{\partial^m B(X)}{\partial X^m} e^{(X-\bar X)\frac\lambda\epsilon}\\
	&=\epsilon^{k+\ell+1}\sum_{m=0}^\infty \frac{ (-1)^m}{m!(-k-\ell-1)!}\frac{\partial^m B(X)}{\partial X^m}({X-\bar X})^{-k-\ell+m-1}
\end{aligned}
\]
and the right-hand side is equal to
\[
\begin{aligned}
\epsilon^{k+\ell+1}&\res_{\partial_X} \, \partial_X^{k+\ell}B(X)\sum_{n=0}^\infty \frac{(u\partial_X)^n}{n!}\big|_{u=X-\bar X}=\\
&=
\epsilon^{k+\ell+1}\sum_{m0}^\infty \begin{pmatrix} k+\ell\\ m \end{pmatrix}\frac1{(-k-\ell+m-1)!}\frac{\partial_X^m B(X)}{\partial_X X^m}(X-\bar X)^{-k-\ell+m-1}\, .
\end{aligned}
\] 
Since
\[
\frac{ (-1)^m}{m!(-k-\ell-1)!}= \begin{pmatrix} k+\ell\\ m \end{pmatrix}\frac1{(-k-\ell+m-1)!}\, ,
\]
we obtain the desired result.
\end{proof}
\begin{lemma}
If the following equality holds:
\begin{equation}
 \res\limits_{\lambda=\infty}d\lambda\, P(X,\lambda)  Q(\bar X,-\lambda) e^{(X-\bar X)\frac\lambda\epsilon}=\sum_{j} R_j(X)S_j(\bar X)\, ,
\end{equation}
then
\begin{equation}
\label{fl2}
\left(P(X,\epsilon\partial_X) Q^*(X,\epsilon\partial_X)\right)_-=\sum_j R_j(X)(\epsilon\partial_X)^{-1} S_j( X)\, .
\end{equation}
\end{lemma}
\begin{proof}
We use  Taylor's formula again and the above lemma:
\begin{multline}
\sum_{j} R_j(X)S_j(\bar X)=\sum_j R_j(X)\sum_{k=0}^\infty (-1)^k\frac{\partial^k S_j(  X)}{\partial X^k}\frac{ (X-\bar X)^k}{k!}=\\
=\epsilon
\res\limits_{\partial}\sum_j R_j(X)(\epsilon\partial)^{-1} S_j( X)e^{u\partial_X}\big|_{u=X-\bar X} \,.
\end{multline}
Thus formula \eqref{fl2} holds.
\end{proof}

\begin{corollary}
The equality
\begin{equation}
\left[ P(X, \lambda) Q(\bar{X}, -\lambda) e^{(X-\bar{X})\frac{\lambda}{\epsilon}}\right]_{\leq0} =\left[\tilde{P}(X,\lambda) \tilde{Q}(\bar{X},\lambda) \right]_{\leq0}
\end{equation}
implies the following identity of pseudo-differential operators
\begin{equation}
\left[P(X,\epsilon\partial_X) (\epsilon\partial_X)^{n-1} Q^*(X, \epsilon\partial_X)\right]_- 
= \res_\lambda  \tilde{P}(X,\lambda) \lambda^{n-1} (\epsilon\partial_X)^{-1} \tilde{Q}(X,\lambda)\, d\lambda, \quad  n\geq0.
\end{equation}
\end{corollary}

\subsubsection{The Hirota equation in pseudo-differential operator form}
Using the previous Corollary we find here a version of the HQE in terms of $\partial_x$-valued pseudo-differential operators in the variable $X$.

\begin{proposition}
The HQE~\eqref{HQE} is equivalent to
\begin{multline} \label{HQEpsi}
\Big[ 
P(q,\epsilon\partial_X) 
\exp \Big( \frac 1\epsilon \sum_{\ell\geq 0} \frac{(\epsilon\partial_X)^{\ell+1}}{(\ell+1)!} (q^1_\ell -\bar q^1_\ell)
- \frac 2\epsilon \sum_{\ell\geq 1}  \frac{(\epsilon\partial_X)^\ell}{\ell!} \ch(\ell)  (q^2_\ell-\bar q^2_\ell) \Big)
\times \\
\qquad \times \exp\Big(  \sum_{\ell\geq 1}  \frac{(\epsilon\partial_X)^\ell}{\ell!} (q^2_\ell-\bar q^2_\ell) \partial_x   \Big)
e^{-(k-1)\epsilon \partial_x}(\epsilon \partial_X)^{n+k-1} \tilde{P}(\bar{q},-\epsilon \partial_X  )^* \Big]_- = \\
 = e^{-k \psi} \res_\lambda\Big[ 
Q(q) e^{\epsilon \partial_x} \tilde{P}(q,\lambda)
  \exp\Big(  \sum_{\ell\geq 1}  \frac{\lambda^\ell}{\ell!} (q^2_\ell-\bar q^2_\ell) \partial_x   \Big)
e^{-(k+1)\epsilon \partial_x}\lambda^{n-k-1}\times  \\
\qquad \times (\epsilon \partial_X)^{-1} 
P(\bar{q},\lambda) Q^{-1}(\bar{q})e^{\epsilon \partial_x} \Big] d\lambda 
\end{multline}
for $n\geq0$ and $k\in \Z$.
\end{proposition}

\begin{remark}
Note that here $P$, $\tilde{P}$, $Q$ depend on $x$ and $X=\bar{X}$, which we have suppressed for simplicity. Moreover recall that the variables $q_0^2$ and $\bar{q}_0^2$ are identified.
\end{remark}

\subsubsection{First consequences}
For $q=\bar{q}$ the HQE becomes
\begin{multline} \label{HQEqq}
\Big[ P(\epsilon\partial_X) e^{-(k-1)\epsilon \partial_x} (\epsilon \partial_X)^{n+k-1} \tilde{P}(-\epsilon \partial_X  )^* \Big]_- = \\
= e^{-k \psi} \res_\lambda \left[ \lambda^{n-k-1}  Q \, \tilde{P}(x+\epsilon,\lambda)e^{-(k-1)\epsilon\partial_x} (\epsilon\partial_X)^{-1} 
 Q^{-1}(x-\epsilon)  P(x-\epsilon,\lambda) \right] d\lambda.
\end{multline}
For notational simplicity in this equation we have suppressed the dependence on $q$ variables and we explicitly indicated the dependence on $x$ only when this variable is shifted. 

Let us consider some consequences of this equation for small values of $k$ and $n$. Notice that the right-hand side vanishes for $k > n$. 

\noindent$\underline{n=0, \ k=1}$ :
we have 
\begin{equation}
\Big[ P(\epsilon\partial_X)  \tilde{P}(-\epsilon \partial_X  )^* \Big]_- = 0,
\end{equation}
hence $ \tilde{P}(-\epsilon \partial_X  )^*  = P(\epsilon\partial_X)^{-1}$.

\noindent$\underline{n=k\geq0}$ :
Equation~\eqref{HQEqq} becomes
\begin{equation} \label{HQEqqnk}
\Big[ P(\epsilon\partial_X) e^{-(k-1)\epsilon \partial_x} (\epsilon \partial_X)^{2k-1} P(\epsilon\partial_X)^{-1}  \Big]_- =e^{-k \psi}Q \, e^{-(k-1)\epsilon\partial_x}(\epsilon\partial_X)^{-1}  Q^{-1}(x-\epsilon).
\end{equation}

\noindent$\underline{n=0, \ k=0}$ :
let us define $S(\epsilon\partial_X, \epsilon \partial_x) := P(\epsilon \partial_X)  e^{\epsilon \partial_x} (\epsilon \partial_X)^{-1} P(\epsilon\partial_X)^{-1}$. The previous equation for $k=0$ gives
\begin{equation}
S(\epsilon\partial_X, \epsilon\partial_x)^{-1} = Q \, e^{\epsilon\partial_x}(\epsilon\partial_X)^{-1}  Q^{-1}(x-\epsilon),
\end{equation}
hence we have that 
\begin{equation}
S(\epsilon\partial_X, \epsilon\partial_x) = R^{-1} \epsilon\partial_X R e^{-\epsilon\partial_x}.
\end{equation}
Notice that $S(\epsilon\partial_X, \epsilon\partial_x) = \tilde{S}(\epsilon\partial_X) e^{-\epsilon\partial_x}$, where 
\begin{equation}
\tilde{S}(\epsilon\partial_X) = R^{-1} \epsilon\partial_X R = \epsilon \partial_X - \phi.
\end{equation}

\noindent$\underline{n= 1, \ k=1 }$ :
defining $\cL(\epsilon \partial_X) = P(\epsilon\partial_X) \epsilon \partial_X P(\epsilon\partial_X)^{-1}$, we get from Equation~\eqref{HQEqqnk} that
\begin{equation}
\Big[ \cL(\epsilon \partial_X) \Big]_- =e^{- \psi}Q \, (\epsilon\partial_X)^{-1}  Q^{-1}(x-\epsilon),
\end{equation}
therefore
\begin{align}
 \cL(\epsilon \partial_X) &= \epsilon \partial_X + Q ( e^\psi \epsilon \partial_X)^{-1} R \\
 &= \epsilon\partial_X + \rho (\epsilon\partial_X - \phi)^{-1}\\
 &=R^{-1} \big( \epsilon\partial_X + \phi + \rho (\epsilon \partial_X)^{-1} \big) R.
\end{align}

\noindent$\underline{n=1, \ k=2}$ :
 Let us define $T(\epsilon\partial_X, \epsilon \partial_x) := P(\epsilon \partial_X)  e^{-\epsilon \partial_x} (\epsilon \partial_X)^{2} P(\epsilon\partial_X)^{-1}$. Equation~\eqref{HQEqq} for $n=1$, $k=2$ gives $T(\epsilon\partial_X, \epsilon \partial_x)_-=0$. We have that $T(\epsilon\partial_X, \epsilon\partial_x) = \tilde{T}(\epsilon\partial_X) e^{-\epsilon\partial_x}$, where $\tilde{T}(\epsilon\partial_X)$ is a differential operator in the variable $X$. By definition we have 
 \begin{align}
  \cL(\epsilon \partial_X)&=T(\epsilon\partial_X, \epsilon\partial_x)  S(\epsilon\partial_X, \epsilon\partial_x)^{-1} = S(\epsilon\partial_X, \epsilon\partial_x)^{-1}T(\epsilon\partial_X, \epsilon\partial_x)\\
  &= \tilde{T}(\epsilon\partial_X) \tilde{S}(\epsilon\partial_X)^{-1} =\tilde{S}(x+\epsilon,\epsilon\partial_X)^{-1} \tilde{T}(x+\epsilon,\epsilon\partial_X).
 \end{align}
 Notice that  $S(\epsilon\partial_X, \epsilon\partial_x)$ and $T(\epsilon\partial_X, \epsilon\partial_x)$ commute, therefore commute with $\cL$, while $ \tilde{S}(\epsilon\partial_X)$ and $\tilde{T}(\epsilon\partial_X)$ don't. 
In particular we have 
\begin{align}
\tilde{T}(\epsilon\partial_X) &= R^{-1} \big( (\epsilon\partial_X)^2 + \phi \epsilon\partial_X + \rho \big) R \\
&= (\epsilon\partial_X - \phi)^{2} + \phi (\epsilon\partial_X - \phi) + \rho .
\end{align} 

\noindent$\underline{n\ge 1, \ k=1 }$ :
from \eqref{HQEqq} we deduce that
\begin{equation}
\Big[ \cL(\epsilon \partial_X) ^n\Big]_- =e^{- \psi} \res_\lambda \left[ \lambda^{n}  Q \, \tilde{P}(x+\epsilon,\lambda) (\epsilon\partial_X)^{-1} 
 Q^{-1}(x-\epsilon)  P(x-\epsilon,\lambda) \right] d\lambda.
\end{equation}

\begin{remark} Notice that all remaining constraints from~\eqref{HQEqq} for $k > n$ are automatically satisfied since the left-hand side of that equations is equal to $[ T^n S^{k-n-1} ]_-$ which vanishes since $T$ and $S$ are differential in $X$.
 \end{remark}

\subsubsection{Sato equations}
Differentiating~\eqref{HQEpsi} with respect to $q_\ell^1$ for $\ell \geq0$ and setting $q=\bar{q}$ and $k=1$, $n=0$ we get the Sato equations for the $q_\ell^1$ flows
\begin{equation} \label{sato1}
\epsilon \frac{\partial P(\epsilon\partial_X)}{\partial q_\ell^1}  = -\left(\frac{ \cL(\epsilon\partial_X)^{\ell+1}}{(\ell+1)!} \right)_- P(\epsilon \partial_X) .
\end{equation}

Differentiating~\eqref{HQEpsi} with respect to $q_\ell^2$ for $\ell>0$ and setting $q=\bar{q}$ and $k=1$, $n=0$ we get
\begin{multline} \label{satooo2}
\epsilon\frac{\partial P(\epsilon \partial_X)}{\partial q_\ell^2}
P(\epsilon \partial_X)^{-1}
= \left( \frac{{\mathcal L}(\epsilon\partial_X)^{\ell}}{\ell !}	\left(\epsilon\frac{\partial P(\epsilon \partial)}{\partial x}P(\epsilon \partial)^{-1}
+2 \ch(\ell)\right )
\right)_- +
\\ 
+ e^{-\psi} \res_{\lambda} \Big[ \frac{\lambda^{\ell-2}}{\ell!} Q \tilde{P}(x+\epsilon,\lambda) (\epsilon \partial_X)^{-1} 
\epsilon\frac{\partial  Q(x-\epsilon)^{-1} P (x-\epsilon,\lambda)}{\partial x}
\Big] d\lambda\,.
\end{multline}
Notice that two terms proportional to $\partial_x$ cancelled in this expression thanks to~\eqref{HQEqq}.

Let us define an operator $\log_+ \cL$ by dressing  $\epsilon \partial_x$
\begin{equation}
\log_+ \cL(\epsilon \partial_X) := P(\epsilon \partial_X) \epsilon \partial_x P(\epsilon\partial_X)^{-1}  =\epsilon\partial_x - \epsilon \frac{\partial P(\epsilon\partial_X)}{\partial x} P(\epsilon\partial_X)^{-1}.
\end{equation}

Notice that $\log_+ \cL$ is given by the sum of $\epsilon \partial_x$ and a pseudo-differential operator 
\begin{equation} \label{step1}
- \epsilon \frac{\partial P(\epsilon\partial_X)}{\partial x} P(\epsilon\partial_X)^{-1} =
\sum_{k\leq-1} 2\hat{w}_k (\epsilon\partial_X)^k
= \sum_{k\leq -1} 2w_k e^{k\epsilon \partial_x} S(\epsilon \partial_X)^k 
\end{equation}

Notice that in this case we don't have a second  dressing operator so we cannot directly define a second logarithm of $\cL(\epsilon \partial_X)$. To avoid this problem we proceed to define $\log \cL(\epsilon \partial_X)$ directly from the coefficients of $\log_+ \cL(\epsilon \partial_X)$. We define the coefficients $w_k$ for $k\ge 0$ as follows (cf. \eqref{wk-k}, \eqref{w0})
\begin{equation}
w_0 = \frac{ \epsilon}2 Q^{-1} \frac{\partial Q}{\partial x}, \qquad 
w_{-k}   = e^{-k \psi} Q e^{-k\epsilon \partial_x}   w_{k} Q^{-1} e^{k\epsilon \partial_x},
\end{equation}
and define
\begin{equation}
\log \cL(\epsilon\partial_X) :=  \sum_{k\leq-1} w_k e^{k\epsilon \partial_x}  S(\epsilon \partial_X)^k +  \sum_{k\geq0 } S(\epsilon \partial_X)^k w_{k}(x-\epsilon)  e^{k\epsilon \partial_x}  .
\end{equation}

\begin{remark}
Note that the first part of $\log \cL(\epsilon \partial_X)$ coincides $\log_+ \cL - \epsilon\partial_x$ and in particular with the first part of the operator  $\log \cL$  defined in the previous section. The second part is reminiscent of the second part of~\eqref{logdef} but the explicit equivalence of the two expressions could not be proved.
\end{remark}

\begin{proposition}
The operator $P$ satisfies the Sato equations
\begin{equation}
\epsilon \frac{\partial P(\epsilon\partial_X)}{\partial q_\ell^i} = -(A_\ell^i)_- P(\epsilon\partial_X)
\end{equation}
\begin{equation} \label{deftal1}
{A}_\ell^1 = \frac{\cL(\epsilon\partial_X)^{\ell+1}}{(\ell +1 ) !} ,\qquad 
{A}_\ell^2 = \frac2{\ell !} \cL(\epsilon\partial_X)^{\ell} ( \log \cL(\epsilon\partial_X) - \ch(\ell)).
\end{equation}
\end{proposition}

\begin{proof}
We just need to consider the case $i=2$. For $\ell =0$ it simply follows from~\eqref{step1}. For $\ell >0$ we need to manipulate the right hand-side of~\eqref{satooo2}. Notice that 

\begin{equation}
\epsilon \frac{\partial Q^{-1} P(\epsilon\partial_X)}{\partial x} 
= \epsilon \frac{\partial Q^{-1}}{\partial x} P(\epsilon \partial_X) 
-2 \sum_{k\leq -1} Q^{-1} w_k e^{k\epsilon \partial_x} S(\epsilon \partial_X)^k P(\epsilon\partial_X),
\end{equation}
therefore
\begin{align}
\epsilon \frac{\partial Q^{-1} P(\epsilon\partial_X)}{\partial x} 
&= -2 \sum_{k\geq0}e^{-k\psi}  e^{-k\epsilon \partial_x}  Q^{-1} w_k S(\epsilon\partial_X)^{-k}  P(\epsilon\partial_X) \\
&=  - 2\sum_{k\geq0}e^{-k\psi}  e^{-k\epsilon \partial_x}  Q^{-1} w_k 
P(\epsilon\partial_X) e^{k\epsilon \partial_x}
(\epsilon \partial_X)^{-k},
\end{align}
which implies
\begin{equation}
\epsilon \frac{\partial Q^{-1} P(\lambda)}{\partial x} 
=  - 2\sum_{k\geq0}e^{-k\psi}  e^{-k\epsilon \partial_x}  Q^{-1} w_k 
P(\lambda) e^{k\epsilon \partial_x}
\lambda^{-k}.
\end{equation}
Substituting this in  equality~\eqref{satooo2} we get that the second term on the right hand-side is equal to 
\begin{multline}
- \frac2{\ell!} \sum_{k\geq0} e^{-(k+1) \psi} \res_\lambda 
\big[ \lambda^{\ell-k-2} Q \tilde{P}(x+\epsilon,\lambda) (\epsilon \partial_X)^{-1} e^{-k\epsilon \partial_x}
 \cdot \\ \cdot Q(x-\epsilon)^{-1} P (x-\epsilon,\lambda)
\Big] d\lambda \
w_k(x-\epsilon)e^{k\epsilon \partial_x} ,
\end{multline}
which can be written using pseudo-differential operators, using the Hirota equation in the form~\eqref{HQEqq}, obtaining
\begin{equation}
-2 \big[ \frac{\cL^\ell}{\ell!} \ \sum_{k\geq0}  S(\epsilon \partial_X)^k w_k(x-\epsilon)e^{k\epsilon \partial_x}
\big]_- .
\end{equation}
The result is proved.
\end{proof}

The Lax equations follow as usual.

\appendix

\end{document}